\documentclass{aa}  
\usepackage{graphicx}
\usepackage{txfonts}
\usepackage{graphicx,longtable,lscape,stfloats,natbib,amssymb,multirow,sidecap,amssymb}
\bibpunct{(}{)}{;}{a}{}{,} % to follow the A&A style
\newcommand{\ltsima} {$\; \buildrel < \over \sim \;$}  
\newcommand{\gtsima} {$\; \buildrel > \over \sim \;$}  
\newcommand{\lta} {\lower.5ex\hbox{\ltsima}}  
\newcommand{\gta} {\lower.5ex\hbox{\gtsima}}  
\newcommand{\Ha} {H$\alpha$}  
\newcommand{\Hb} {H$\beta$}  

\newcommand{\ergs}{\>{\rm erg}\,{\rm s}^{-1}}
\newcommand{\ergscm}{\>{\rm erg}\,{\rm s}^{-1}\,{\rm cm}^{-2}}
\newcommand{\ergscmA}{\>{\rm erg}\,{\rm s}^{-1}\,{\rm cm}^{-2}\,{\rm \AA}^{-1}}
\newcommand{\kms}{$\rm{\,km \,s}^{-1}$}

\newcommand{\forb}[2]{\mbox{$[{\rm #1\, #2}]$}}
\newcommand{\oiii}{\forb{O}{III}}
\newcommand{\oi}{\forb{O}{I}}
\newcommand{\sii}{\forb{S}{II}}
\newcommand{\nii}{\forb{N}{II}}

\newcommand{\fwhm}{FWHM}
\usepackage{color}

\begin{document}

\title{The HST view of the broad line region in low luminosity AGN}
\subtitle{} \titlerunning{The HST view of the broad line region in LLAGN} 
\authorrunning{Balmaverde \& Capetti}

\author{B.~Balmaverde\inst{1}
\and A.~Capetti\inst{1}}

\institute {INAF - Osservatorio Astrofisico di Torino, Via Osservatorio 20,
  I-10025 Pino Torinese, Italy}

\offprints{balmaverde@oato.inaf.it} 

\abstract{We analyze the properties of the broad line region (BLR) in low
  luminosity AGN by using HST/STIS spectra. We consider a sample of 24 nearby
  galaxies in which the presence of a BLR has been reported from their Palomar
  ground-based spectra. Following a widely used strategy, we used the \sii\
  doublet to subtract the contribution of the narrow emission lines to the
  \Ha+\nii\ complex and to isolate the BLR emission. Significant residuals that
  suggest a BLR, are present. However, the results
  change substantially when the \oi\ doublet is used. Furthermore, the spectra
  are also reproduced well by just including a wing in the narrow \Ha\ and \nii\
  lines, thus not requiring the presence of a BLR. We conclude that complex
  structure of the narrow line region (NLR) is not captured with this approach
  and that it does not lead to general
  robust constraints on the properties of the BLR in these low luminosity AGN.\\
  Nonetheless, the existence of a BLR is firmly established in 10 objects, 5
  Seyferts, and 5 LINERs. However, the measured BLR fluxes and widths in the
  5 LINERs differ substantially with respect to the ground-based data. \\
  The BLR sizes in LINERs, which are estimated by using the virial formula from the line
  widths and the black hole mass, are clustered between $\sim$ 500 and 2,000
  Schwarzschild radii (i.e., $\sim 5 - 100$ light days). These values are
  $\sim$ 1 order of magnitude greater than the extrapolation to low
  luminosities of the relation between the BLR radius and AGN luminosity
  observed in more powerful active nuclei. We found BLR in objects with
  Eddington ratios as low as $L_{\rm bol}/L_{\rm Edd}\sim 10^{-5}$, with the
  faintest BLR having a luminosity of $\sim 10^{38}$ $\ergs$. This contrasts
  with theoretical models that predict the BLR disappearance at low luminosity. \\
  We ascribe the larger BLR radius to the lower accretion rate in LINERs when compared
  to the Seyfert, which causes the formation of an inner region dominated by an
  advection-dominated accretion flow (ADAF). The estimated BLR sizes in LINERs
  are comparable to the radius where the transition between the ADAF and the
  standard thin disk occurs due to disk evaporation. 
  We suggest that BLR clouds cannot coexist with
  the hot inner region and that they only form in the correspondence with a
  thin accretion disk.}


\maketitle
\section{Introduction}
\label{intro}

The clearest signature of an active galactic nucleus (AGN) is the presence of
broad emission lines in its spectrum, produced by dense clouds of ionized gas
located in the broad line region (BLR).  The substantial line widths and rapid
variability indicate that the BLR is located very close to the central
source. Therefore, the BLR represents a unique laboratory for exploring the
process of accretion onto supermassive black holes (SMBH). The BLR has been
intensively studied for many years, but it is far from being completely
understood. Several questions concerning, for instance, the origin of the BLR clouds,
their dynamics, and the way they are related to the overall properties of the
AGN are still waiting for an answer.

\begin{figure*}
\includegraphics[width=18cm]{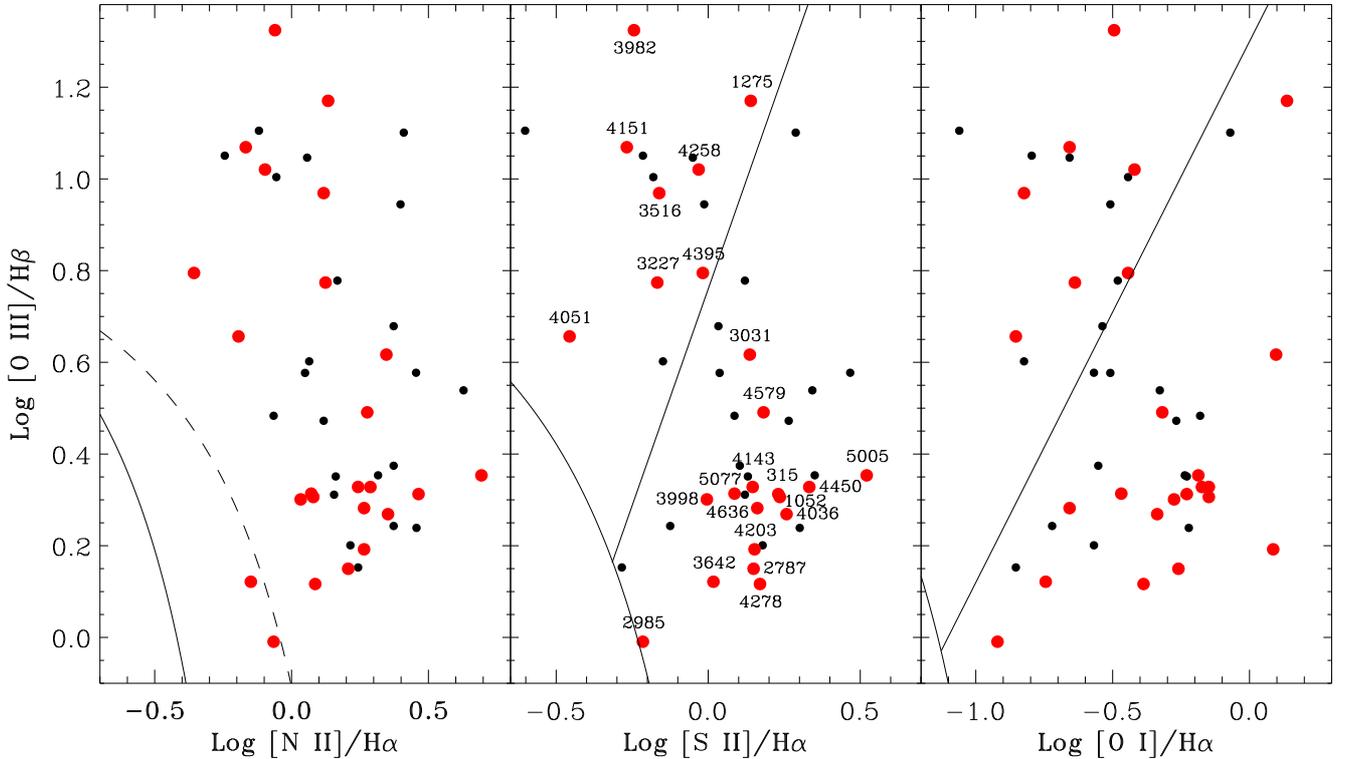}
\caption{Spectroscopic diagnostic diagrams for the galaxies of the Palomar
  sample with reported broad lines. The red (black) dots represent galaxies
  with (without) available HST spectra. The solid lines are from
  \citet{kewley06} and separate star-forming galaxies, LINER, and Seyfert; in
  the first panel, the region between the two curves is populated by the
  composite galaxies.}
\label{dd}
\end{figure*}

The BLR is spatially unresolved in even the nearest AGNs, and the only
information about its structure can be inferred from the properties of the
broad emission lines. The most powerful method of exploring the geometry and
kinematics of the BLR is the reverberation mapping technique, based on the
time lag between the changes in the broad emission lines in response to the
variation in  the ionizing radiation \citep{blandford82}. Among other results,
reverberation mapping constrains the BLR size and offers the possibility of
estimating the mass of the SMBH (e.g., \citealt{peterson00}). The relation
linking the BLR radius and the AGN luminosity (e.g., \citealt{kaspi05,bentz13})
can then be used to estimate black hole masses from single-epoch measurements
of luminosities and line profiles in all broad lined AGN, even at high
redshift \citep{vestergaard02}. Clearly, any improvement in the physical
picture of the BLR would increase confidence in this extremely powerful tool.

\begin{table*}
  \caption{Log of the HST/STIS observations.}
\begin{tabular}{l|c|c|c|l|r|r}
\hline
Name & Slit width
& Grism & Obs. Date& Obs. Id & Exp. Time [s] & Scale [\AA/pix]\\
\hline
NGC~0315  &0.1$\arcsec$ &G750M  &2000-06-18 &O5EE02060 to 70   & 500 \,\,\,\,\,\,\,\,  & 1.11 \,\,\,\,\,\,\,\,\\
NGC~1052  &0.2$\arcsec$ &G750M  &1999-01-02 &O57203050         &1974 \,\,\,\,\,\,\,\,  & 0.55 \,\,\,\,\,\,\,\,\\
NGC~1275  &0.2$\arcsec$ &G430L  &2000-08-18 &O62O05010	       & 720 \,\,\,\,\,\,\,\,  & 5.50 \,\,\,\,\,\,\,\,\\
NGC~2787  &0.2$\arcsec$ &G750M  &1998-12-05 &O4E002010 to 30   &2012 \,\,\,\,\,\,\,\,  & 0.55 \,\,\,\,\,\,\,\,\\
NGC~2985  &0.1$\arcsec$ &G750M  &2011-07-02 &OBIB02010 to 70   &5322 \,\,\,\,\,\,\,\,  & 0.55 \,\,\,\,\,\,\,\,\\
NGC~3031  &0.1$\arcsec$ &G750M  &1999-07-14 &O51301010	       &1000 \,\,\,\,\,\,\,\,  & 0.55 \,\,\,\,\,\,\,\,\\ 
NGC~3227  &0.2$\arcsec$ &G750M  &1999-01-31 &O57204040	       &1890 \,\,\,\,\,\,\,\,  & 0.55 \,\,\,\,\,\,\,\,\\ 
NGC~3516  &0.2$\arcsec$ &G750M  &2000-06-18 &O56C01020 to 40   &2116 \,\,\,\,\,\,\,\,  & 0.55 \,\,\,\,\,\,\,\,\\ 
NGC~3642  &0.2$\arcsec$ &G750M  &2000-10-13 &O5H720030 to 40   & 864 \,\,\,\,\,\,\,\,  & 0.55 \,\,\,\,\,\,\,\,\\
NGC~3982  &0.2$\arcsec$ &G750M  &1998-04-11 &O4E006010 to 30   &2997 \,\,\,\,\,\,\,\,  & 0.55 \,\,\,\,\,\,\,\,\\ 
NGC~3998  &0.1$\arcsec$ &G750M  &2002-04-07 &O6N902010 to 40   & 520 \,\,\,\,\,\,\,\,  & 1.11 \,\,\,\,\,\,\,\,\\
NGC~4036  &0.2$\arcsec$ &G750M  &1999-03-25 &O57206030         &2896 \,\,\,\,\,\,\,\,  & 0.55 \,\,\,\,\,\,\,\,\\
NGC~4051  &0.2$\arcsec$ &G750M  &2000-03-12 &O5H730030 to 40   & 864 \,\,\,\,\,\,\,\,  & 1.11 \,\,\,\,\,\,\,\,\\ 
NGC~4143  &0.2$\arcsec$ &G750M  &1999-03-20 &O4E009010 to 30   &2856 \,\,\,\,\,\,\,\,  & 0.55 \,\,\,\,\,\,\,\,\\
NGC~4151  &0.1$\arcsec$ &G430L  &1998-02-10 &O42303050         & 720 \,\,\,\,\,\,\,\,  & 2.75 \,\,\,\,\,\,\,\,\\ 
NGC~4203  &0.2$\arcsec$ &G750M  &1999-04-18 &O4E010010 to 30   &2779 \,\,\,\,\,\,\,\,  & 0.55 \,\,\,\,\,\,\,\,\\
NGC~4258  &0.2$\arcsec$ &G750M  &2001-03-16 &O67104030	       &1440 \,\,\,\,\,\,\,\,  & 0.55 \,\,\,\,\,\,\,\,\\ 
NGC~4278  &0.1$\arcsec$ &G750M  &2000-05-11 &O57207030         &3128 \,\,\,\,\,\,\,\,  & 0.55 \,\,\,\,\,\,\,\,\\
NGC~4395  &0.2$\arcsec$ &G430M  &2011-05-25 &OBGU04010 to 20   &1386 \,\,\,\,\,\,\,\,  & 0.28 \,\,\,\,\,\,\,\,\\ 
NGC~4450  &0.2$\arcsec$ &G750M  &1999-01-31 &O4E016010 to 30   &2697 \,\,\,\,\,\,\,\,  & 0.55 \,\,\,\,\,\,\,\,\\
NGC~4579  &0.2$\arcsec$ &G750M  &1999-04-21 &O57208040	       &2692 \,\,\,\,\,\,\,\,  & 0.55 \,\,\,\,\,\,\,\,\\ 
NGC~4636  &0.2$\arcsec$ &G750M  &2001-04-02 &O5L204020 to 30   &3590 \,\,\,\,\,\,\,\,  & 0.55 \,\,\,\,\,\,\,\,\\
NGC~5005  &0.2$\arcsec$ &G750M  &2000-12-24 &O5H741030         & 734 \,\,\,\,\,\,\,\,  & 0.55 \,\,\,\,\,\,\,\,\\
NGC~5077  &0.1$\arcsec$ &G750M  &1998-03-12 &O4D305020         & 418 \,\,\,\,\,\,\,\,  & 1.11 \,\,\,\,\,\,\,\,\\ 
\hline
\end{tabular}
\label{table1}
\end{table*}

The purpose of our study is to gain a deeper understanding of the BLR and of
its link to the central engine by exploring its properties in low luminosity
AGNs (LLAGNs). While most of the observational and theoretical effort has been
devoted to studying the BLR in luminous AGNs, LLAGNs can offer a different
perspective. It is indeed becoming increasingly clear that LLAGNs are not
simply scaled down versions of more powerful active galaxies, because at the lowest
accretion levels, the mechanism of black hole feeding is likely to be
substantially different. Given the strong connection between the accretion
disk and the BLR predicted by most models, we might expect that the broad
lines in LLAGNs should also be affected. For example, it has been proposed that
the BLR disappears in these objects (e.g.,
\citealt{laor03,nicastro00,elitzur06}), and it is then essential to establish
whether this is indeed the case and at which level of activity this occurs.

A significant improvement in our knowledge of the BLR properties in LLAGNs has
been achieved with the ground-based spectroscopic study of a complete sample
of 486 nearby galaxies by \citet{ho97,ho97b}, hereafter, the Palomar
survey. These data are a fundamental reservoir of measurements for statistical
studies of the nuclear properties of LLAGNs.  However, the detection and the
measurement of broad lines in the spectra of LLAGN is particularly
difficult. Nonetheless, in their survey, the presence of a BLR has been
reported in 46 objects, at various levels of confidence, of the 211 galaxies
classified as Seyferts or LINERs \citep{ho97d}. Based mainly on these results,
\citet{wang03} and \citet{zhang07} argue that the BLR size in LLAGNs is
apparently larger than what it would have been expected given their
luminosity. They suggest that this is due to a lower ionization (and/or a lower
density) of the BLR clouds in these ``dwarf'' AGNs with respect to Seyfert 1
galaxies and QSOs.

However, the BLR properties of LLAGN obtained from observations with the STIS
spectrograph onboard HST differ, often dramatically, from what is seen in
ground-based spectra (see, e.g., \citealt{ho00,shields00,
  barth01,defrancesco06}). This might have been expected since ground-based
spectra are affected by a strong stellar continuum level and by narrow lines
contamination. HST/STIS is better suited to studying faint broad emission
lines, thanks to the strongly reduced aperture size that enhances the contrast
between weak, broad emission lines against the bulge starlight and the narrow
lines.  We here analyze the available HST/STIS spectra of the galaxies for
which the presence of a broad \Ha\ emission has been reported in the Palomar
survey.

The paper is organized as follows. In Sect. \ref{sample} we define the sample
of LLAGNs studied, while in Sect. \ref{analysis} we present the analysis of the
HST spectra and the main observational results. The properties of the detected
BLR and the comparison with the ground-based results are presented in Sect.s.
\ref{blrprop} and \ref{gbcfr}. We explore the detectability of the BLR in the
objects where it is not seen in Sect. \ref{undetected}. In Sect.
\ref{scaling} we discuss whether the BLR scaling relations derived for
luminous AGN can also be applied to LLAGNs. In Sect. \ref{models} the
properties of the BLR in LLAGN are compared to various models that predict its
disappearance at low luminosities.  In Sect. \ref{summary} we provide a
summary and our conclusions.

\section{The sample and the archival observations}
\label{sample}

\begin{table*}
\caption{Parameters of the sample galaxies.}
\begin{tabular}{l|c|c|c|c|c||c|c||c|c|c|c}
\hline\hline
Name & T & D & $\sigma_{\star}$ & L$_x$ & L$_{\oiii}$&\multicolumn{2}{|c}{Ground measurements} & \multicolumn{4}{|c}{HST measurements} \\   
\hline
     &  & & & & & F(\Ha$_{b}$) &  FWHM&F(\Ha$_{n}$)& F(\Ha$_{b}$)&
     L(\Ha$_{b}$) &  FWHM \\
%  & &[Mpc]&  [km s$^{-1}$] & [erg s$^{-1}$]&[erg s$^{-1}$]&[erg cm$^{-2}$ s$^{-1}$] & [erg cm$^{-2}$ s$^{-1}$] &[erg s$^{-1}$] &[erg s$^{-1}$] & [km s$^{-1}$]& [km s$^{-1}$]
%& [erg cm$^{-2}$ s$^{-1}$] & [km s$^{-1}$]\\
(1)&(2)&(3)&(4)&(5)&(6)&(7)&(8)&(9)&(10)&(11)&(12)\\
\hline
NGC~0315   &L   & 65.8  & 303.7  &  41.63  &  39.43 & -13.85  & 2000 &  -14.95  &   $<$-14.05 &   $<$39.66  &    --  \\
NGC~1052   &L   & 17.8  & 215.0  &  41.53  &  40.10 & -12.32  & 1950 &  -13.43  &   $<$-13.72 &   $<$38.86  &    --  \\
NGC~1275   &A   & 70.1  & 258.9  &  42.86  & 41.61  &  -12.05 & 2750 &  -13.01$^c$&   $<$-13.21$^c$ &   $<$40.56$^c$  &    --  \\
NGC~2787   &L   & 13.0  & 202.0  &  38.79  &  38.37 & -13.56  & 2050 &  -14.26  &   $<$-13.47 &   $<$38.83  &    --  \\
NGC~2985   &L   & 22.4  & 140.8  &39.46$^a$&  38.69 & -13.98  & 2050 &  -14.94  &   $<$-14.22 &   $<$38.56  &    --  \\
NGC~3031   &L   &  1.4  & 161.6  &  39.38  & 37.72  &  -11.94 & 2650 &  -13.03  &      -12.08 &      38.29  &    4200\\
NGC~3227   &S   &  20.6 & 136.0  &  41.70  & 40.68  &  -11.50 & 2950 &  -13.10  &      -11.87 &      40.83  &    2934\\
NGC~3516   &S   &  38.9 & 181.0  &  42.39  & 40.80  &  -11.54 & 3850 &      --  &      -11.48 &      41.78  &    4236\\
NGC~3642   &L   & 27.5  & 85.00  &  39.84  &  38.96 & -13.59  & 1250 &  -14.52  &   $<$-13.35 &   $<$39.61  &    --  \\
NGC~3982   &S   &  17.0 &  73.0  &  38.76  & 39.83  &  -13.85 & 2150 &  -14.10  &   $<$-14.62 &   $<$37.92  &    --  \\
NGC~3998   &L   & 21.6  & 304.6  &  41.34  &  39.62 & -12.59  & 2150 &  -13.05  &      -12.48 &      40.27  &    5200\\
NGC~4036   &L   & 24.6  & 215.1  &  39.96  &  39.16 & -13.70  & 1850 &  -14.64  &   $<$-14.26 &   $<$38.60  &    --  \\
NGC~4051   &S   &  17.0 & 89.0   &  42.07  & 40.18  &  -11.77 & 1000 &      --  &      -11.91 &      40.63  &    760  \\
NGC~4143   &L   & 17.0  & 204.9  &  40.03  &  38.81 & -13.39  & 2100 &  -13.99  &   $<$-13.31 &   $<$39.23  &     -- \\ 
NGC~4151   &S   &  20.3 & 97.0   &  43.07  & 41.74  &  -10.95 & 3250 &  -12.40$^c$  &      -10.85$^c$ &      41.84$^c$  &    4465\\
NGC~4203   &L   & 9.7   & 167.0  &  39.69  &  38.53 & -13.45  & 1500 &  -13.96  &      -12.86 &      39.19  &    7200\\
NGC~4258   &S   &  6.8  & 148.0  &  40.89  & 38.76  &  -13.09 & 1700 &  -13.87  &   $<$-13.14 &   $<$38.60  &    --  \\
NGC~4278   &L   & 9.7   & 261.0  &  39.64  &  38.88 & -13.09  & 1950 &  -14.32  &   $<$-14.07 &   $<$37.98  &    --  \\
NGC~4395   &S   &  3.6  & 30.0   &  39.58  & 38.35  &  -12.88 & 442  &  -13.28$^c$  &      -12.85$^c$ &      38.34$^c$  &    786 \\
NGC~4450   &L   & 16.8  & 135.0  &  40.02  &  38.78 & -13.70  & 2300 &  -13.56  &      -12.83 &      39.70  &    7700\\
NGC~4579   &L   &  16.8 & 165.0  &  41.15  & 39.42  &  -13.12 & 2300 &  -14.52  &      -13.03 &      39.50  &    8000\\
NGC~4636   &L   & 17.0  & 202.7  &  39.38  &  38.09 & -14.16  & 2450 &  -16.32  &      --     &      --     &    --  \\
NGC~5005   &L   & 21.3  & 172.0  &  39.94  &  39.41 & -12.69  & 1650 &  -14.57  &   $<$-14.11 &   $<$38.62  &    --  \\
NGC~5077   &L   & 40.6  & 254.6  &39.74$^b$&  39.52 & -14.07  & 2300 &  -14.40  &   $<$-14.96 &   $<$38.33  &    --  \\
\hline
\end{tabular}
\label{bigtable}

\medskip
\small{ (1) Object name;
  (2) spectroscopic classification based on the diagnostic diagram
  of \citet{kewley06} (S=Seyfert, L=Liners, A=ambiguous galaxy);  (3)
  distance of the source in Mpc from \citet{ho95};  (4) stellar
  velocity dispersion in \kms\ from \citet{ho09a};  (5) Logarithm of
  the un-absorbed nuclear X-ray luminosity in the 2-10 keV range in
  $\ergs$\ from \citet{ho09}; $^b$ from \citet{gultekin12}; $^a$ we fitted
  Chandra data Obs.Id 11669 with the model
  phabs(powerlaw+mekal), obtaining $N_{\rm H}$=2.2e20 cm$^{-2}$ freezed,
  $\Gamma$=1.8, kT=1.1 keV, Cstat=10.68(9 d.o.f.);
  (6) Logarithm of the nuclear [OIII] emission line luminosity  in $\ergs$\ from  \citet{ho97};
  (7)  Logarithm of the flux of the broad \Ha\ line in $\ergscm$\ from Palomar measurement;
  (8) \fwhm\ in \kms\ from the Palomar measurement (\citealt{ho97}).
  (9) Logarithm of the flux of the narrow \Ha\ emission line in $\ergscm$\ from this work; for the objects marked with $^c$ the 
  \Ha\ luminosities and fluxes are derived from the observed \Hb\ line.
  (10) Logarithm of the flux of the broad \Ha\ emission line in $\ergscm$;
  (11) Logarithm of the luminosity of the broad \Ha\ emission line in $\ergs$;
  (12) \fwhm\ in \kms; \\
  Note: The upper limit measurement for the flux of broad \Ha\ emission line reported 
  depends on the assumption about the broad emission line profile (we here assume a gaussian profile with FWHM reported in Tab.\ref{tab4b}).}
\end{table*}
\medskip

The Palomar survey consists of optical spectroscopic observations, performed
with the Palomar 5 m Hale telescope, of 486 bright (B$_T$ $\leq$ 12.5 mag)
galaxies, located in the northern sky \citep{filippenko85,ho95}. In these
ground-based spectra, 46 galaxies show definite or probable evidence of broad
\Ha\ emission \citep{ho97d}. We searched the Hubble Legacy Archive (HLA) for
HST/STIS spectra of these sources covering the \Ha\ line\footnote{The spectral resolution of the G750L grism is insufficient for our purposes
  so we selected only data obtained with the medium-resolution grism
  G750M.} or, when not available, the \Hb\ region. We found data for 24
galaxies, as reported in Table \ref{table1}. 

When available, we combined multiple observations to remove cosmic rays and
bad pixels.  From the fully calibrated data we extracted the nuclear spectrum
from a synthetic aperture of 0\farcs15\ and applied the proper aperture
corrections.

\begin{figure}
\includegraphics[width=9cm]{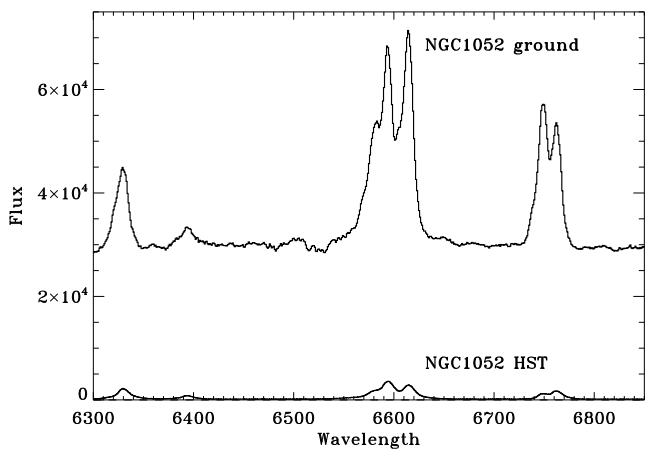}
\caption{Comparison of HST and Palomar spectra of NGC~1052. The fluxes are in
  unit of 10$^{-18}$$\ergscmA$, while wavelengths are in \AA.}
\label{ngc1052cfr}
\end{figure}

The HST aperture is thus 0\farcs2 (or even 0\farcs1)$\times$0\farcs15,
which is significantly smaller than the 2$\arcsec\times$4$\arcsec$ aperture of the
Palomar survey. This reduces the contamination of starlight and of narrow
lines, favoring the detection of any broad line component. The dramatic change
in our view of LLAGN when using HST spectra is described well by the
comparison with the corresponding ground based spectrum (see
Fig. \ref{ngc1052cfr}).

We explored the location of these sources in the spectroscopic diagnostic
diagrams (see Fig. \ref{dd}) by using the ground-based emission line
ratios. According to the criteria given by \citet{kewley06}, 16 of them are
LINERs, 7 Seyfert, and one (namely NGC~1275) is an ambiguous source, since it
moves from the Seyfert region in the middle panel to the LINERs region, in
the righthand panel (see Table \ref{bigtable}).

\section{Analysis of the spectra}
\label{analysis}

\begin{figure*}
\centerline{
\includegraphics[scale=0.37,angle=0]{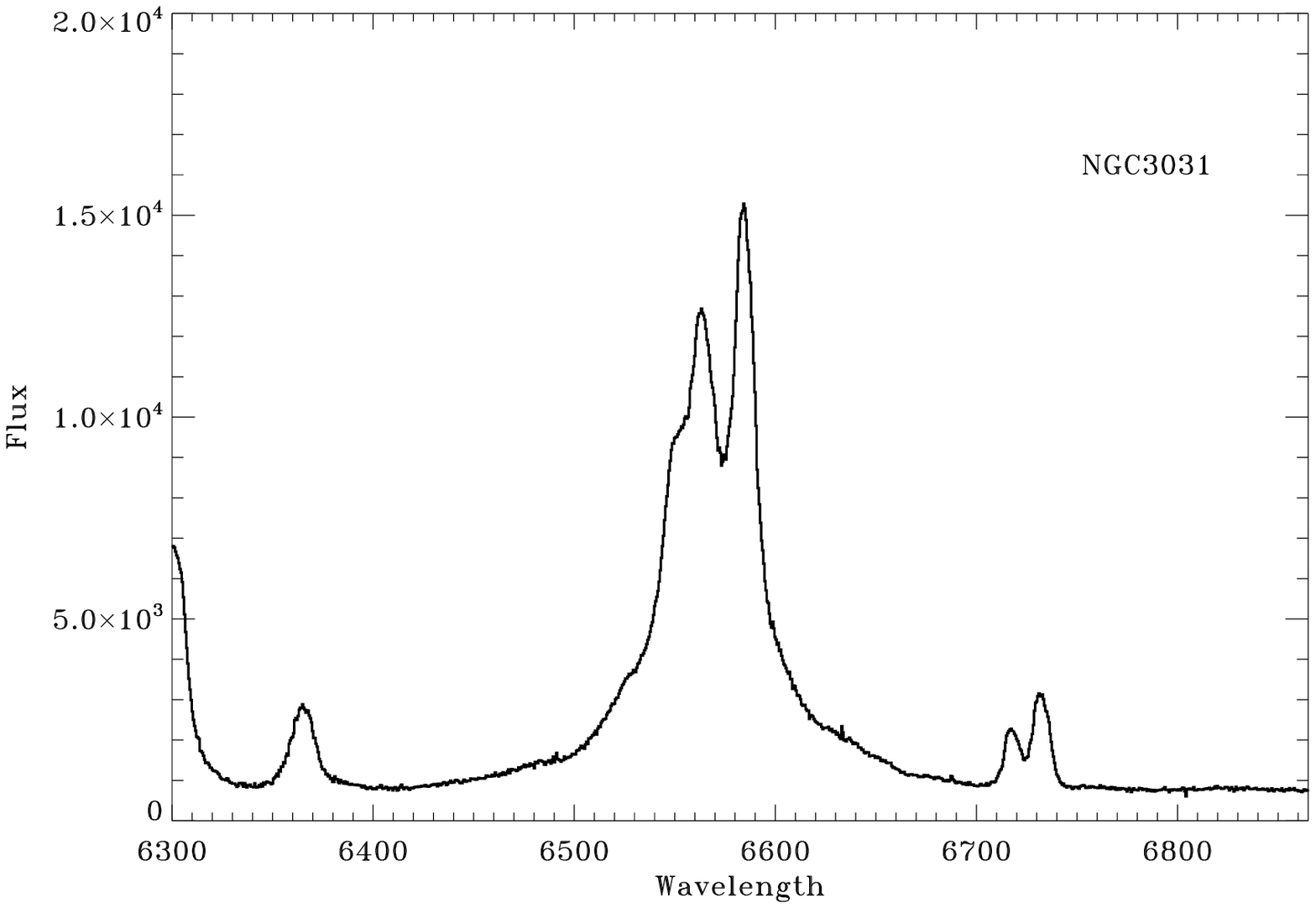}
\includegraphics[scale=0.37,angle=0]{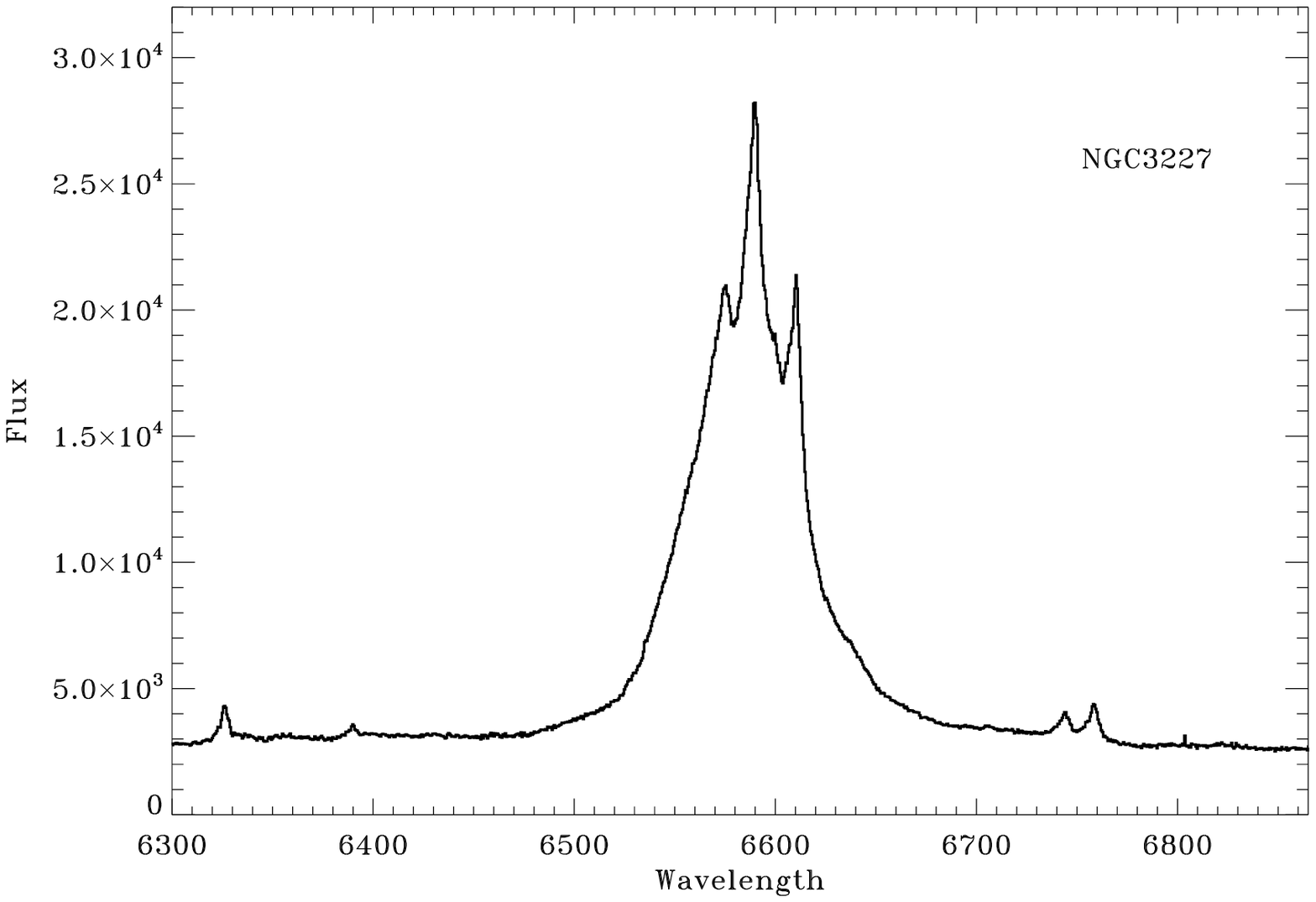}
}
\centerline{
\includegraphics[scale=0.37,angle=0]{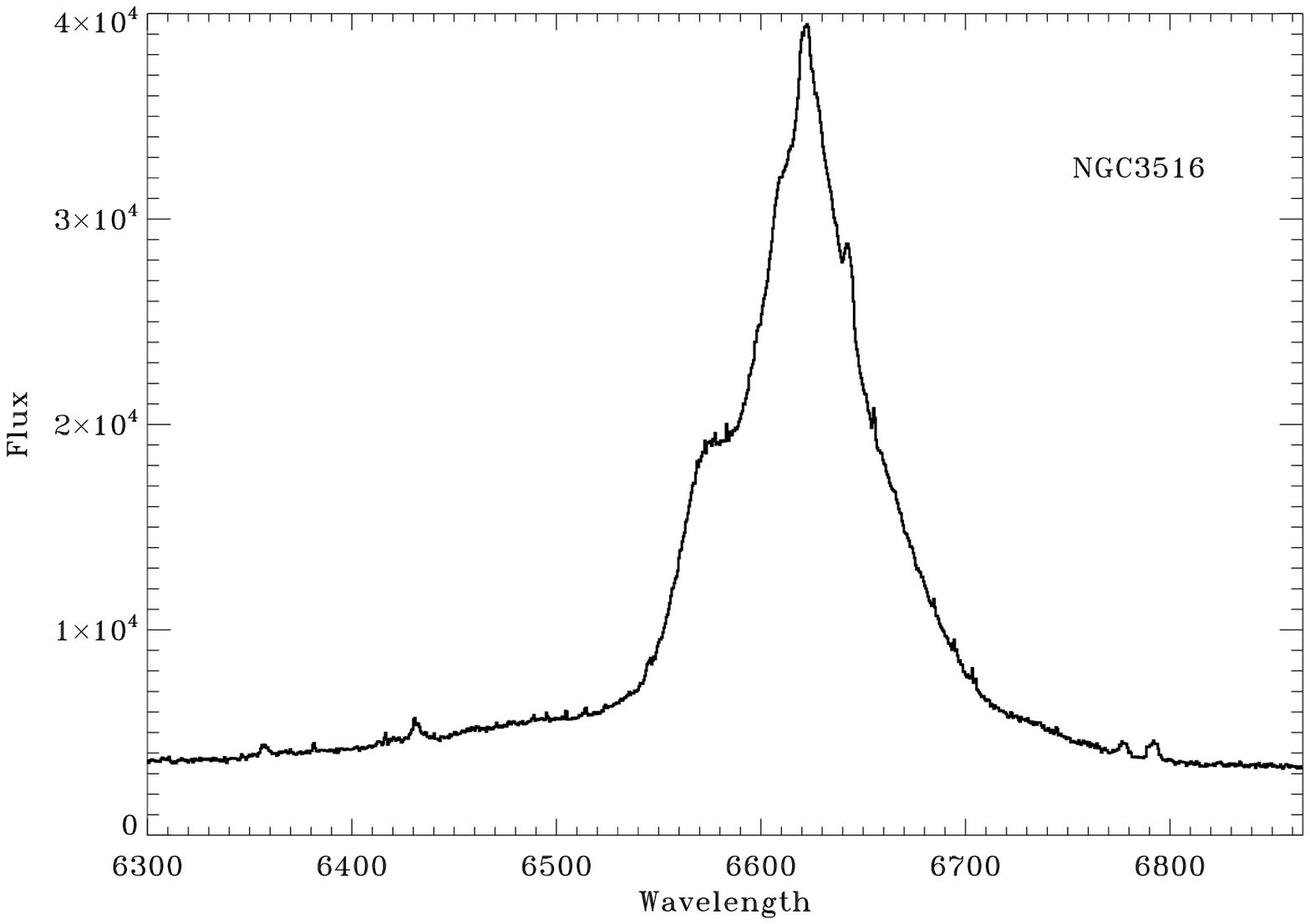}
\includegraphics[scale=0.37,angle=0]{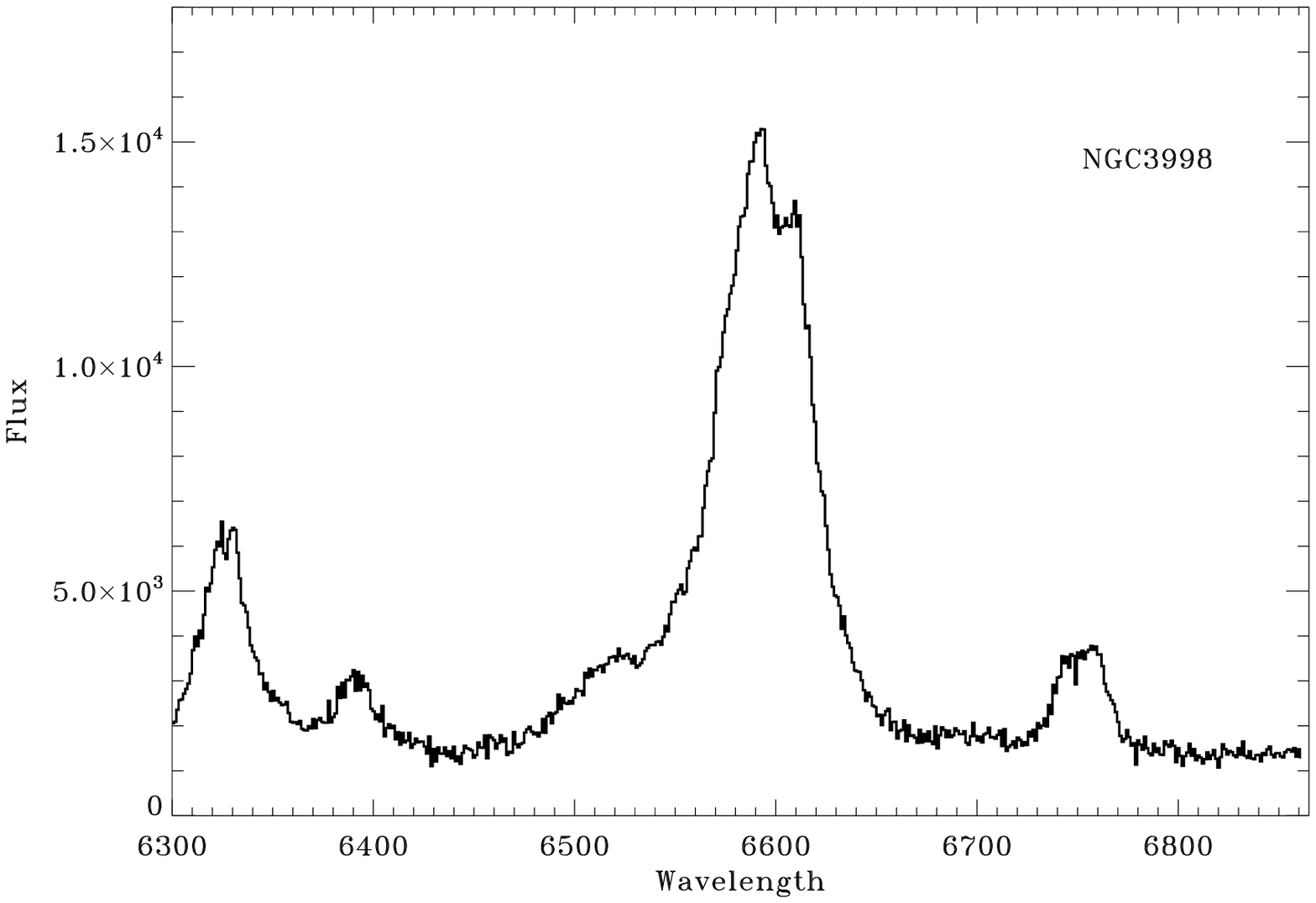}
}
\centerline{
\includegraphics[scale=0.37,angle=0]{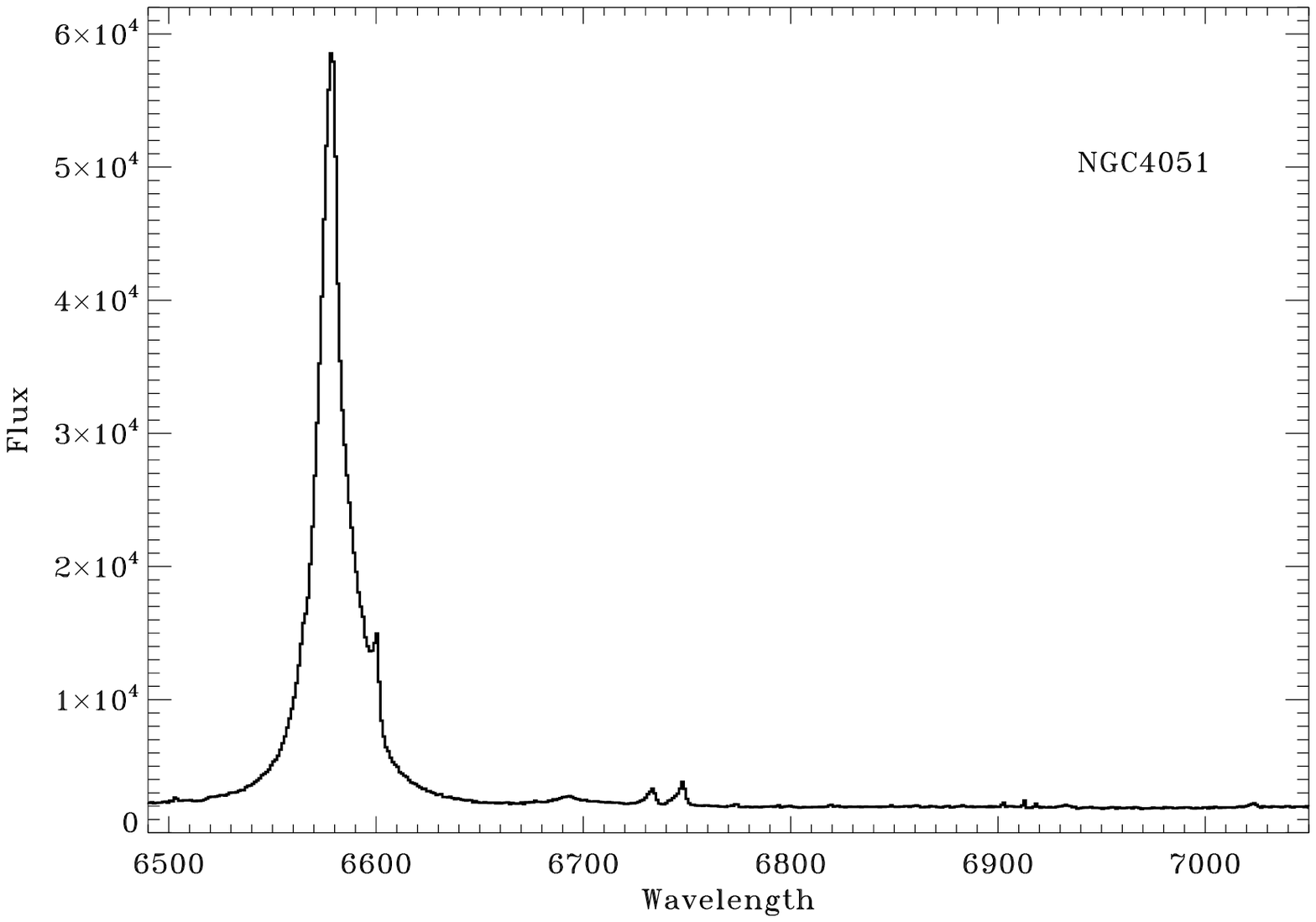}
\includegraphics[scale=0.37,angle=0]{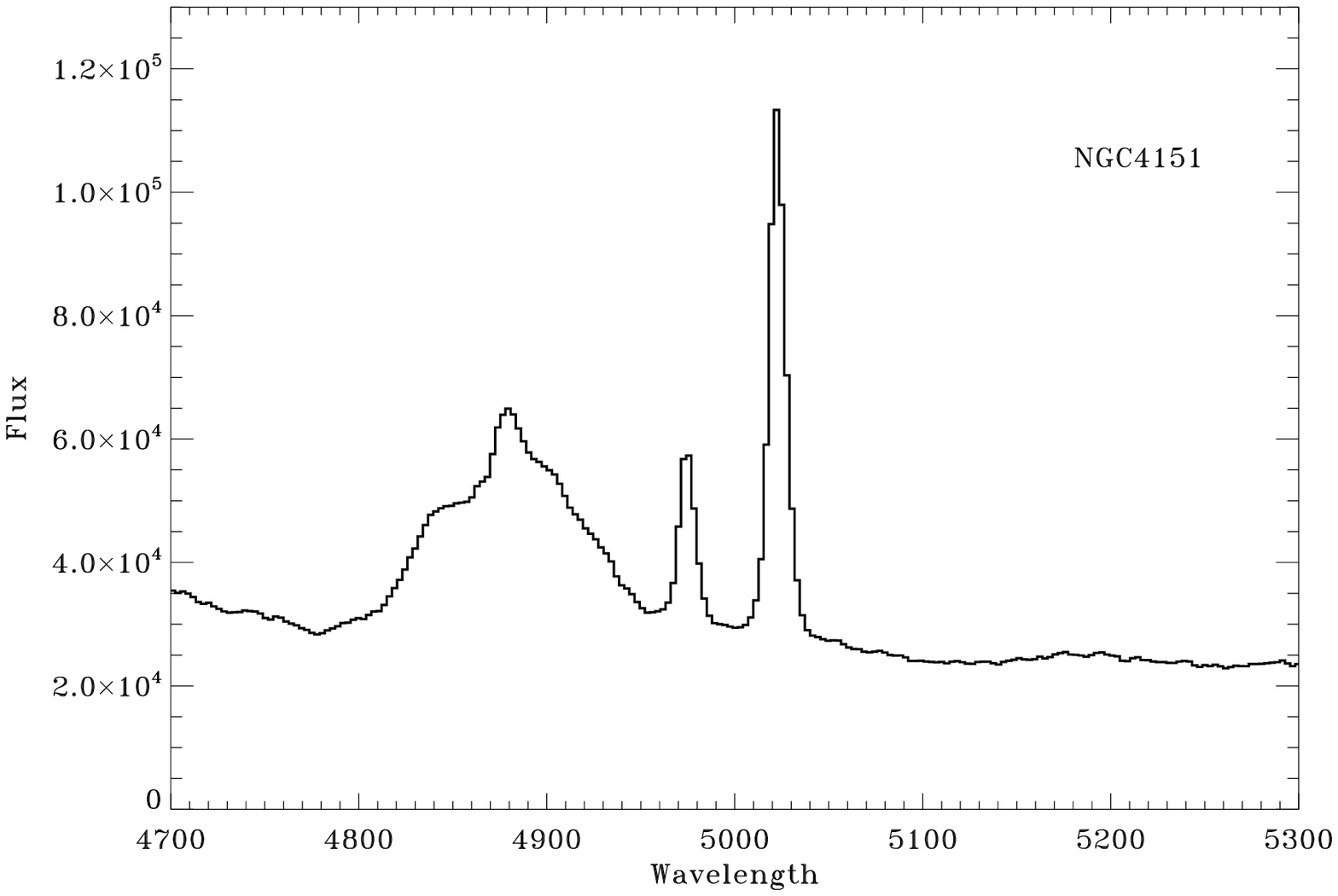}
}
\centerline{
\includegraphics[scale=0.37,angle=0]{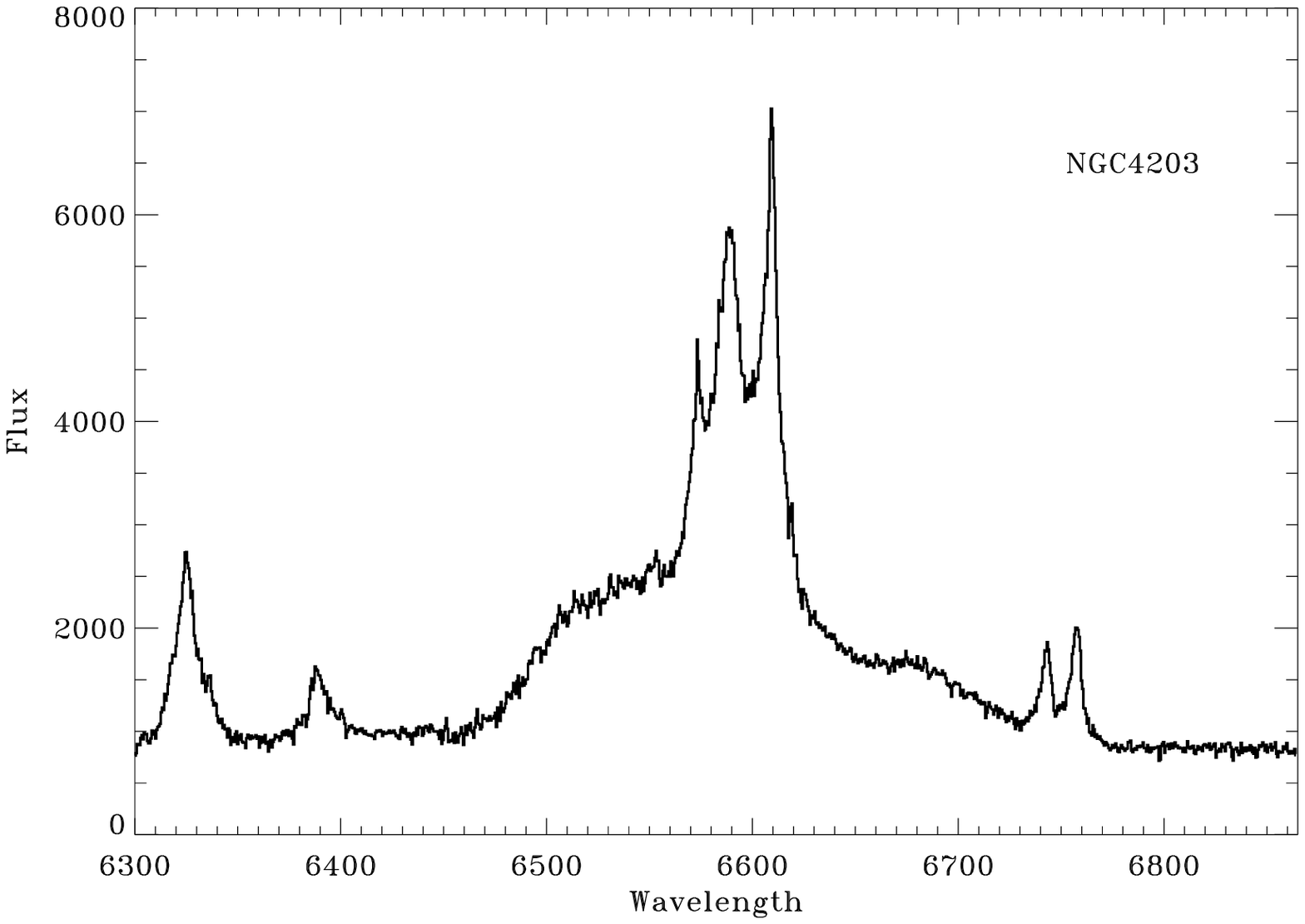}
\includegraphics[scale=0.37,angle=0]{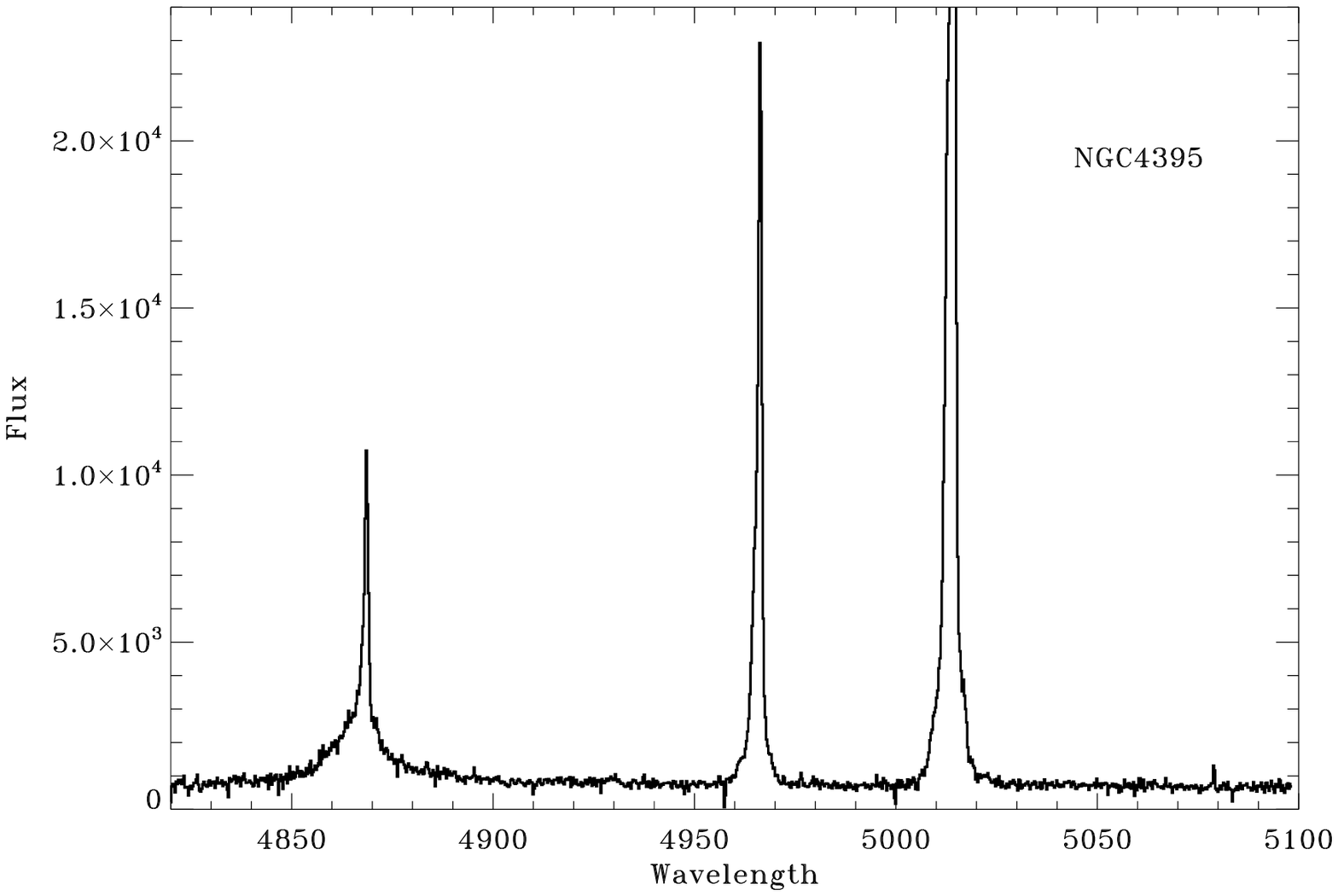}
}
\centerline{
\includegraphics[scale=0.37,angle=0]{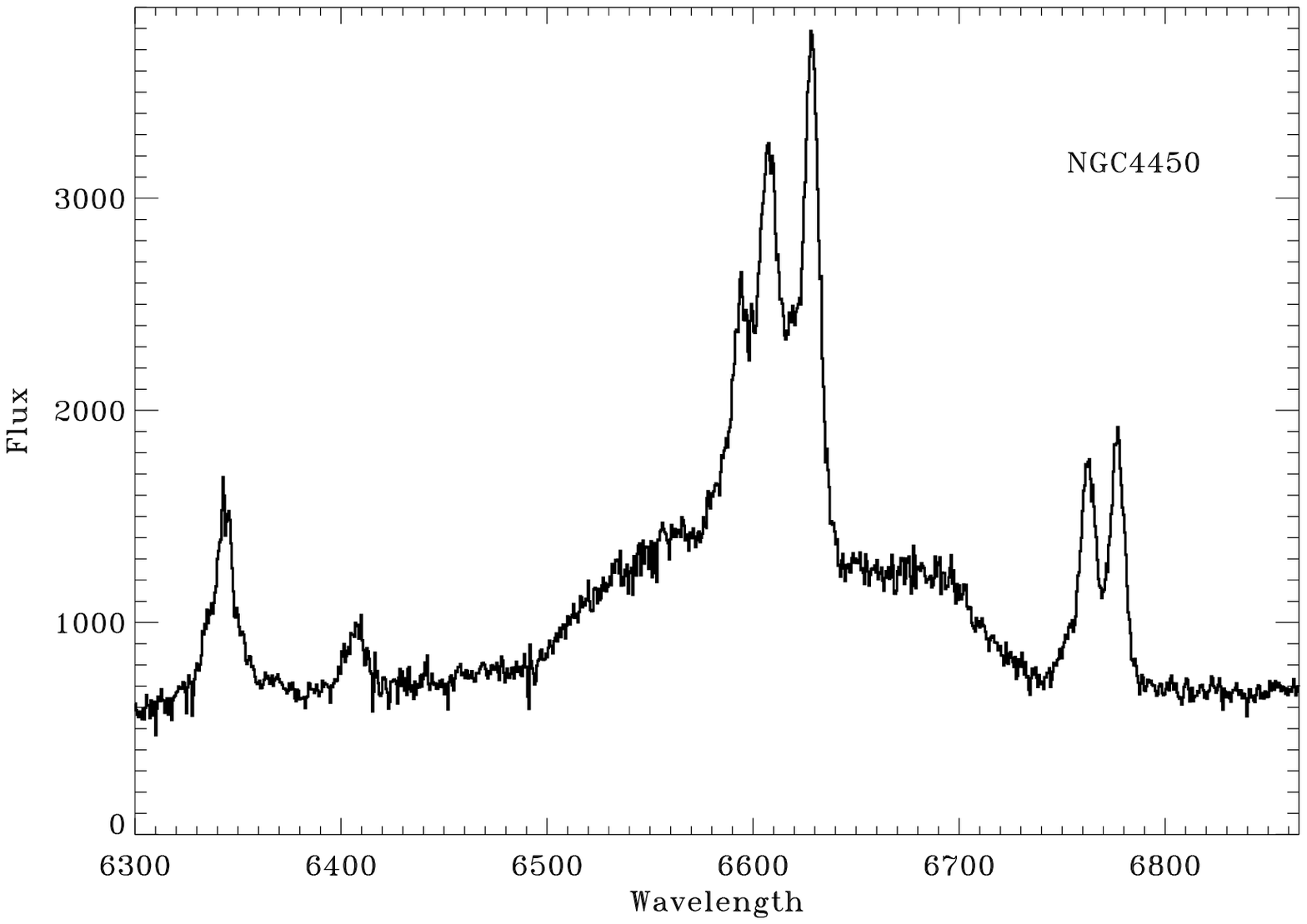}
\includegraphics[scale=0.37,angle=0]{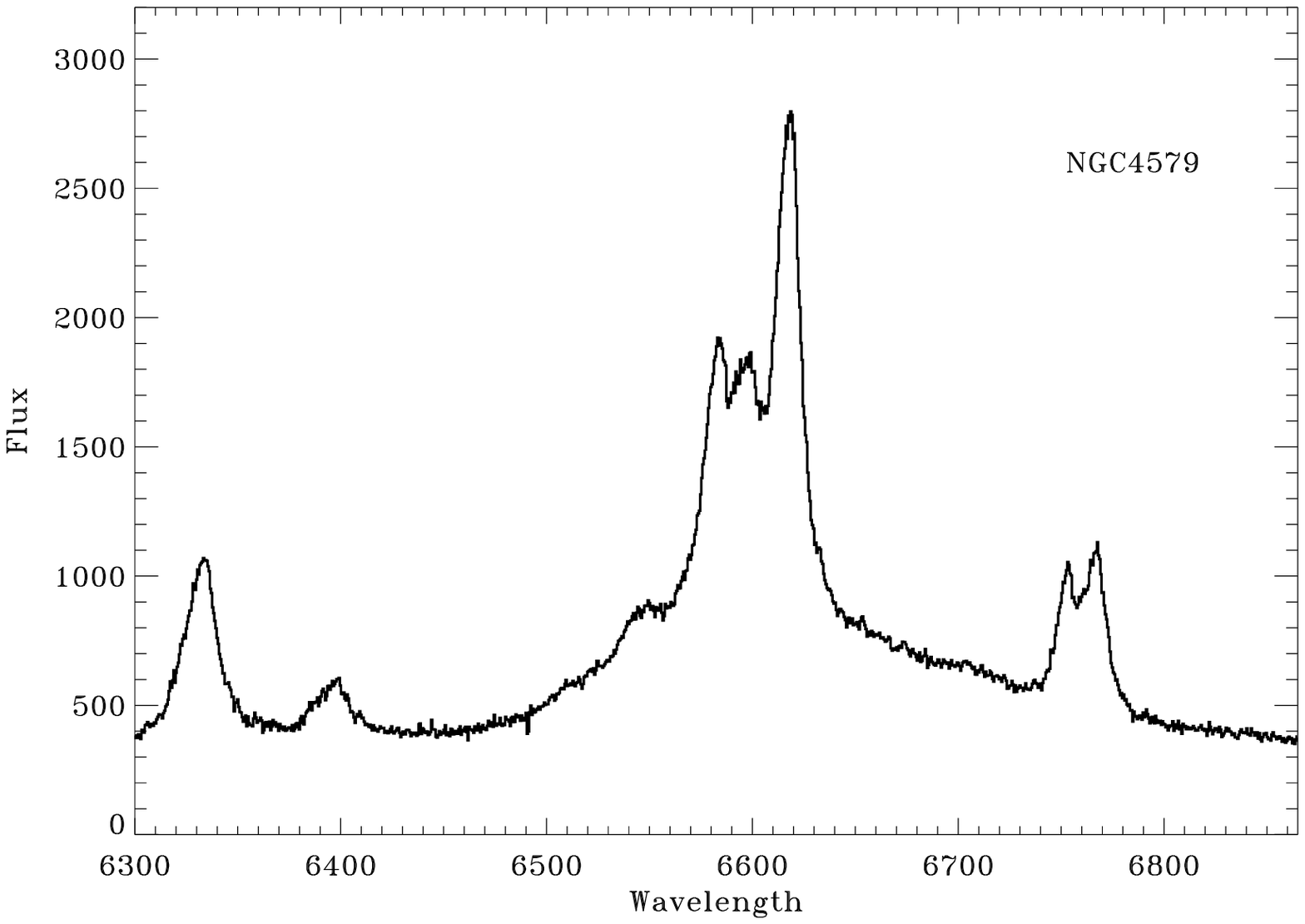}
}
\caption{Spectra of the objects in which a broad Balmer line is readily
  visible with just a visual inspection.}
\label{clearblr}
\end{figure*}

In this section we describe the analysis of the spectra used to assess
the presence of any broad component in the Balmer lines. In ten cases,
a BLR is readily visible with just a visual inspection of the
spectra. This is the case for all Seyfert galaxies (with the exception
of NGC~3982 and NGC~4258, which is probably not a true Seyfert, as we discuss in
Sect. 8) and of five LINERs.  Their spectra are shown in
Fig. \ref{clearblr} and are analyzed in more detail toward the end
of this section.

The spectra of the remaining sources require a more accurate analysis before
we can conclude whether a BLR is present and, in this case, derive its
properties. The method we adopted is based on the assumption that the lines of
the \sii\ doublet provide a good representation of the shape of the \nii\ and
of the \Ha\ narrow line component, similar to the method used by
\citet{ho97d}. The \sii\ lines are reproduced with a Gaussian profile but,
when needed, two Gaussians for each line are used.

\begin{figure*}
\centerline{
\includegraphics[scale=0.5,angle=0]{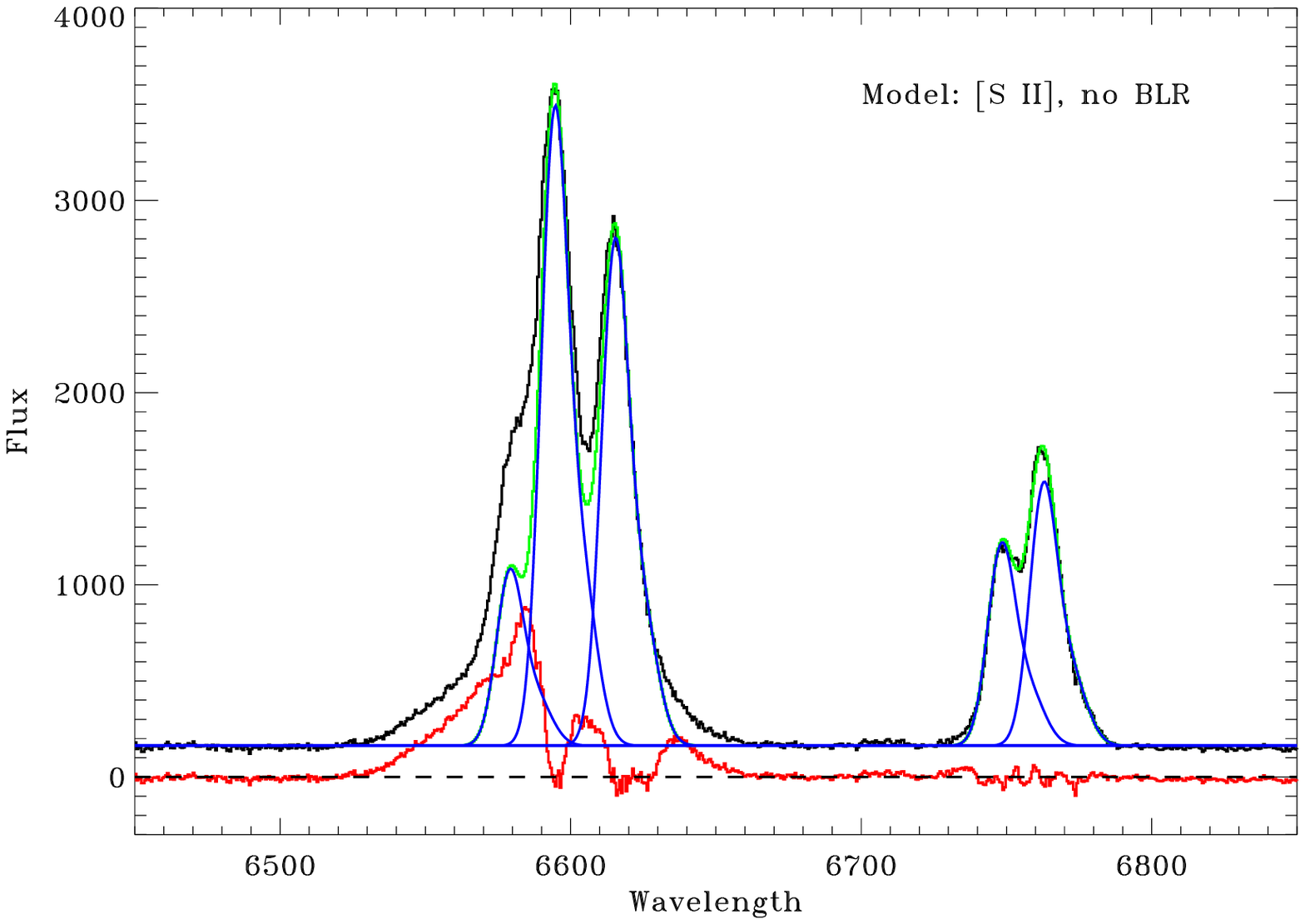}
\includegraphics[scale=0.5,angle=0]{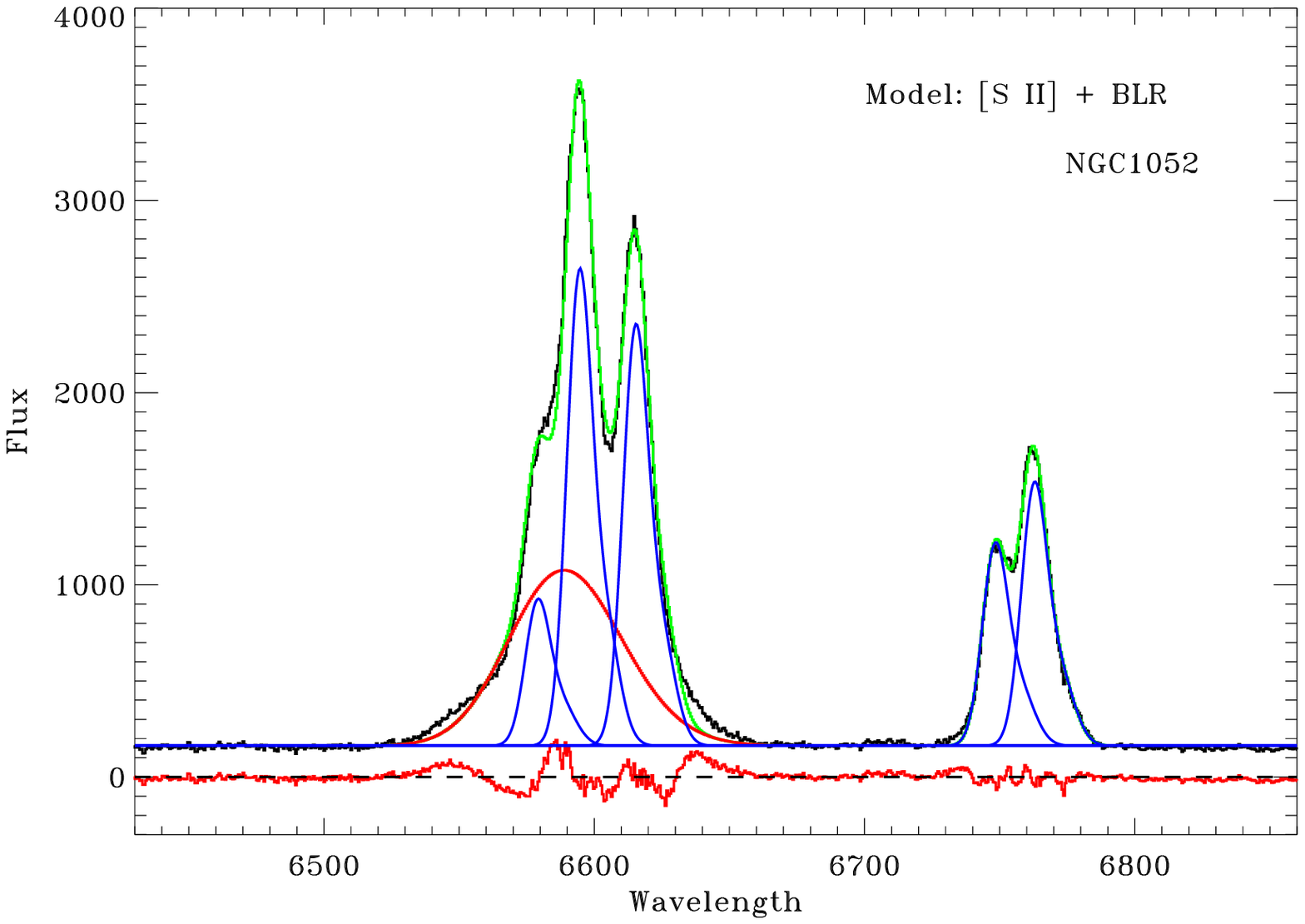}
}
\centerline{
\includegraphics[scale=0.5,angle=0]{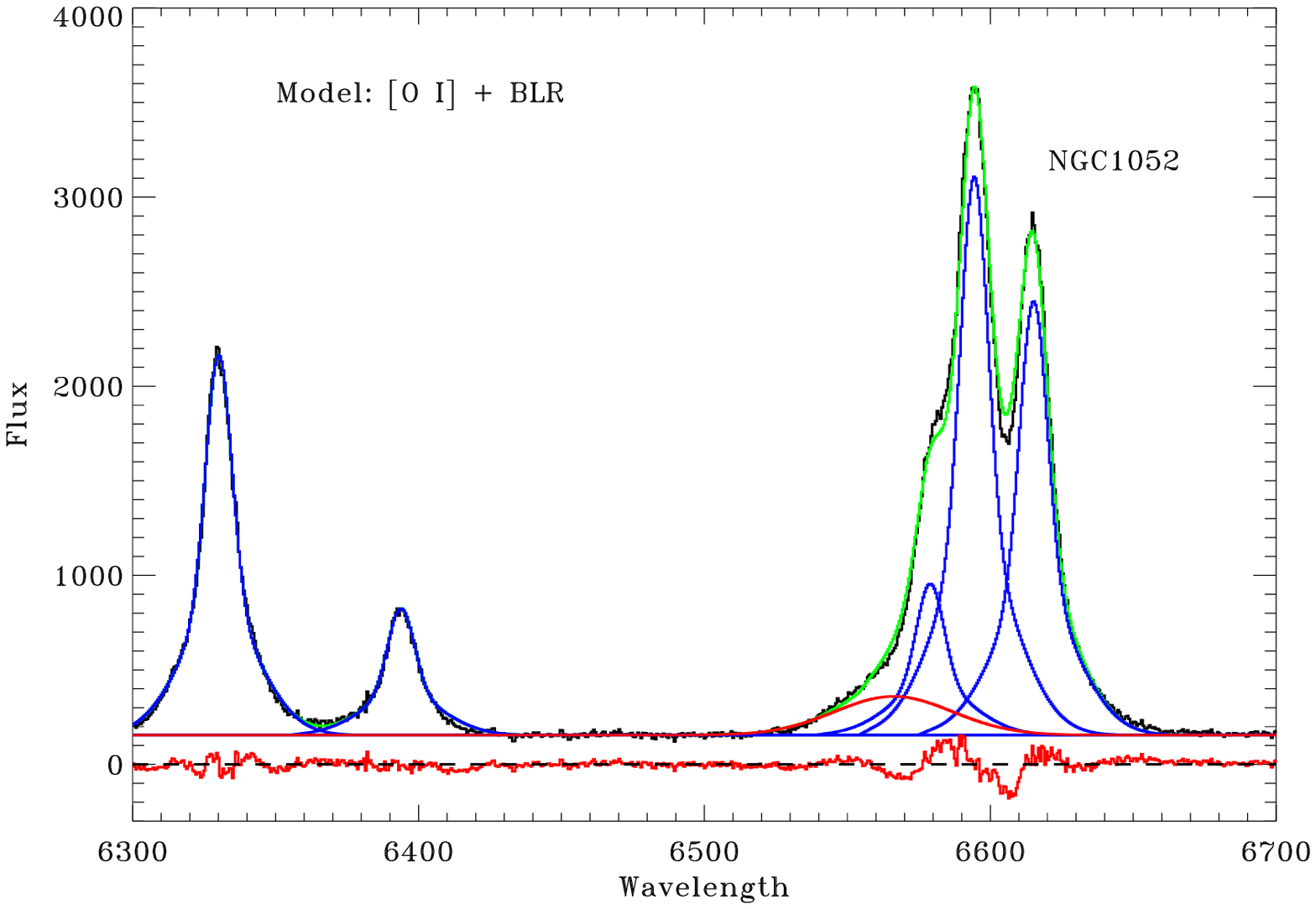}
\includegraphics[scale=0.5,angle=0]{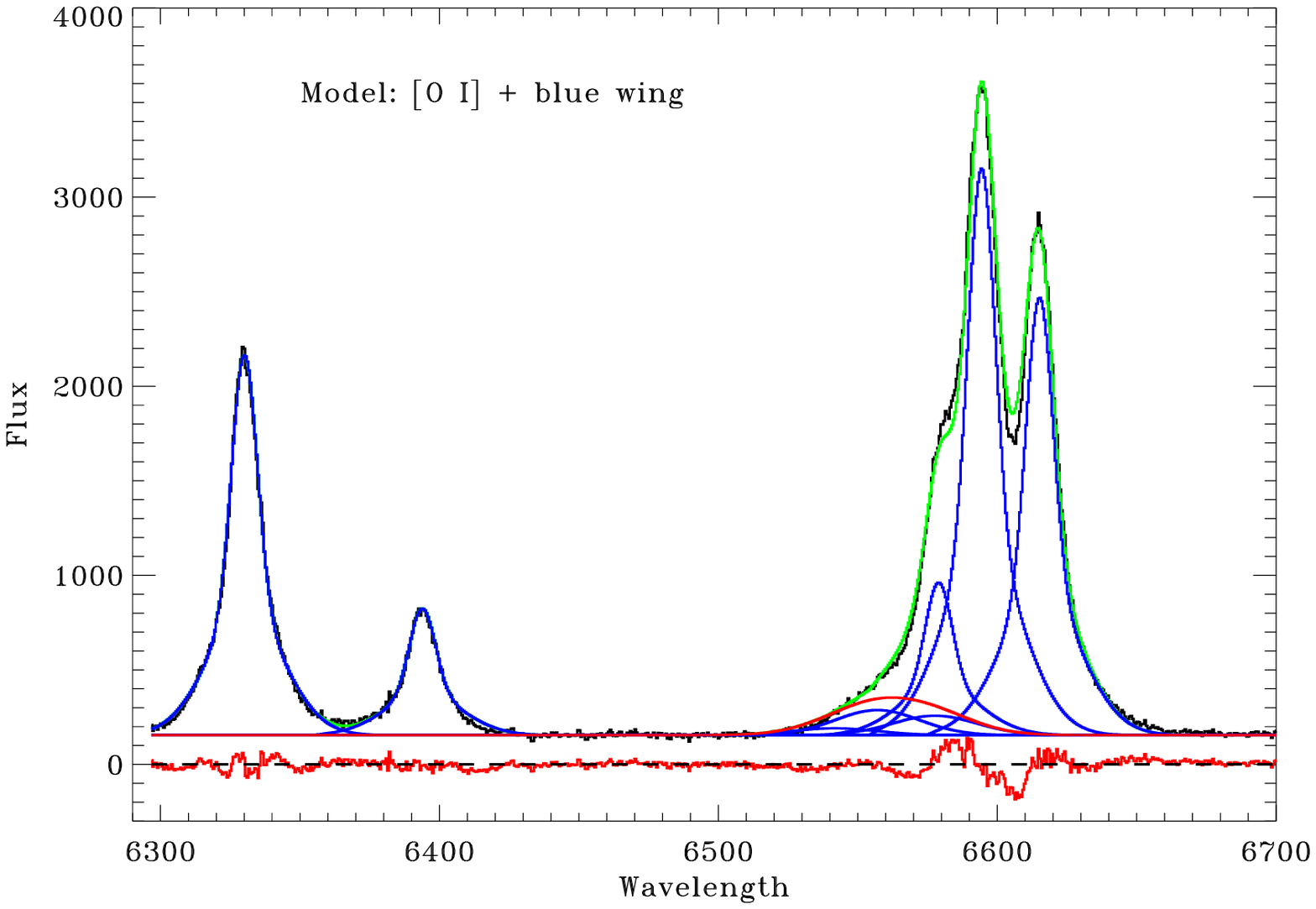}
}
\caption{Top left: spectrum of NGC~1052 modeled by adopting the same shape obtained from the Gaussian fitting of the
  \sii\ doublet for the \Ha\
and
  \nii\ lines. The \Ha\ and \nii\ intensities are set by matching their
  respective emission peaks.  The original spectrum is in black, the
  individual lines are in blue, the total line emission in green, the
  residuals in red. Top right: same as in the left panel but including a broad
  \Ha\ component (red). Bottom left: same as in the top right panel, but using
  the \oi\ doublet as template. Bottom right: same as in the bottom left
  panel, but including blue wings to the narrow lines instead of a
  BLR. Wavelengths are in \AA, while fluxes are in units of
  $10^{-18}$$\ergscmA$.}
\label{ngc1052blr}
\end{figure*}

For each source, we initially attempted to reproduce the \Ha+\nii\ complex by
fixing the line's shape to the one derived for the \sii\, without including
any broad \Ha\ component by just scaling the line intensities to
match the \Ha\ and \nii$\lambda6584$ line peaks (we recall that the ratio
between the two \nii\ lines is fixed).  The result of this procedure is shown
in the top lefthand panel of Fig. \ref{ngc1052blr}, by using NGC~1052 again as an
example. Significant residuals are present in the form of blue and red wings
around the \Ha+\nii\ complex, apparently a clear indication of the presence of
a broad \Ha\ component. We then include such a component in our analysis
(Fig. \ref{ngc1052blr}, top right panel) by adopting a skewed
Gaussian shape in this case. The free parameters (the intensities of all lines and the
width of the broad component) are derived by minimizing the residuals. For the
BLR in NGC~1052, we obtain a full width at half maximum (FHWM) of 2240 \kms, a
flux of 8.7$\times\,10^{-14} \ergscm$, and a skewness parameter of 0.06.

Rather different results are instead obtained when we use the \oi\ doublet as
template for the shape of the \nii\ and \Ha\ lines (Fig. \ref{ngc1052blr},
bottom left panel). The \oi\ lines are in fact significantly broader than the
\sii\ ones (the FWHM are 660 \kms \  for \oi and 570 for \sii,
see Table \ref{tab3}). They also show much more developed wings on both the
blue and red sides. As a result, the red end of the \Ha+\nii\ complex is
perfectly reproduced by the shape of the \oi, while a broad component is still
needed to account for its blue side. By comparing with the results obtained
from the \sii\ template, we note that while the FWHM of BLR is similar ($\sim$
2200 \kms), the BLR flux is reduced by a factor of 5 (see Table \ref{tab4b})
and it is blue-shifted by $\sim$ 1200 \kms.

As a final test, we performed a fit in which a Gaussian blue wing is added to
\oi\ shape for each narrow component of the complex (see Fig. \ref{ngc1052blr}, bottom right
panel), keeping the \nii/\Ha\ ratio fixed to the same value measured for the
main component. This model reproduces the spectrum equally well as the fit
that includes a BLR. The contribution of the blue wing increases only
marginally the FW20 of the \nii\ line, from 1320 \kms\ to 1370 \kms.

Summarizing, in NGC~1052, including a broad \Ha\ line improves the
spectral fit obtained by using the the \sii\ (or the \oi) lines as templates
for the shape of \nii\ and \Ha; however, the results differ depending on
which line is actually used, owing to the mismatch between the profiles of \sii\
and \oi. An equally satisfactory fit is achieved by adding a blue wing to the
emission lines of the narrow \Ha+\nii\ complex.

How general is this behavior? Such an analysis can be performed only when both
the \sii\ and \oi\ doublets are covered by the HST spectrum and when their
signal-to-noise is sufficient for a reliable fit of their profiles.  This is
possible for the nine sources listed in Table \ref{tab3}. The main results
obtained for NGC~1052 are confirmed:
\begin{itemize}
\item 
i) the three fitting methods reproduce the data with similar accuracy;

\item
ii) the \oi\
lines are generally broader than the \sii\ lines (see Table \ref{tab3}, where
the width at half maximum and at 20\% intensity are compared);

\item iii) the \nii\ and \Ha\ wings required to reproduce the spectra increase
  the FW20 of these lines by less than 30\%, and usually by less than 15
  \%;\footnote{The only partial exception is NGC~4143. In this source the \oi\
    lines are broader than the \nii\ and the required additional Gaussian
    component is a narrow core of the line, statistically preferred to a broad
    wing. In this case the red end of the \nii + \Ha\ complex should be
    reproduced with a further component.}

\item iv) excluding the objects presented in Fig. \ref{clearblr} showing a
  clear BLR, the BLR flux obtained from cloning the \sii\ are larger than when
  using the \oi\ (see Table \ref{tab4b} where we compare the flux and width
  obtained in the two cases), while the widths are similar.

\end{itemize}
Another common result is the width of the broad \Ha\ component. In all cases
the fitting procedure returns a value confined in a rather narrow range, i.e.,
1360 < FWHM$_{\rm H\alpha}$ < 2890 \kms. In Appendix \ref{appA} we show
the fit obtained applying the three methods described above.

\begin{table}
\caption{Forbidden line widhts}
\begin{tabular}{l||r|r||r|r}
\hline
Name & \multicolumn{2}{|c||}{\sii}  & \multicolumn{2}{|c}{\oi} \\   
\hline
             & FWHM & FW20 & FWHM  & FW20   \\   
\hline
NGC~1052     & 570  &  990 & 660   & 1310   \\  
NGC~4036     & 520  &  760 & 510   &  740   \\
NGC~4143     & 420  &  810 & 840   & 1310   \\
NGC~4258     & 420  &  690 & 470   &  870   \\
\hline                                                                    
NGC~3031     & 420  &  620 & 470   &  660   \\  
NGC~3998     & 840  & 1280 & 1410  & 2530   \\  
NGC~4203     & 320  &  470 & 580   & 1050   \\  
NGC~4450     & 390  &  840 & 630   & 1000   \\  
NGC~4579     & 520  &  860 & 920   & 1500   \\  
%\hline                                                                    
%NGC~3227     & 250  &  420 & 240   & 390     \\  
%NGC~3516     & 240  &  340 & 180   &  240    \\  
\hline                                                                    
\end{tabular}
\label{tab3}

\medskip
\small{(1) Object name; 
  (2) - (3): FWHM and FW20 of the \sii\ line, (4) - (5) FWHM and FW20 of the
  \oi\ line, all in \kms. We report separately the LINERs with and without a clear BLR.}
\end{table}

\begin{table}
%\caption{Comparison of the broad line parameters}
\caption{Broad line parameters from different templates}
\begin{tabular}{l||r|r||r|r}
\hline
Name & \multicolumn{2}{|c||}{\sii}  & \multicolumn{2}{|c}{\oi} \\   
\hline
             & FWHM & log F & FWHM  & log F   \\   
\hline
NGC~1052     & 2240  & -13.06 & 2210   & -13.72   \\   
NGC~4036     & 2080  & -14.27 & 1720   & -14.26   \\ 
NGC~4143     & 2100  & -13.07 & 2890   & -13.31   \\ 
NGC~4258     & 1390  & -12.99 & 1360   & -13.14   \\ 
\hline                                                                    
NGC~0315     & 2590  &  -14.05  & & \\   
NGC~2787     & 2200  &  -13.47  & & \\  
NGC~2985     & 1010  &  -14.22  & & \\  
NGC~3642     & 1330  &  -13.35  & & \\  
NGC~4278     & 2940  &  -14.07  & & \\  
NGC~5005     & 2310  &  -14.11  & & \\  
NGC~5077     & 1550  &  -14.96  & & \\  
\hline                                                                    
\hline                                                                    
\end{tabular}
\label{tab4b}

\medskip
\small{FWHM (in \kms) and logarithm of the flux (in $\ergscm$) 
  of the broad \Ha\ line required by the fit using as template for the 
  narrow component the \sii\ lines (column 2 and 3) and the \oi\ lines (column 4 and 5).}
\end{table}

For the five LINERs with a clear BLR, the situation is similar, once this
broad component is taken into account. We treat
NGC~4203 in detail as an example. In this source there is a well defined broad \Ha\ component,
extending over $\sim 15,000$ \kms. We reproduce this component with a skewed
Gaussian (the resulting parameters are given in Table \ref{bigtable}), obtained
after masking the spectral region covered by the narrow \nii+\Ha\ complex.

\begin{figure*}[htp]
\centerline{
\includegraphics[scale=0.30,angle=0]{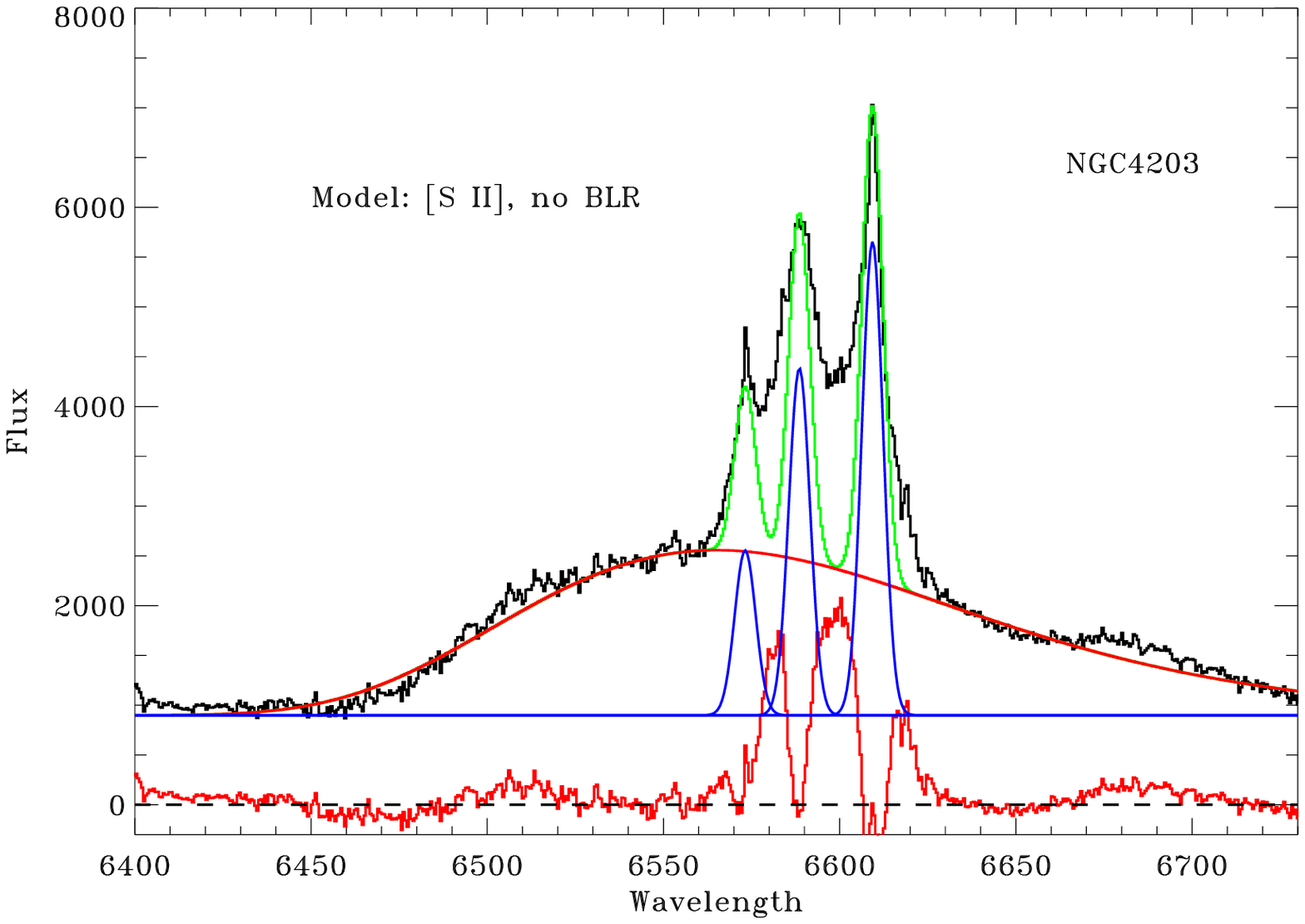}
\includegraphics[scale=0.30,angle=0]{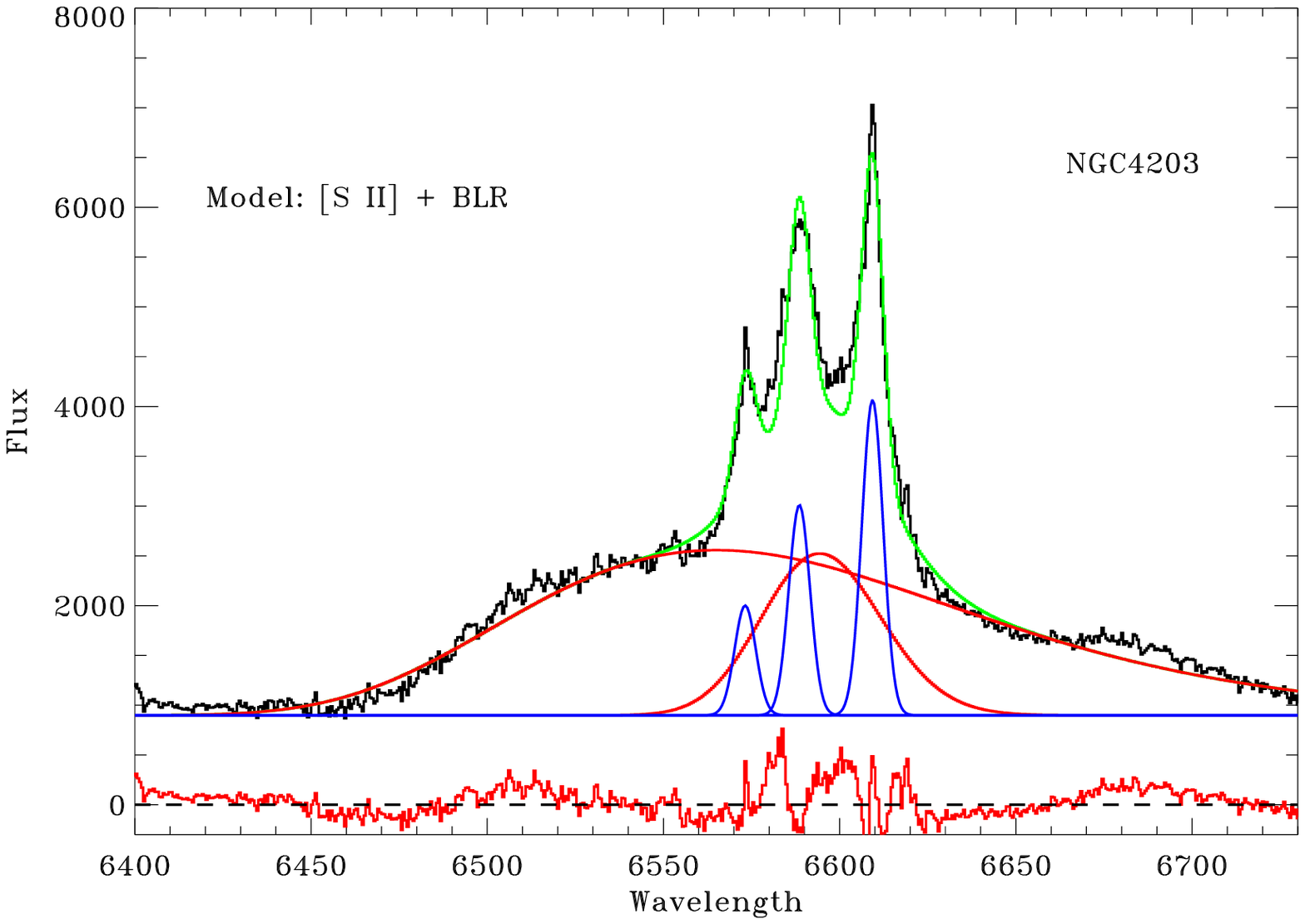}
\includegraphics[scale=0.30,angle=0]{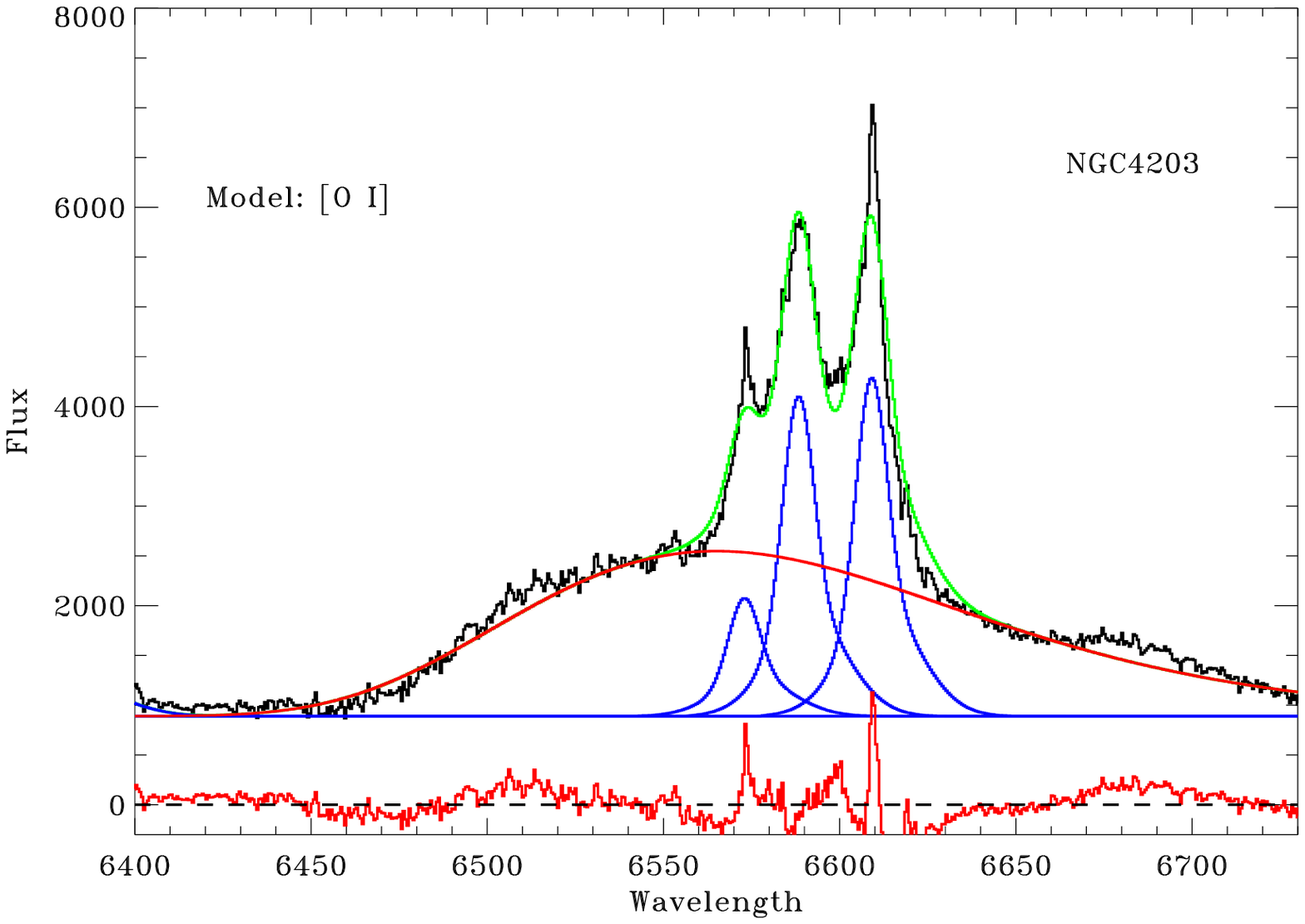}
}
\caption{Spectrum of a LINER with a clear BLR (NGC~4203) reproduced with a
  skewed Gaussian (red curve). The residual \nii+\Ha\ complex is modeled with
  3 different methods: by cloning the \sii\ doublet (left), by including a
  second broad \Ha\ component (middle), or by using the \oi\ doublet
  (right). The original spectrum is in black, the individual lines are in
  blue, the total line emission in green, the residuals in red. Wavelengths
  are in \AA, while fluxes are in units of $10^{-18}$$\ergscmA$.}
\label{ngc4203}
\end{figure*}

\begin{figure*}[hbp]
\centerline{
\includegraphics[scale=0.5,angle=0]{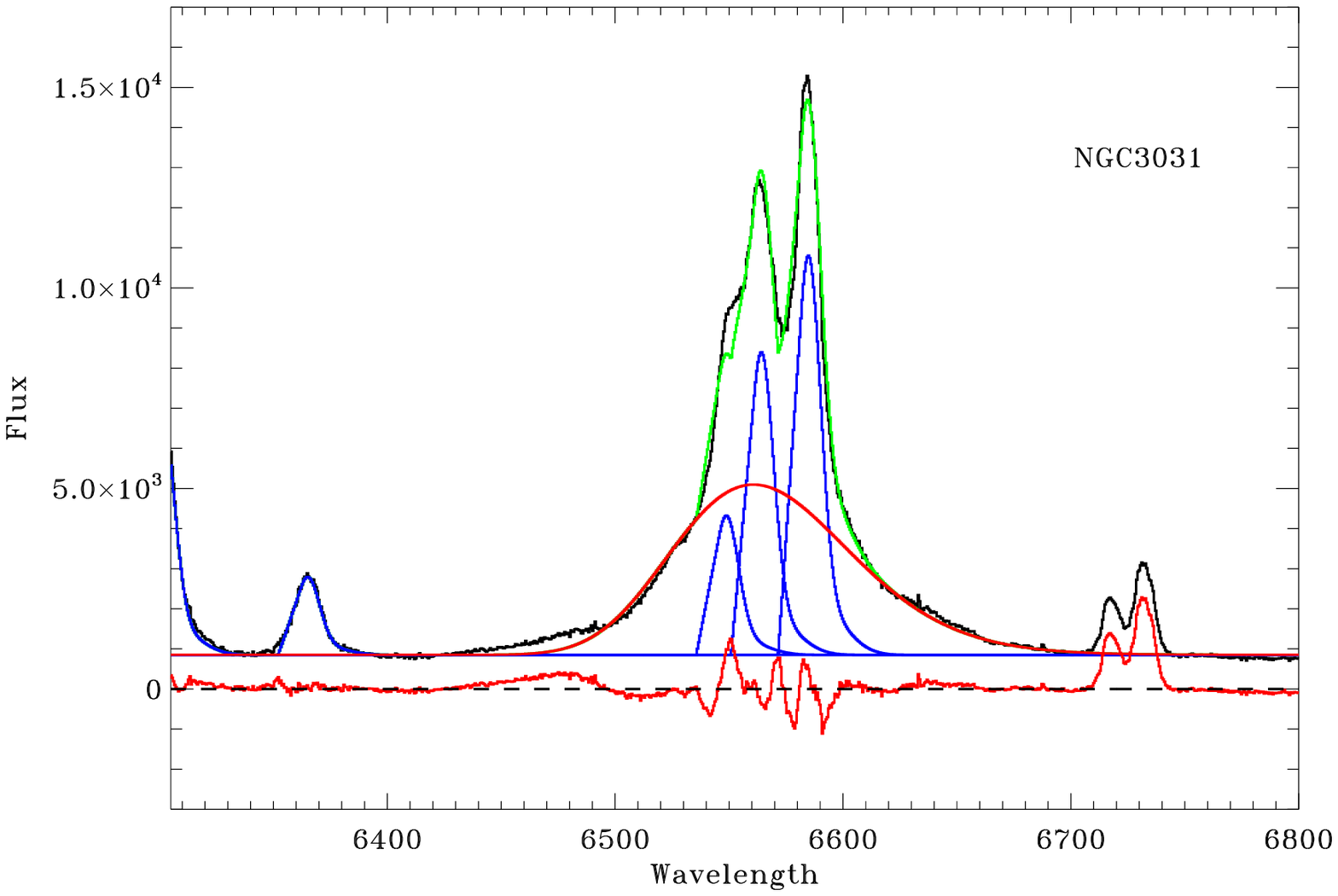}
\includegraphics[scale=0.5,angle=0]{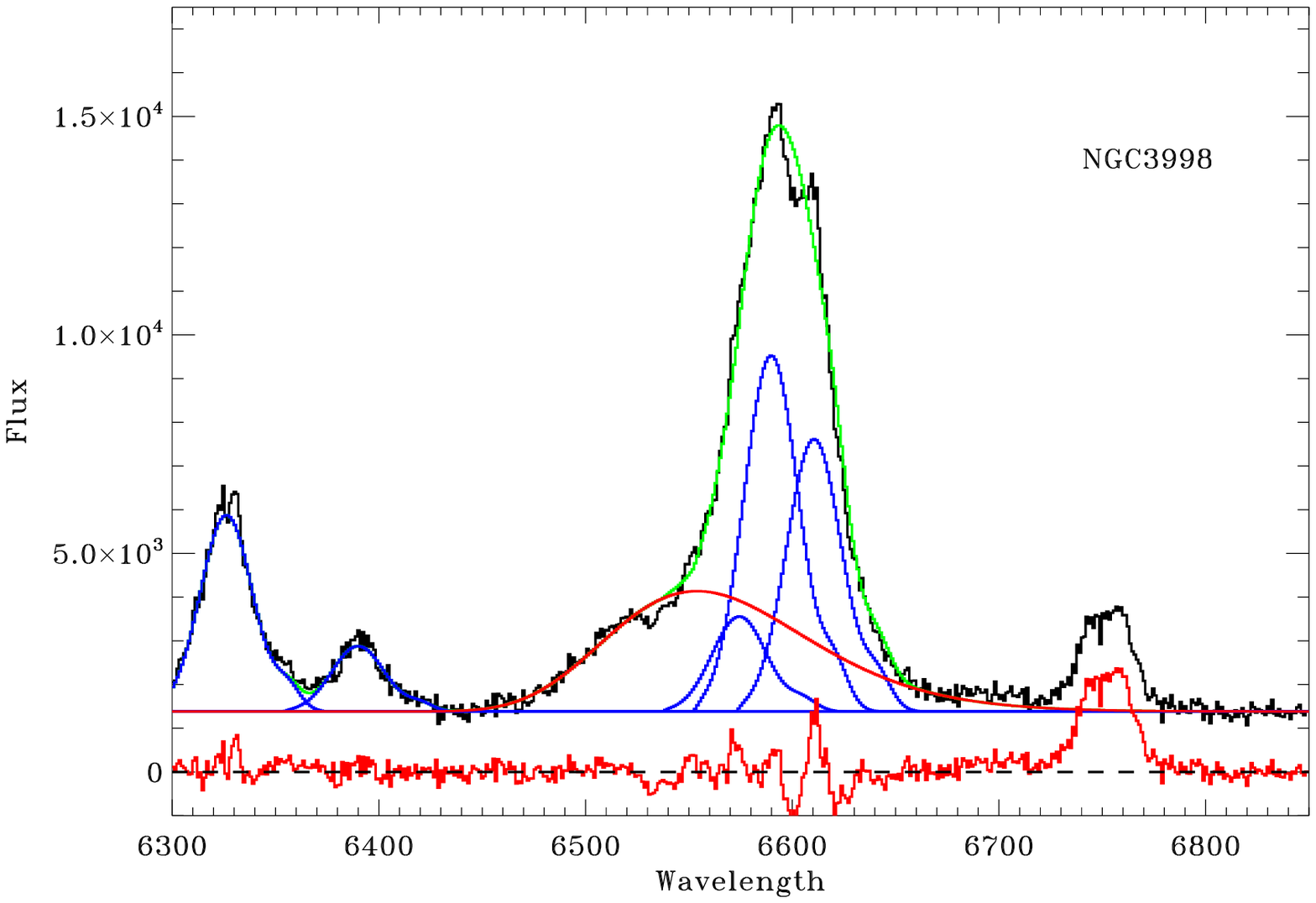}
}
\centerline{
\includegraphics[scale=0.5,angle=0]{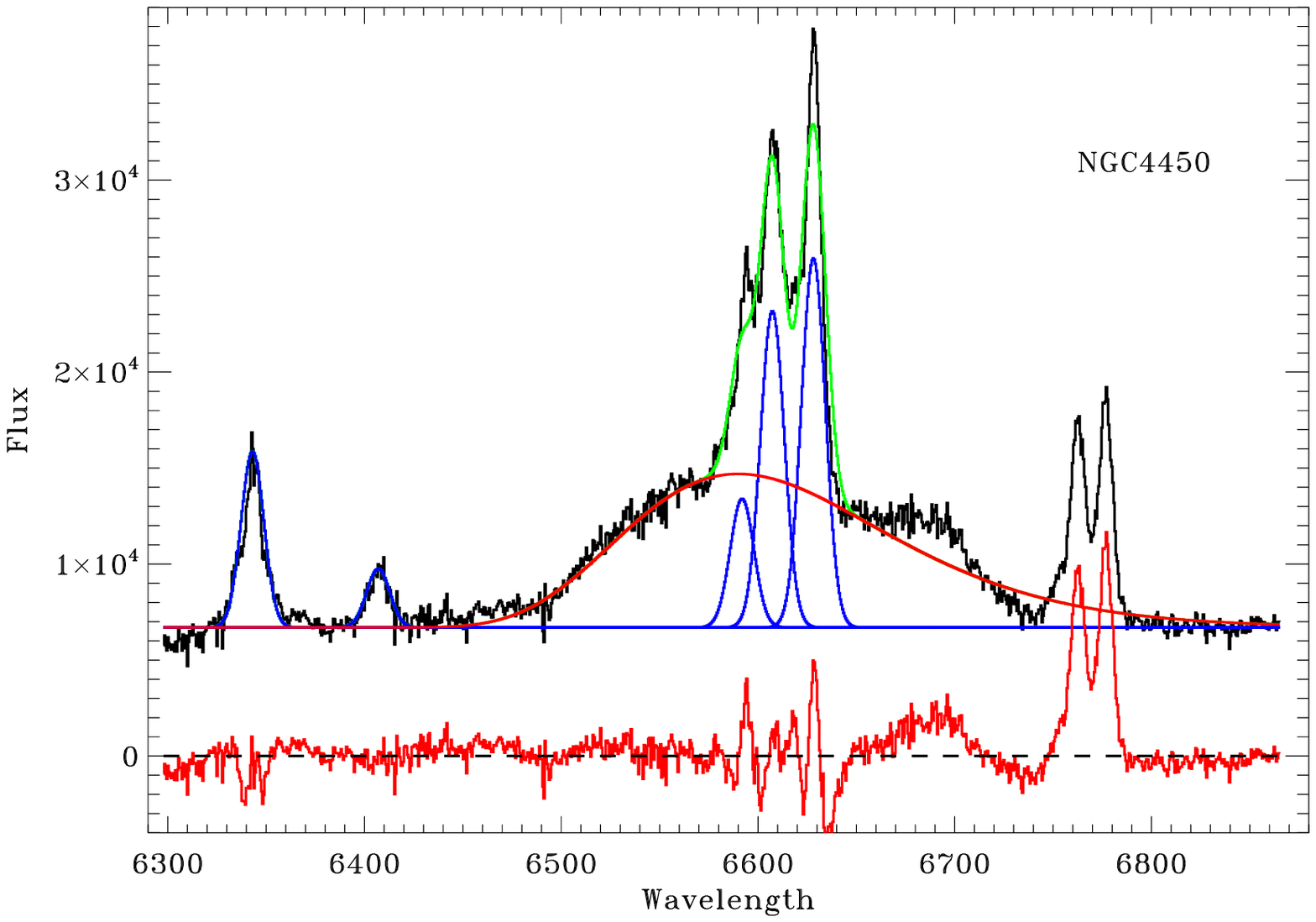}
\includegraphics[scale=0.5,angle=0]{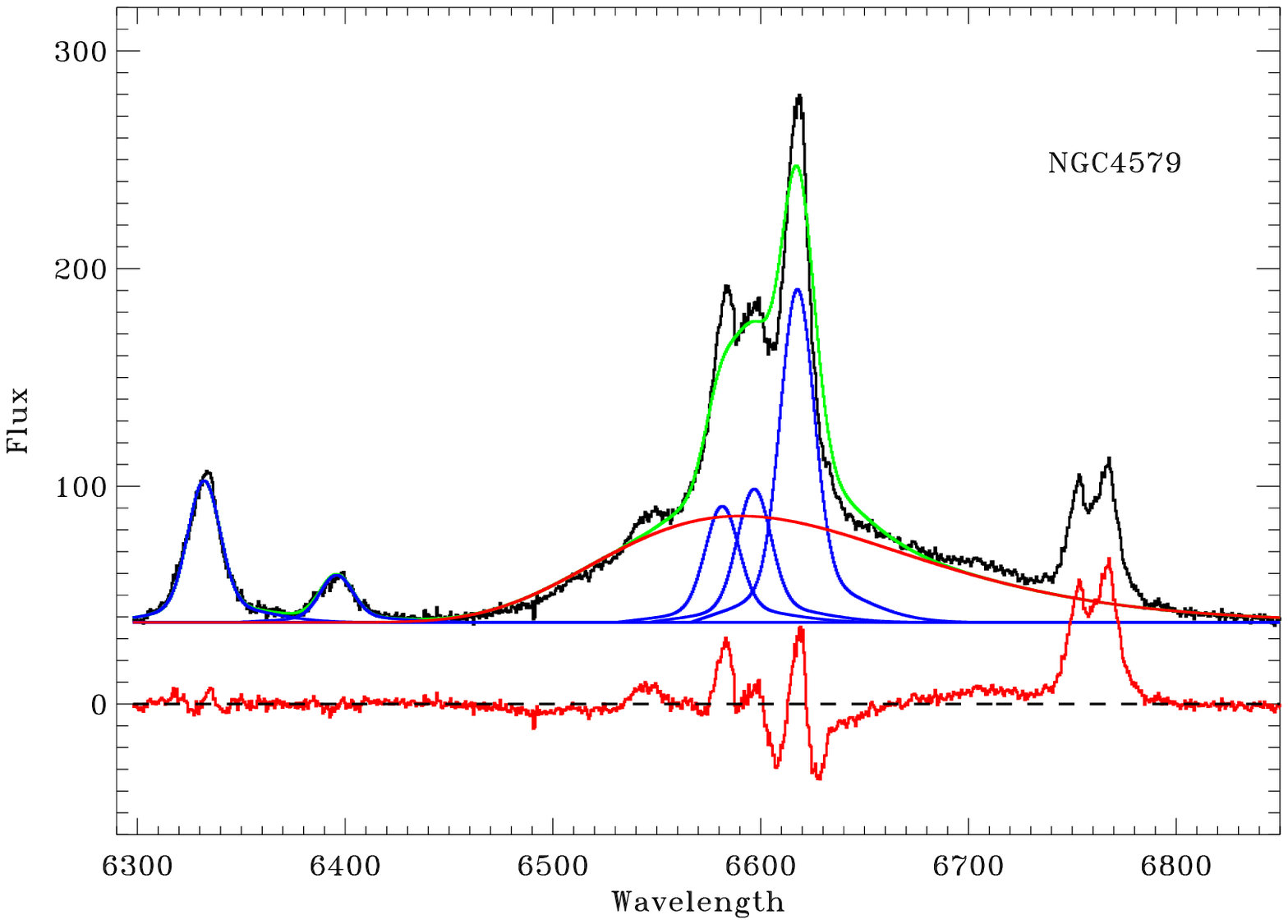}
}
\caption{Spectra of the LINERs with a clear BLR, in addition to NGC~4203
  presented above, modeled using \oi\ as template for the narrow lines.  The
  broad \Ha\ (red curve) is reproduced with a skewed Gaussian. The original
  spectrum is in black, the individual narrow lines are in blue, the total
  line emission in green, the residuals in red. Wavelengths are in \AA, while
  fluxes are in units of $10^{-18}$$\ergscmA$.}
\label{4clearblr}
\end{figure*}

We then attempt to reproduce the remaining emission in three different ways:
i) using the \sii\ as template without any further broad component
(Fig. \ref{ngc4203}, right panel), ii) by adding a second broad \Ha\ component
(middle panel), and iii) by using the \oi\ line as template (left
panel). While a second broad component is clearly needed from the \sii\
template, the fit obtained with the \oi\ is already satisfactory. This is due
to the fact that the \oi\ lines are broader than the \sii, similar to the
cases presented above. In the Fig. \ref{4clearblr} we show the analysis for
the four other  LINERs with clear BLR. It leads to analogous results; i.e., no
broad \Ha\ line (in addition to its main component) is required to reproduce
the data.

\begin{figure}[htp]
\includegraphics[scale=0.5]{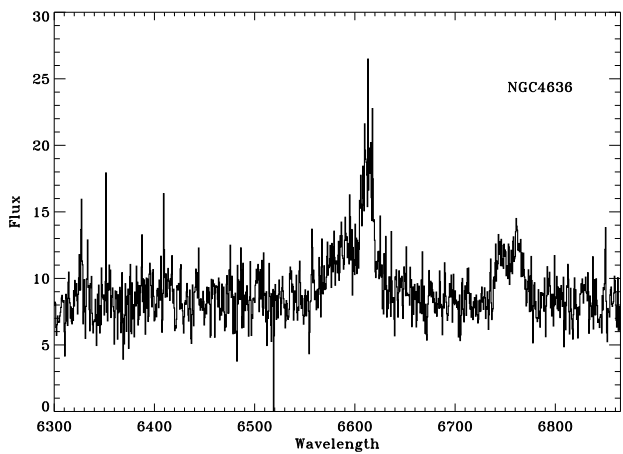}
\caption{Spectrum of NGC~4636. The signal-to-noise ratio is insufficient for
  any reliable analysis of this source.}
\label{ngc4636}
\end{figure}

In seven galaxies the \oi\ line is either not covered by the HST spectrum or is
too faint to be modeled accurately. In these cases we can only rely on the
\sii\ lines as template. In Table \ref{tab4b} we report FWHM and fluxes of the
broad \Ha\ component. The relative fit are shown in Fig. \ref{s2only2}. The
results obtained share several aspects with those presented before. In
particular an accurate fit is also obtained by just adding a broad wing in both \nii\ and \Ha\ to the \sii\
template. Furthermore, the widths of the
broad lines are again clustered around a FWHM of $\sim$ 2000 \kms, being
included between $\sim$ 1000 and $\sim$ 3000 \kms.
For NGC~4636 the data quality is insufficient to proceed to any reliable 
analysis. Its spectrum is shown in Fig. \ref{ngc4636}.

In three cases (namely NGC~4151, NGC~4395, and NGC~1275), there are no medium-resolution HST spectra covering the \Ha\ region, and we then used spectra
including the \Hb\ line. The separation between the emission lines in this
spectral region is large enough that data from the low-resolution
grism G430L can also be used. We adopted the same method described above, but by
using the \oiii$\lambda$5007 as template for the shape of the \Hb\ line. This
is an even less accurate assumption than the similarity of the \sii\ and \nii\
profiles, since \oiii\ is a high ionization line that might be produced in
regions that are significantly different from the Balmer lines. However,
this is not a significant issue since in two galaxies (namely NGC~4151 and
NGC~4395), the presence of the broad \Hb\ is clearly visible and its properties
are not significantly affected by the de-blending procedure. In the third case
(i.e., the ambiguous galaxy NGC~1275), we failed to detect a broad \Hb\
component.

\section{The properties of the BLR in LLAGN}
\label{blrprop}
The results presented in the previous section indicate that there is a
substantial mismatch between the profiles of the different emission lines
considered. This is very likely due to a stratification in density and ionization
within the NLR that causes differences in the location of the emitting region
for the various lines and, consequently, differences in the lines
profiles. For example, the various lines considered are associated with
different critical densities, with the [S~II] lines having the lowest
value.\footnote{The logarithms of critical densities, in cm$^{-3}$ units, are
  $\sim$ 3.2, 3.6, 6.3, and 4.9 for [S~II]$\lambda6716$, [S~II]$\lambda6731$,
  [O~I]$\lambda6300$, and [N~II]$\lambda6584$, respectively
  \citep{appenzeller88}.} Several indications for the presence of a dense,
compact emitting region within the NLR, located within a radius of a few pc
from the central black hole are emerging
\citep{capetti05b,peterson13,baldi14}. It can be envisaged that this region is
poorly represented in the \sii\ profile, while it is more prominent for the
\oi, owing to its higher critical density. On the other hand, the possible
presence of blue wings in the \nii\ profile, with respect to \oi, might be due
to an ionization stratification of the NLR.

The complex structure of the NLR is not completely captured with the approach
usually followed to model the \nii+\Ha\ complex by using different forbidden
emission lines as templates. In this situation, the properties of the BLR
obtained with this procedure, including its very detection, must be treated
with caution. In particular, the need for a BLR based only on the improvement
in the fit when such a component is added is highly questionable: it might not
imply that a broad component is really present, but it could be just the
consequence of the complex NLR structure not fully represented by the
templates. The similarity in the quality of the fit with the three methods
presented (by using the \sii\ or \oi\ as templates, or adding a broad
base wing to them) casts doubts on the reliability of the BLR detections.

The other indication against the reality of the broad \Ha\ is the strong
clustering of the derived values for their widths.  The values derived for the
LINERs considered are typically of $\sim$ 2000 \kms. This is intriguingly
similar to the value that corresponds to the separation of the two lines of
the \nii\ doublet, 1650 \kms. Indeed, any positive residual left after the
subtraction of a template will be distributed over a spectral region
corresponding to the observed values. We then adopt the fluxes of the BLR
obtained with the \sii\ (or \oi) template as conservative upper limits, rather
than genuine detections.

  The values reported in Table \ref{bigtable} are strictly valid
  only for the FWHM quoted in Table \ref{tab4b}, i.e., $\sim$ 2000 \kms. We
  then also explored the dependence of this limit on the BLR width, by
  repeating the fit procedure with the FWHM fixed at various values, ranging
  from 3000 to 30,000 \kms. We find that the limit on the BLR flux has a
  complex relation with its width, but with some common feature from source to
  source. Despite the increasing width, the limit initially decreases because
  when the broad line wings exceed the width of the \nii+\Ha\ complex, the BLR
  would become readily visible even for relatively low intensities. For
  higher values, FWHM $\sim$ 4,000 - 15,000, it increases almost
  linearly. However, for FWHM $\gtrsim$ 15,000 \kms, the spectral gradient
  becomes too small to separate any BLR from the continuum and the BLR flux
  sharply increases. We return to this issue in Sect. \ref{undetected}.

For the LINERs in which the BLR is instead visible, there is a similar
effect. There is not a disappearance of the whole BLR, but only of its
relatively narrow core (again with FWHM $\sim$ 2000 \kms) superposed on a
broad \Ha\ base. 

The HST spectra of these objects have already been presented and analyzed
(NGC~3031, \citealt{devereux03}; NGC~3998, \citealt{defrancesco06}; NGC~4203
\citealt{shields00}; NGC~4450, \citealt{ho00}; NGC~4579, \citealt{barth01},
but see also \citep{walsh08,rice06,shields07}.
These authors used slightly different fitting strategies to deblend the broad
line emission: for example \citealt{ho00} and \citealt{shields00} reproduce
the narrow the \nii+\Ha\ complex with a synthetic \sii\ profile. They
interpret the residual of the fit as a combination of a ``normal'' broad line
(with a width of $\sim$ 2000 \kms, similar to those observed from the ground)
superposed on a double-peaked emission with high velocity wings. As explained
above, we argue that the ``normal'' BLR is actually due to the mismatch
between the lines in the \nii+\Ha\ complex and the templates used.
While this affects the BLR profiles, the BLR fluxes they obtained are similar
to ours. 

In Table \ref{bigtable} we report fluxes (or upper limits) and widths (in case
of detections) obtained for the broad Balmer lines.

\begin{figure*}[hbp]
\includegraphics[width=6.5cm,angle=90]{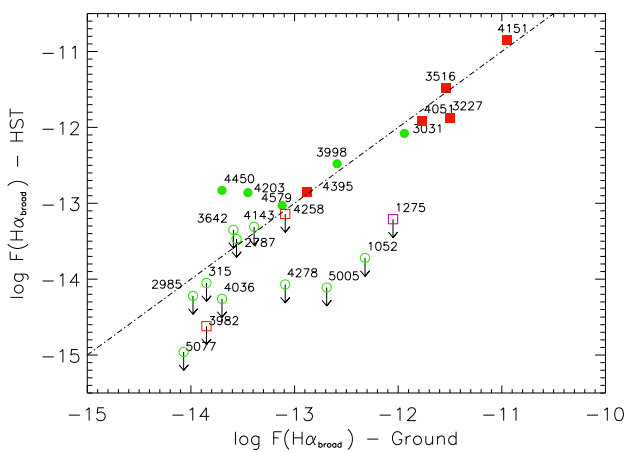}
\includegraphics[width=6.5cm,angle=90]{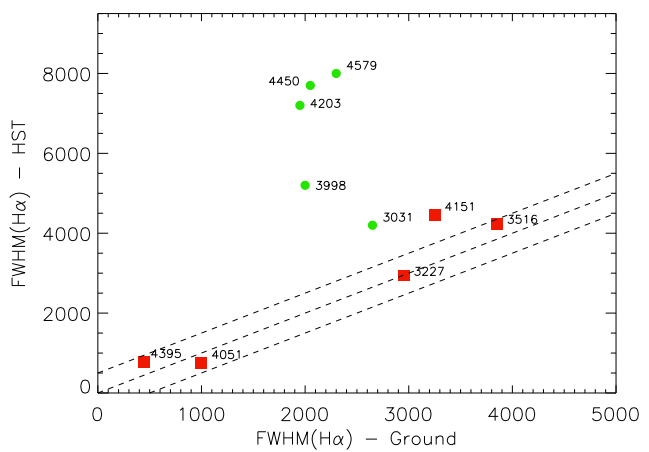}
\caption{Logarithmic fluxes (left, in $\ergscm$) and full width half maximum
  (right, in \kms) of the \Ha\ broad line emission obtained from the HST/STIS
  spectra compared to the results of the Palomar spectroscopic survey. The
  dashed line has a slope equal to one.  The green symbols represent the
  LINERs of the sample, while the red squares are the Seyferts.  NGC~1275, the
  ambiguous galaxy, is marked with a purple square. Empty symbols refer to the
  undetected BLR.}
\label{compare}
\end{figure*}

\section{Comparison with ground-based measurements}
\label{gbcfr}

In Fig. \ref{compare} we compare the BLR fluxes measured from HST and ground-based spectra. We confirm the presence in the HST spectra of a broad Balmer
line emission in all Seyferts but one. Their HST line fluxes, ranging from
$\sim$$10^{-13}$ to $\sim$$10^{-11}\,\ergscm$, are all very similar to the
ground-based measurements, typically differing by less than a factor of
1.5. The only exception is NGC~3982, the Seyfert galaxy with the largest
absorbing column density in X-rays (${\rm N_{\rm H}} \sim 4.3 \times
10^{23}$cm$^{-2}$, \citealt{akylas09}) of the sample. Most likely, the lack of
a BLR is due to nuclear obscuration. From the point of view of broad lines
widths, in the righthand panel of Fig. \ref{compare} we compare the FWHM of the broad \Ha\ emission line obtained from the HST and ground-based spectra. For the Seyferts, the differences are all less than 500 \kms,
with the only exception of NGC~4151. However, for this object, we show
the width of the \Hb\ line, properly corrected.\footnote{We convert the \Hb\
  luminosity and width tabulated in this catalog to \Ha\ luminosity using
  L(\Ha)=3.5$\times$L(\Hb) and FWHM(\Hb)=1.07$\times$FWHM(\Ha)$^{1.03}$,
  \citet{green05}.}  Nonetheless, inspection of the Palomar spectrum shows
that in this object the \Hb\ is indeed significantly broader than the \Ha\ and
in good agreement with the HST measurement.

\begin{figure}[ht]
\includegraphics[width=9cm]{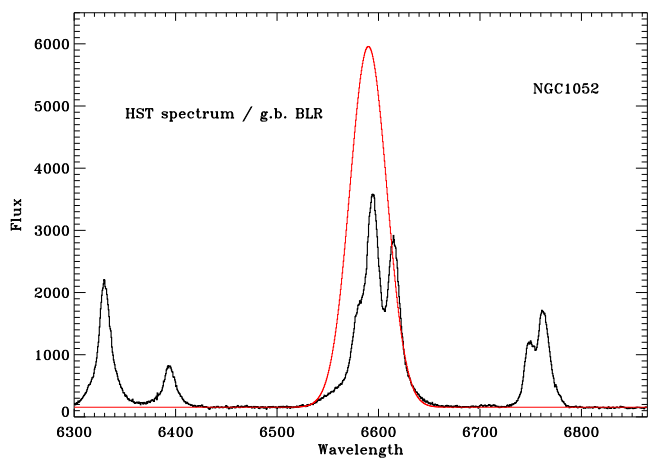}
\caption{The broad \Ha\ line derived from the ground based data of NGC~1052
  (red gaussian) is compared to the HST spectrum.}
\label{ngc1052blr-gb}
\end{figure}

The situation for the LINERs is quite different. We initially concentrate on
the five objects with a clear BLR. First of all their BLR are generally of lower
fluxes than in Seyferts, from $\sim$$10^{-13}$ to
$\sim$$10^{-12}\,\ergscm$. The spread of the ratios between HST and ground
measurements for the detected sources is much larger, ranging from 0.3 to 20.
All HST measurements of the FWHM are in strongly excess with respect to the
ground-based data. While the ground-based data indicate widths between 1500
and 3000 \kms, much broader lines are seen in the HST data with FWHM reaching
8000 \kms. As explained above, we argue that the ground-based BLR detections
are spurious. The genuine broad emission lines in these sources (as previously
noted by \citealt{ho00} for NGC~4203, \citealt{shields00} for NGC~4450, and
\citealt{defrancesco06} for NGC~3998) are not visible in ground-based spectra,
owing to the higher continuum and narrow line fluxes.

\begin{figure*}
\centerline{
\includegraphics[width=8.5cm,angle=90]{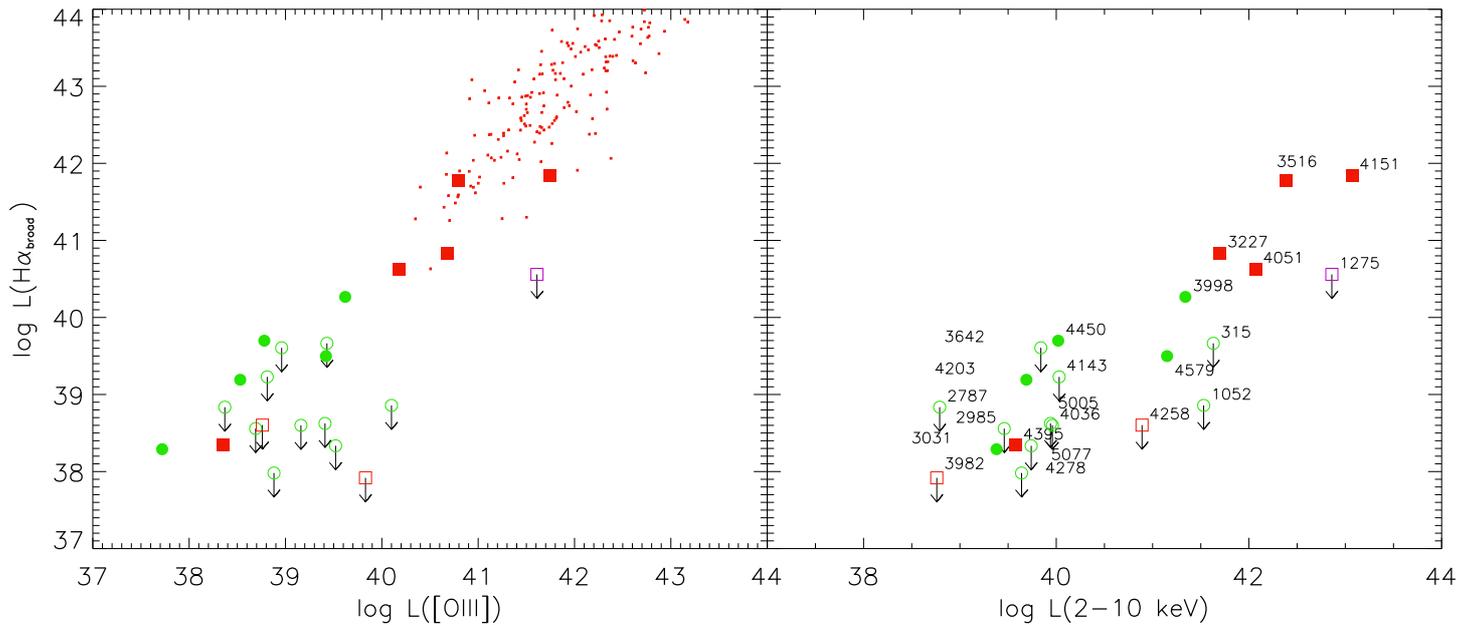}
}
\caption{BLR luminosity against 2 AGN power estimators: the \oiii\ narrow line
  luminosity (left panel) and the intrinsic X-ray luminosity corrected for
  absorption in the 2-10 keV band (right panel). The green circles represent
  LINERs, the red squares are the Seyferts (empty symbols refer to the
  undetected BLR), the red dots are the objects from \citet{marziani03}.}
\label{bol}
\end{figure*}

In the galaxies without a clear BLR, the ground-based measurements are usually
higher than the flux obtained from the HST data, by a factor of up to
10. Particularly instructive is the comparison between the broad \Ha\ line
derived from the ground- based data of NGC~1052, shown in
Fig. \ref{ngc1052blr-gb}, that indicates that the ground-based BLR is
inconsistent with the data, regardless of the modeling scheme, since
it exceeds
the whole flux of the \nii+\Ha\ complex.

\section{On the detectability of BLR in LLAGN}
\label{undetected}

In most of the LINERs we considered, the presence of a BLR is not readily visible
and it is not required by the analysis of their spectra. Is the BLR really
missing, or they are just too faint to be detected? 

To answer to this question in Fig.\ref{bol} we compare the broad \Ha\
luminosity with two estimators of the AGN power, i.e., the luminosity of the
\oiii\ narrow line and the unabsorbed nuclear X-ray luminosity in the 2-10
keV band, see Table \ref{bigtable}. A strong trend is clearly present in all
comparisons. This might have been expected since both the NLR and the BLR are
photoionized by the AGN continuum, that is estimated well by its X-ray. The
LINERs smoothly extend the relation seen in more powerful AGN, representing
the low-luminosity end of the distributions in all diagrams. The median ratio
between the BLR and the X-ray luminosities measured in the ten objects with a
clear BLR is $L_{\rm BLR} / L_{\rm X} \sim 0.08$ (and $L_{\rm BLR} / L_{\rm
  [O~III]} \sim 3$). All ten data points fall within a factor of 10 of the
median value (and within a factor of three considering the relation linking
$L_{\rm BLR}$ and $L_{\rm [O~III]}$). We then predict the BLR luminosity in
the objects apparently without broad lines by assuming that it follows these
multiwavelength trends.

The BLR widths can be instead derived from the virial
formula, by adopting the scaling relations of more powerful AGN and the BLR
luminosity just derived. We find that such BLR could not be detected in any
source, because of the extremely high line widths whose median is of
$\sim$ 30,000 \kms.

\begin{figure*}
\centerline{
\includegraphics[width=6.3cm]{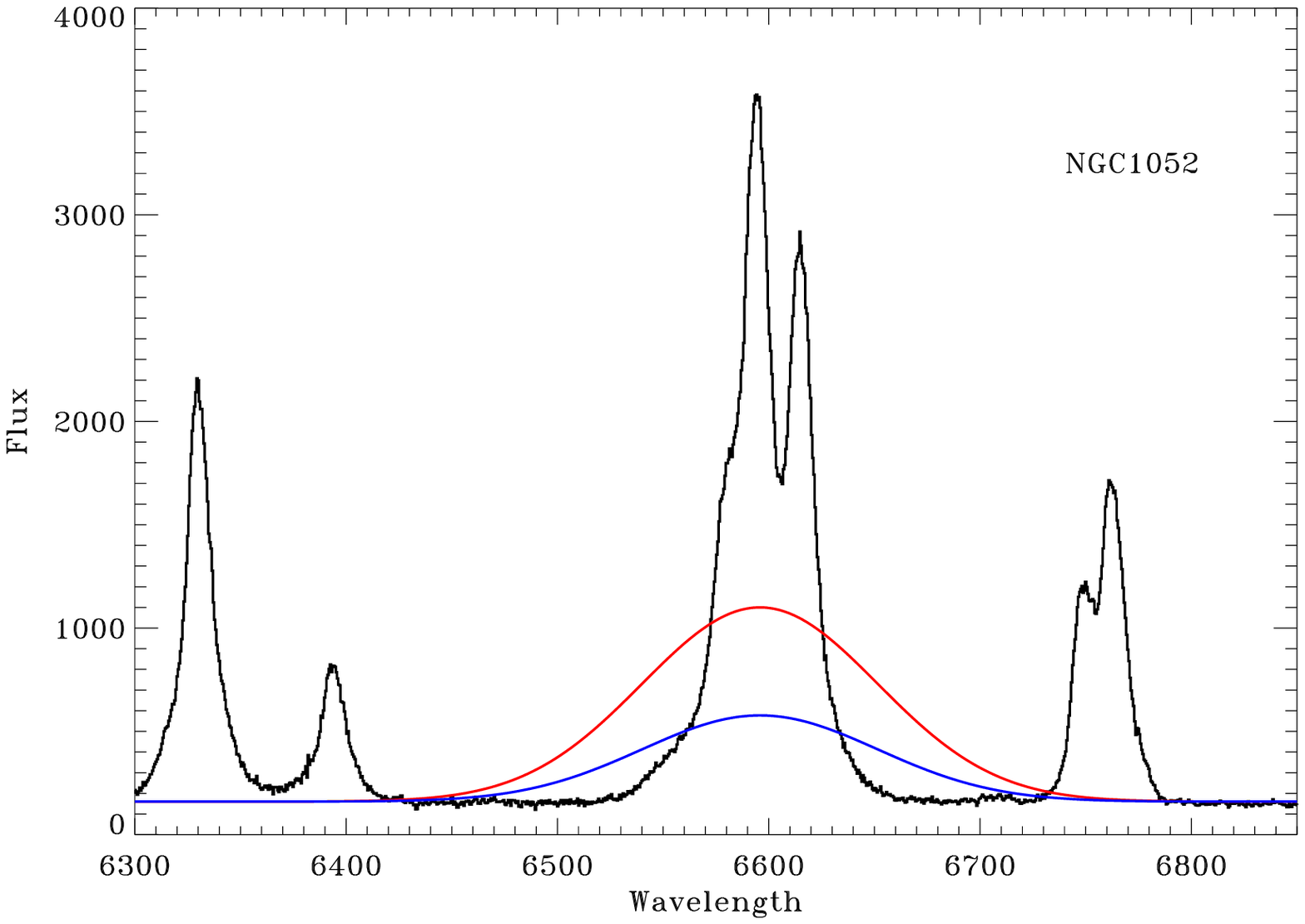}
\includegraphics[width=6.3cm]{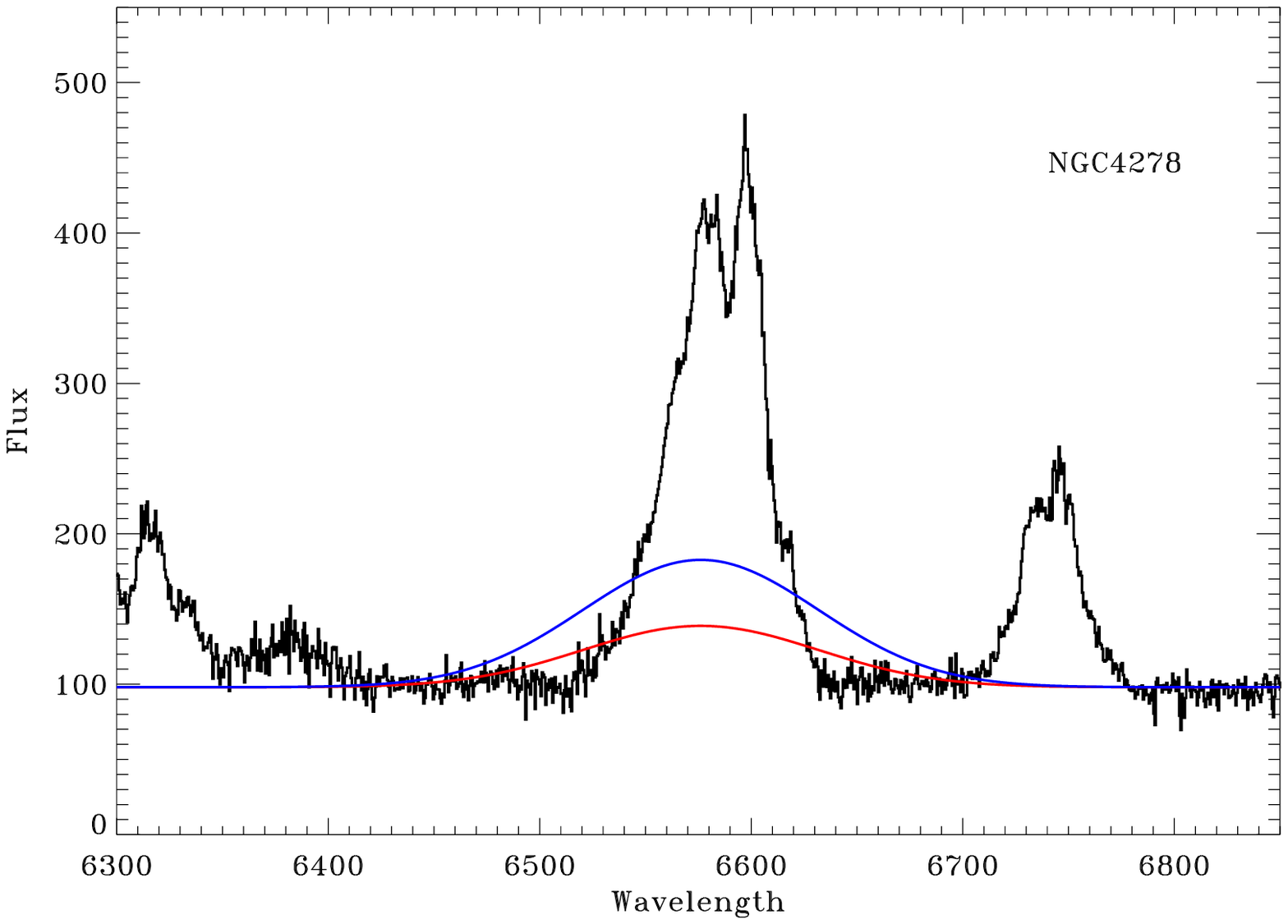}
\includegraphics[width=6.3cm]{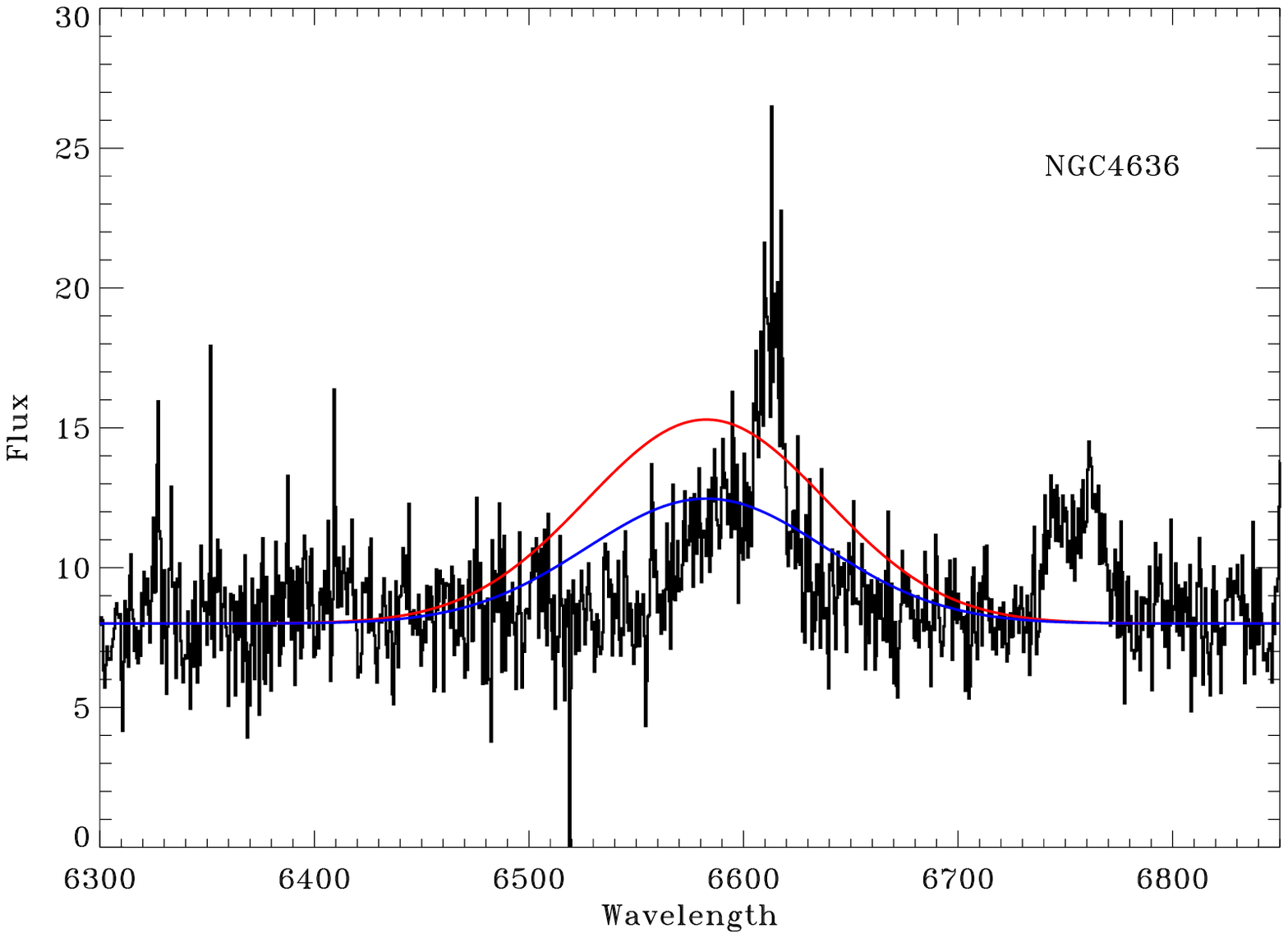}
}
\caption{The blue curve represents a synthetic BLR with flux derived from the
  X-ray flux and a width of 6500 \kms, conservatively reduced by a factor 10
  to explore its detectability in the HST spectrum for NGC~1052 (left),
  NGC~4278 (center), and NGC~4636 (right). For the red curve the
  normalization is derived from the \oiii\ flux, in this case reduced by a
  factor of 3.}
\label{ngc1052fakeblr}
\end{figure*}

We also consider another possibility, i.e., that the BLR radius has the same
value, $R_{\rm BLR} \sim 1,000 r_s$, shown by the 5 LINERs with a clear
BLR. This simply corresponds to widths similar to those observed and we adopt
the average value of 6500 \kms. In this case the BLR should be readily visible
in 5 of the 12 LINERs considered. In order to account for the observed scatter
in the ratios used to estimate the BLR luminosity, we reduced conservatively
all the predicted BLR fluxes by a factor of 10 for those derived from the
$L_{\rm BLR} / L_{\rm X}$ ratio and by a factor of 3 for those estimated from
the $L_{\rm BLR} / L_{\rm [O~III]}$ ratio. The BLR should be still detected in
three objects, namely NGC~1052, NGC~4278, and NGC~4636, according to both
estimates, see Fig. \ref{ngc1052fakeblr}.

\section{The BLR scaling relations in LLAGN}
\label{scaling}

Reverberation studies of AGN revealed that the BLR properties are strongly
linked to those of the AGN and of the central black hole.  Time lag
measurements between changes in the continuum and broad lines lead to estimates
of the BLR radius, $R_{\rm BLR}$. The SMBH mass can thus be derived from the
virial formula, relating the SMBH mass with the BLR width and its radius in
the form $M_{\rm BH}=f\,G^{-1}\,{\rm FWHM}^2 R_{\rm BLR}$. The factor $f$ is
empirically determined to match the SMBH masses derived with the reverberation
technique and the values obtained from the measurements of stellar velocity
dispersions $\sigma_{\star}$ \citep{ferrarese00,gebhardt00}.\footnote{
  \citet{onken04} estimated a correction factor $f = 5.5$ when using the
  second moment of the line profile in the virial formula. Since
  \citet{peterson04} found a mean ratio between FWHM and the second moment of
  2.03, we then adopt $f=5.5/2.03^2=1.3$.}  Furthermore, reverberation
mapping measurements show a close connection between the BLR radius and the
AGN continuum luminosity (${R_{\rm BLR}=33.65\, L_{5100,44}^{0.533}}$ lt-days,
where $L_{5100,44}$ is the luminosity at 5100 \AA\ in units of
$10^{44}$$\ergs$, \citealt{bentz13}).

For LLAGN the situation is somewhat different. In fact, the reduced
contribution of the active nucleus, with respect to more powerful AGN, enables
us to estimate their black hole mass from measurements of the stellar velocity
dispersion $\sigma_{\star}$. Furthermore, the measurements of the optical
continuum are severely contaminated by the starlight, particularly in the
least luminous AGN. This is clearly shown by the strong stellar absorption
features visible in their blue HST spectra (see, e.g.,
\citealt{sarzi05}). Therefore, their optical continuum luminosity cannot be
used to test the validity of the scaling relations obtained for more powerful
AGNs in LLAGN.

\begin{figure}[h]
\centerline{
\includegraphics[width=6.5cm,angle=90]{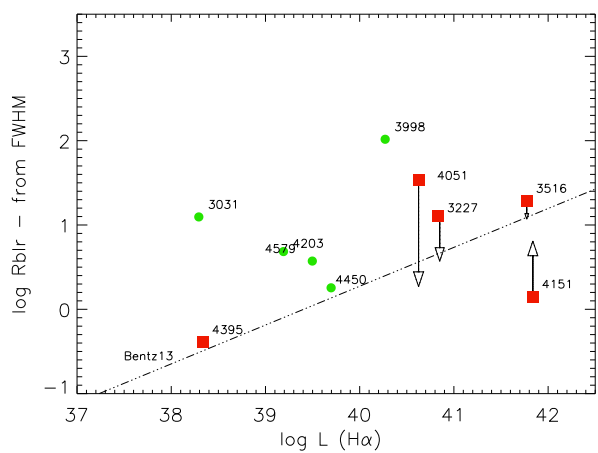}
}
\caption{BLR radius (in light days) against \Ha\ luminosity (in $\ergs$
  units). The dashed lines represent BLR radius-luminosity relations derived
  from the law linking $R_{\rm BLR}$ with the continuum luminosity
  \citep{bentz13}. The green circles are LINERs, the red squares Seyferts. The
  arrows point to the reverberation mapping measurement of $R_{\rm {BLR}}$,
  when available \citep{bentz13}.}
\label{blr}
\end{figure}

Nonetheless, having estimated the SMBH mass\footnote{We adopt the formulation of \citet{tremaine02},
$M_{\rm BH,\star}=1.35\,10^8\,\sigma_{\star,200}^{4.02} \,M_\sun$,
where $\sigma_{\star,200}$ is in 200 \kms\ units.}, we can invert the virial formula
and use it to estimate the BLR radius:
$$R_{\rm BLR} = 5.12\,10^{-6} \,M_{\rm BH,\star} / (f \times {\rm FWHM}^2_3)
\,\,{\rm lt-days.}$$ In Fig.\ref{blr} we compare the BLR radius with its
luminosity\footnote{We use here the total \Ha\ luminosity, including the broad
  and narrow components, although the latter contributes only less than
  $\sim$ 10 \%.}  for the ten LLAGNs with a detected BLR.  The BLR radius is
typically $\sim$10 lt-days, but it reaches for a value of 100
lt-days NGC~3998.

For a comparison with the more powerful AGN, we can rely on the
strong relation between $L_{5100}$ and \Ha\ emission line luminosity
\citep{green05}; i.e., $L_{\rm H\alpha_{,42}}=5.25\,L_{5100,44}^{1.157}$ (in
10$^{42}$ $\ergs$ units), which in this case includes both the broad and narrow
\Ha\ emission. By combining this relation with that derived by
\citeauthor{bentz13} we can derive the law linking the BLR radius to the \Ha\
luminosity; i.e.,
%$ L_{\rm H\alpha} = 7.17 \,10^{39}\,R_{\rm BLR}^{2.17}$ $\ergs$.  
$ R_{\rm BLR} = 15.7 \, L_{\rm H\alpha,42}^{0.46}$ lt-days.  This relation is
reported in Fig. \ref{blr} as a dashed-dotted line.

The 5 Seyferts in the Palomar sample generally follow, with some scatter, the
relation BLR radius-luminosity relation of the sample of Seyfert and quasars
with reverberation mapping measurements (that actually includes all but 1 of
the Seyfert considered here).  For the LINERs galaxies we instead obtained BLR
radii always larger than predicted by the $L_{\rm H\alpha}-R_{\rm BLR}$
relation. The median excess is of $\sim$ 1 order of magnitude. We conclude
that LINERs do not follow the BLR scaling relations defined by the more
luminous AGN. In the following Section we investigate the origin of this
result.

\section{Comparison with BLR theoretical models.}
\label{models}

Several models predict that for accretion rate or bolometric luminosity below
a threshold limit, the BLR and the standard torus cease to exist. We consider the predictions of three such models
in detail and compare them with our
results.  According to \citet{nicastro00} the BLR forms from disk
instabilities in correspondence of the transition radius from density- to
radiation-dominated regions.  For bolometric luminosities lower than
$L_{\rm{bol}} <0.024 \,(M_{\rm{BH}}/M_\sun)^{-1/8} \,L_{\rm{Edd}}$, this radius
is smaller than the last stable orbit and the BLR cannot be present.

According to \citet{laor03}, the BLR clouds are disrupted when the velocity of
BLR clouds exceeds $\Delta v>$ 25000 \kms. For the BLR radius-luminosity
relation, at low luminosity the $R_{{\rm BLR}}$ shrinks and $\Delta v$ exceeds
this threshold, and the BLR cannot form. This occurs at a luminosity of
$L_{\rm{bol}}\,<\,10^{41.8}(M_{\rm{BH}}/10^8M_\sun)^2 $$\ergs$.

In the scenario of \citet{elitzur06}, the torus and the BLR are smoothly
connected, with the BLR formed by an outflow of ionized gas from the accretion
disk that extends outward until the dust sublimation radius, the inner
boundary of the dusty, and clumpy torus. The column density required for the
formation of the BLR and the torus are achieved only for high Eddington ratios
and both the BLR and the torus disappear when the bolometric luminosity falls
below $L_{\rm{bol}}<10^{42}$$\ergs$.
 
\begin{figure*}
\centerline{
\includegraphics[width=7.0cm,angle=90]{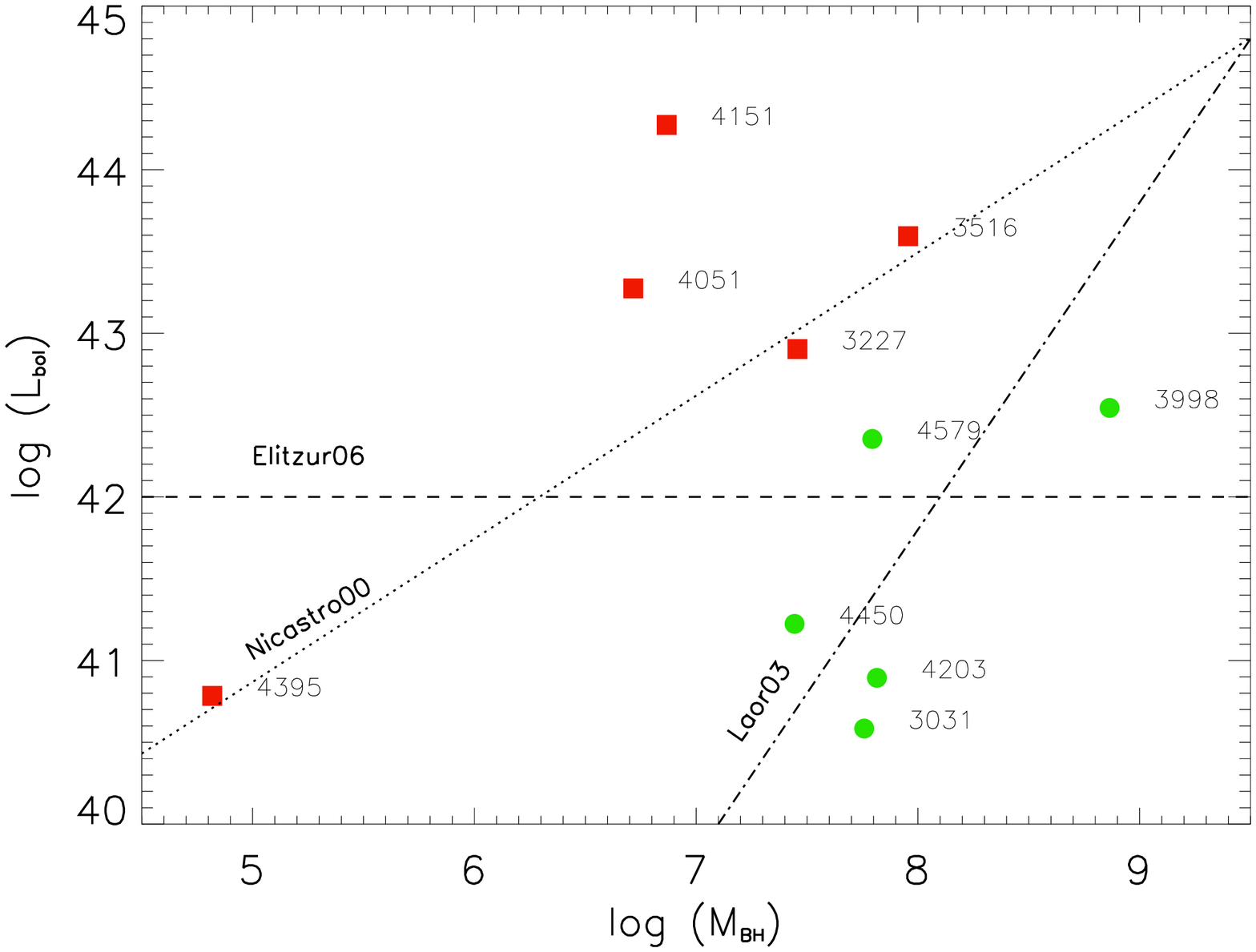}
\includegraphics[width=7.0cm,angle=90]{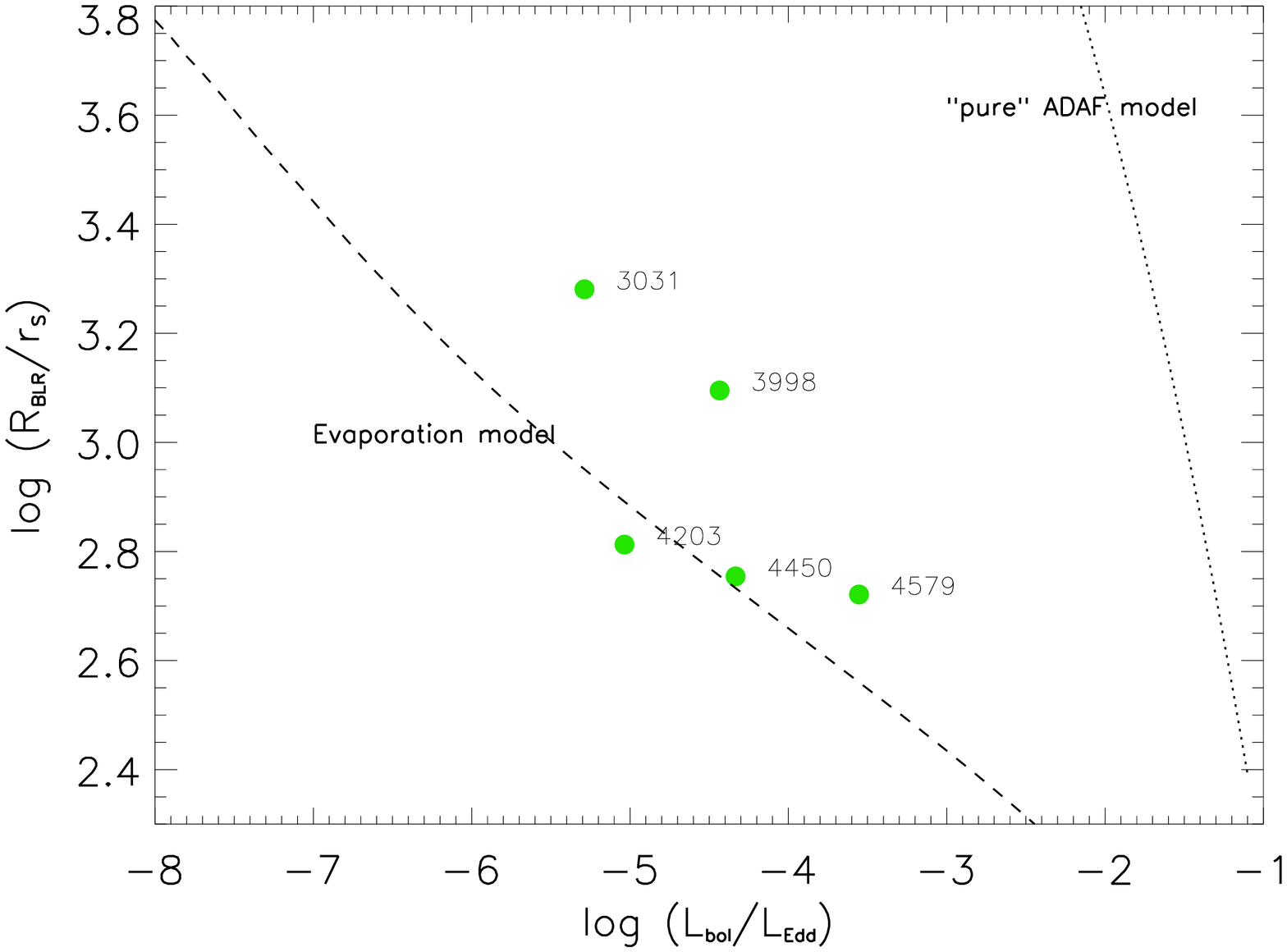}
}
\caption{Left panel: limits for the presence of a BLR from \citet{elitzur06},
  \citet{laor03}, and \citet{nicastro00}. These models predict that objects
  located in the portion of the plane below the dashed lines cannot form a
  BLR. Left panel: predictions on the transition radius between the
  geometrically thick and thin regions of the accretion disk in a pure ADAF
  model and Evaporation model (model A and C from \citealt{czerny04},
adopting $\alpha$=0.1 and $\beta$=0.99, see text.)}
\label{blrmod}
\end{figure*}

These models predict different dependencies on the bolometric luminosity and
the black hole mass (or equivalently on Eddington luminosity), below which a
BLR cannot form, but all agree that BLR cannot exist in object with bolometric
luminosity lower than $L_{\rm{bol}}\lesssim 10^{42}$$\ergs$ and black hole
mass higher than $M_{\rm{BH}}\gtrsim$10$^7M_\sun$--10$^8M_\sun$ (see
Fig.\,\ref{blrmod}). All these models fail to predict the presence of a BLR
in the two least luminous LINERs with a clear BLR. This confirms on different
grounds that the scaling relations between BLR and AGN (implicitly assumed by
all the models) do not properly predict the properties of broad emission lines
in LINERs.

\begin{figure}[htp]
\centerline{
\includegraphics[width=7.0cm,angle=90]{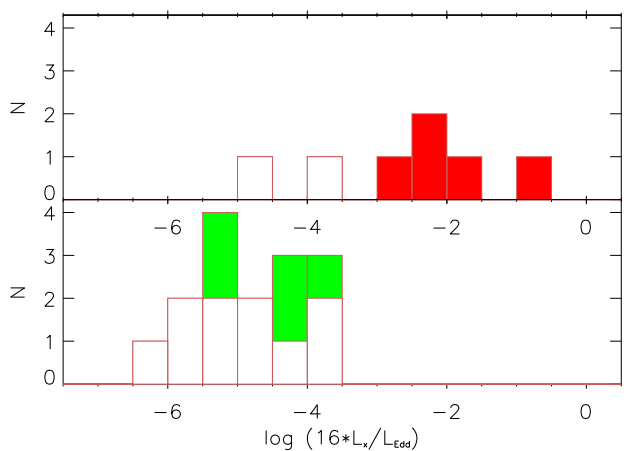}
}
\caption{Distribution of the ratio between bolometric luminosity (estimated as
  $\sim$ 16\,L$_X$) and the Eddington luminosity for LINERs (green) and
  Seyferts (red). The empty regions of the histograms represent objects with
  undetected BLR.}
\label{edd}
\end{figure}

In Sect.~\ref{undetected} we showed that the LINERs smoothly extend the
relations that, in more powerful AGN, link the nuclear and the BLR power,
representing the low end of the luminosity distributions. However, the LINERs
of our sample are not simply scale-downed versions of Seyfert galaxies.  In
Fig. \ref{edd}, we show the distributions of the Eddington ratios for the two
types of AGN. We apply a bolometric correction to the nuclear X-ray luminosity
of 16 (i.e., $L_{\rm bol}$=16\,$L_{\rm X}$, \citealt{ho08}).  LINERs and
Seyferts are well separated, in line with results obtained previously from the
analysis of larger AGN samples (e.g., \citealt{kewley06}). The overlap between
the two classes is only due to NGC~4258 and NGC~3982. As discussed above, the
measurements for NGC~3982 are not reliable because of nuclear obscuration. The
second object is NGC~4258 for which we argue that the identification as
Seyfert is questionable. Indeed, the HST spectrum shows that the \sii/\Ha\ and
\oi/\Ha\ ratios are higher that what is found in the Palomar data. In
particular, we measure \oi/\Ha=0.96, which locates NGC~4258 well into the LINERs
region, regardless of its \oiii/\Hb\ ratio.\footnote{This result is similar to
  what is found for NGC~5252, an object with a LINER nuclear spectrum, and a
  large scale line emission with ratios typical of Seyferts
  \citep{goncalves98}. The separation between the two classes occurs at
  $L_{\rm bol}/L_{\rm Edd} \sim 10^{-3}$.}

Various models predict that below a given threshold in Eddington ratio
($L_{\rm{bol}}/L_{\rm{Edd}}$ $\lesssim$ 0.01), the radiatively efficient
accretion disk (geometrically thin and optically thick), which is typical of powerful
AGNs, changes its structure into an inner hot and radiatively inefficient flow
(optically thin and geometrically thick), possibly an advection-dominated
accretion flow, ADAF, (e.g., \citealt{narayan95}) while only at large radii does the
standard disk survive.  Application of an ADAF model to the SED of low
luminosity AGNs suggests a transition radius (between the inner ADAF and the
outer disk) on the order of $R_{\rm{tr}} \sim 100-1000\,r_{\rm{s}}$,
\citep{ho08},\footnote{Equivalent to $R_{\rm{tr}} \sim 1.14\,(10^{-8}-10^{-7})
  M_{BH}$ in light days} to explain the absence of the so-called "blue bump",
the excess in the UV continuum typical of Seyferts galaxies.

From a theoretical point of view, the hot corona is truncated at the radius
where the ADAF solutions ceases to exist. Following \citet{czerny04}, the
truncation radius is given by $$R_{\rm{BLR}}=2.0\,\left({\cal
    F}^{-1}\,\dot{L}\right)^{-2}\alpha^4_{0.1}\,r_{\rm s}$$ (model A in
\citealt{czerny04}, where $\alpha$ is the viscosity parameter, and ${\cal
  F}=\dot{L}/\dot{m}$ is the accretion radiative efficiency, function of model
parameters and of the accretion rate).

However, taking the process of evaporation of the cold disk in
the ADAF into account, and assuming a two-temperature model of the corona ("Evaporation
model C" in \citealt{czerny04}), one obtains a different prescription for the
transition radius:
$$R_{\rm{BLR}}=19.5\,\left({\cal F}^{-1}\,\dot{L}\right)^{-0.53}\alpha^{0.8}_{0.1}\beta^{-1.08}r_{\rm
  s}$$
where $\beta$ is ratio of the gas pressure to the total (gas+magnetic)
pressure. 

In Fig.\ref{blrmod} we report the dependence of the transition radius on the
Eddington ratio predicted by the two models and compare it with the
estimated BLR radii for LINERs. We find the $R_{\rm BLR}$ values in LINERs are
clustered in the range 500 - 2,000 $r_{{\rm s}}$. This is much lower than the
truncation radii predicted by the pure ADAF model, but they are comparable to
the values of the transition radii obtained from the evaporation
model.\footnote{We adopted a viscosity parameter $\alpha$=0.1 and a magnetic
  parameter $\beta$=0.99. The `model C' curve shifts downward by only
  $\sim$0.1 dex for, e.g., $\alpha$=0.04, and upward by $\sim$0.9 dex for,
  e.g., $\beta$=0.1; therefore, the transition radii are quite sensitive to the
  magnetic field strength.}

This supports earlier suggestions (e.g., \citealt{czerny04,liu09}) that the
broad line clouds do not form (or cannot survive) within the hot part of the
accretion flow, i.e., below the transition radius. Conversely, the BLR appears
to naturally form in the presence of a thin (and relatively `cold') accretion
disk, which, in the case of LINERs, represents the outer portion of the
disk structure.

This offers an explanation for the discrepancies found for LINERs behavior
with respect to more powerful AGN and the violation of the BLR scaling
relations. Indeed the BLR radii estimated for the LINERs from such relations
would be $\sim$ 0.3 and 3 lt-days (i.e., between $\sim$ 30 and 250 $r_s$
for a $10^8 M_{\rm BH,\sun}$ black hole), well below the values obtained
for the transition radii of the hot corona. We suggest that the BLR radius in
these objects is not set by the AGN luminosity but from the presence of an
inner forbidden region.

It is also interesting to estimate the dust sublimation radius. This is given
by $R_{\rm dust} \sim 1.3 \,L_{46}^{0.5}$ pc, assuming a sublimation
temperature of 1500 K \citep{barvainis87}. For the LINERs considered here we
obtain $R_{\rm dust} \sim 0.4 - 10 $ lt-days (and this must be considered as
an upper limit, since our optical luminosities might be overestimated). These
values are well within the ADAF region and much less than the BLR radii.
This suggests that the structure of circum-nuclear tori in these sources (if
at all present) differ profoundly from those seen in more powerful
AGN. The likely
absence of a standard Seyfert-like torus in low luminosity AGNs is also
confirmed by high resolution MIR observations \citep{mason12}.
The authors suggest that LLAGN may have a dust-to-gas ratio lower than
most Seyfert galaxies, consistent with a disk wind scenario, in which
at low Eddington ratio, the torus may be optically thin or may contain
fewer clouds than a standard torus.

\section{Summary and conclusions}
\label{summary}

We explored the properties of the BLR in low luminosity AGNs by using archival
HST/STIS spectra. We considered the objects in which the presence of broad
lines has been reported from their Palomar ground-based spectra. High spatial
resolution data, such as those obtained with HST, are essential to achieving
reliable measurement of the properties of the BLR, especially for low
luminosity AGN. Indeed the smaller aperture of HST, which is more than $\sim$$100$
times smaller than the aperture of the Palomar survey, allows us to decrease
the stellar continuum and reduce the narrow line contamination.

In the HST archive we found data for 16 LINERs, 7 Seyferts, and 1 ambiguous
galaxy. We separated the broad line from the \Ha+\nii\ narrow line complex by
applying the shape of the de-blended lines profile of the \sii\ doublet and by
scaling the line intensity in order to match the \Ha+\nii\ peaks.  

We confirm the presence of broad line emission in all but one Seyfert: the HST
line widths are similar to those derived from the Palomar observations, while
the fluxes typically differ by less than a factor 1.5 from the ground-based
data. The only exception is NGC~3982, a source with a highly absorbed nucleus
in the X-ray, where the BLR is most likely obscurated.  

The discrepancies between HST and ground-based measurements are instead
remarkable for LINERs. Only in five LINERs is a BLR is readily visible in the HST
spectra. However, there is a large spread between the HST and ground-based
data, with the fluxes differing by a factor ranging from 0.3 to 20;
furthermore, the HST measurements of the line widths (between 4200 and 8000
\kms) are substantially larger than the ground-based FWHM (with all widths in
the range 1500 - 3000 \kms).

Conversely, we do not find convincing evidence of a BLR in the remaining 14
sources. A BLR must generally be included to obtain a good fit to their
spectra when using the \sii\ lines as template for the \nii\ and the narrow
component of the \Ha. However, if we instead use the \oi\ lines we obtain
rather different results, because the \oi\ lines are
broader than the \sii\ lines. Furthermore, the spectra are also
reproduced well without any BLR but by just adding a blue wing to the \nii\ and
narrow \Ha\ lines.

This is likely due to a stratification in density and ionization within the
NLR that causes differences in the location of the emitting region for the
various lines and, consequently, differences in the lines profiles.  For
example, the existence of a compact and dense emitting region, located within
a radius of a few pc from the central black hole, is supported by various
studies. This region is poorly represented in the \sii\ profile owing to the low
critical density associated with this transition. 

We conclude that complex structure of the NLR is not captured with the
technique of spectral fitting based on the narrow line templates that it does
not return, in general, robust constraints on the properties of the BLR in
these low luminosity AGN.

For the ten galaxies in which instead a BLR is clearly detected, we estimated
its radius by assuming the dominance of gravitational motions, i.e., by
applying the virial formula, knowing the value of the $\sigma$-derived
black hole mass and the broad line width. The resulting BLR radii in the five
LINERs are clustered around $\sim$ 1,000 Schwarzschild radii (i.e., $\sim 3 -
100$ light days). These values are significantly higher, by a factor of $\sim$
10 to 100, than the extrapolation to low luminosities of the scaling relations
linking radio and luminosity of the BLR.

Our preferred interpretation to account for this inconsistency relies on the
change in the accretion disk structure at low luminosities. LINERs differ from
Seyfert for their lower accretion rates, causing the formation of an inner
region dominated by an advection-dominated accretion flow (ADAF). The location
of the transition radius between the ADAF region and the outer thin disk 
predicted by the evaporation model
compares favorably with the estimated BLR radii for LINERs. This confirms
earlier predictions that the BLR cannot coexist with the hot inner region and
that they form (or survive) only in the presence of a thin accretion disk.  As
a result, LINERs do not obey the scaling relation defined by more powerful
AGN.

The structure of circumnuclear tori, if at all present in these sources, must
also differ profoundly from those seen in more powerful AGN since
the estimated dust sublimation radii are smaller than the BLR size.

An interesting result of this study is that, despite the differences in BLR
structure between LINERs and Seyfert, there is a continuity between the
relations between BLR luminosity and AGN power for the two groups. This
implies that the BLR covering factor is similar for the two classes of AGN.
Nonetheless, a BLR is not readily visible in most LINERs. We used the
relations linking the BLR and the AGN power to predict the broad line
luminosity, hence whether the BLR in these objects should instead be
seen. We find that the BLRs could not be detected if their widths follow the
scaling relation of more powerful AGN, because of the extremely large FWHM,
predicted to have a median value of $\sim$ 30,000 \kms. However, if the BLR
radius in LINERs has a constant value, $R_{\rm BLR} \sim 1,000 r_s$, broad
lines should be easily observed in at least three LINERs, based on their
multiband luminosities. This might suggest that the BLR in LINERs are
transient phenomena. We also note that the three objects where we would have
expected to see a BLR are all radio-loud galaxies. 

An important step forward for a better understanding of the BLR in LLAGNs can
come from variability and reverberation mapping studies of these objects. They
can provide us with a direct estimate of the BLR size, proofing or refuting
our estimates based on the virial formula. Indeed, our assumption of the
dominant role of the gravitational motions should be tested, and the role of
various effects (e.g., related to the presence of winds or to the importance
of radiation pressure) must be assessed. For example, it might be envisaged
that the BLR in LINERs has a pure wind origin, lacking a rotating disk
component, and consequently invalidating our analysis.

\begin{acknowledgements}  
We are grateful to the anonymous referee for usuful comments that significantly improved the paper.
This work was mainly supported by the Italian Space Agency through contract ASI-INAF I/009/10/0 
and ASI/GLAST I/017/07/0. 
\end{acknowledgements}

\appendix
\section{Spectral modeling}
\label{appA}

In this Appendix we report the result of modeling the objects where
the BLR is not clearly visible in their spectra. For four galaxies, the spectra
cover the \oi\ spectral region, and these lines are bright enough to
allow us to use them as templates. In these cases, we modeled the spectra with
three different methods 1) by using the \oi\ lines as templates and including a
broad \Ha\ component, 2) by using the \oi\ template but adding a
wing to the \nii\ and narrow \Ha\ lines and 3) by using the
\sii\ lines as template and including a BLR.

For the eight remaining galaxies, the spectra do not cover the \oi\ spectral region
or these lines are not bright enough to be used as templates.  In these
cases we modeled the spectra with two different methods by using the \sii\
lines as templates and 1) including a broad \Ha\ component or 2) adding a wing to the \nii\ and narrow \Ha\ lines.

\begin{figure*}
\centerline{
\includegraphics[scale=0.37,angle=0]{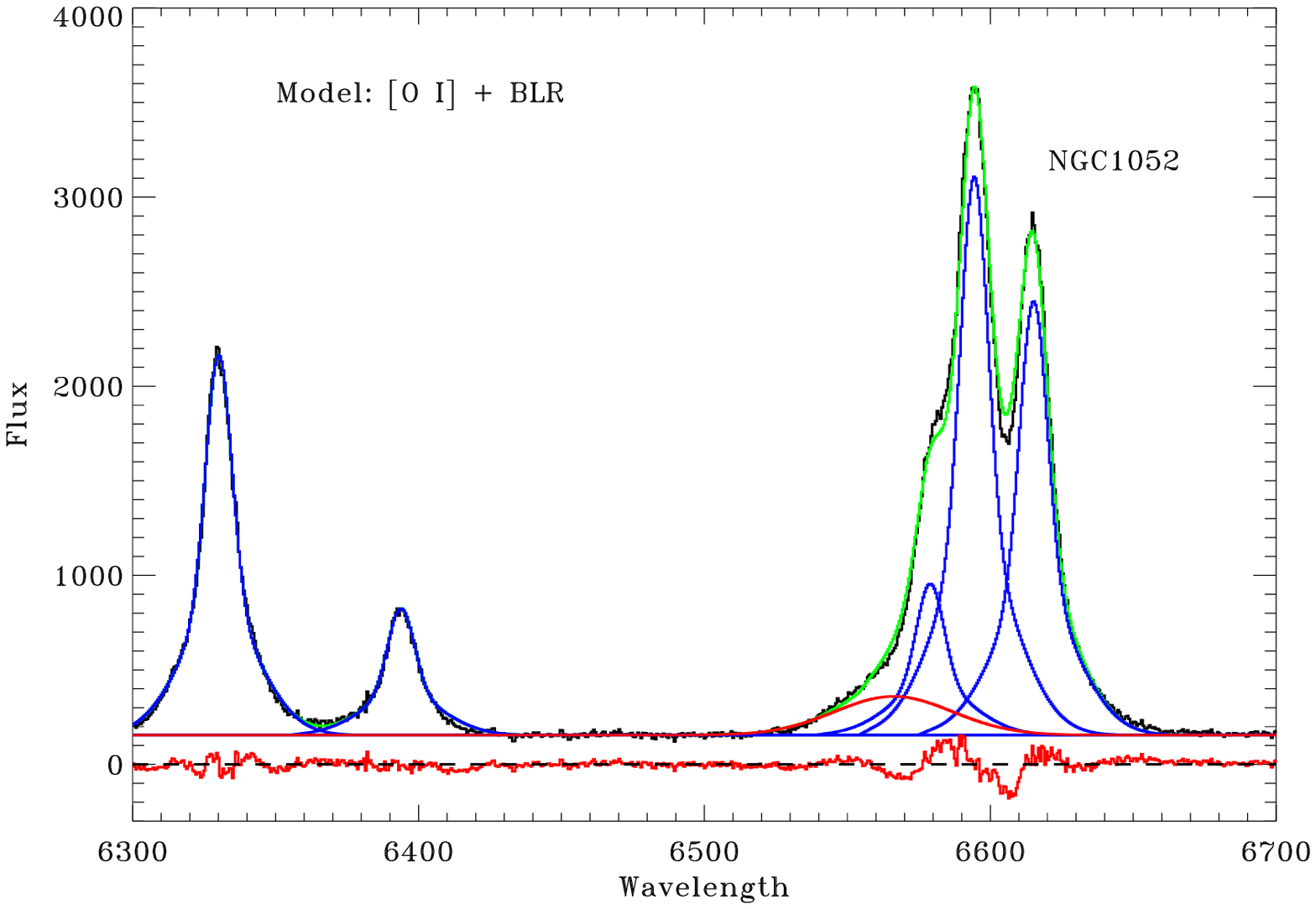}
\includegraphics[scale=0.37,angle=0]{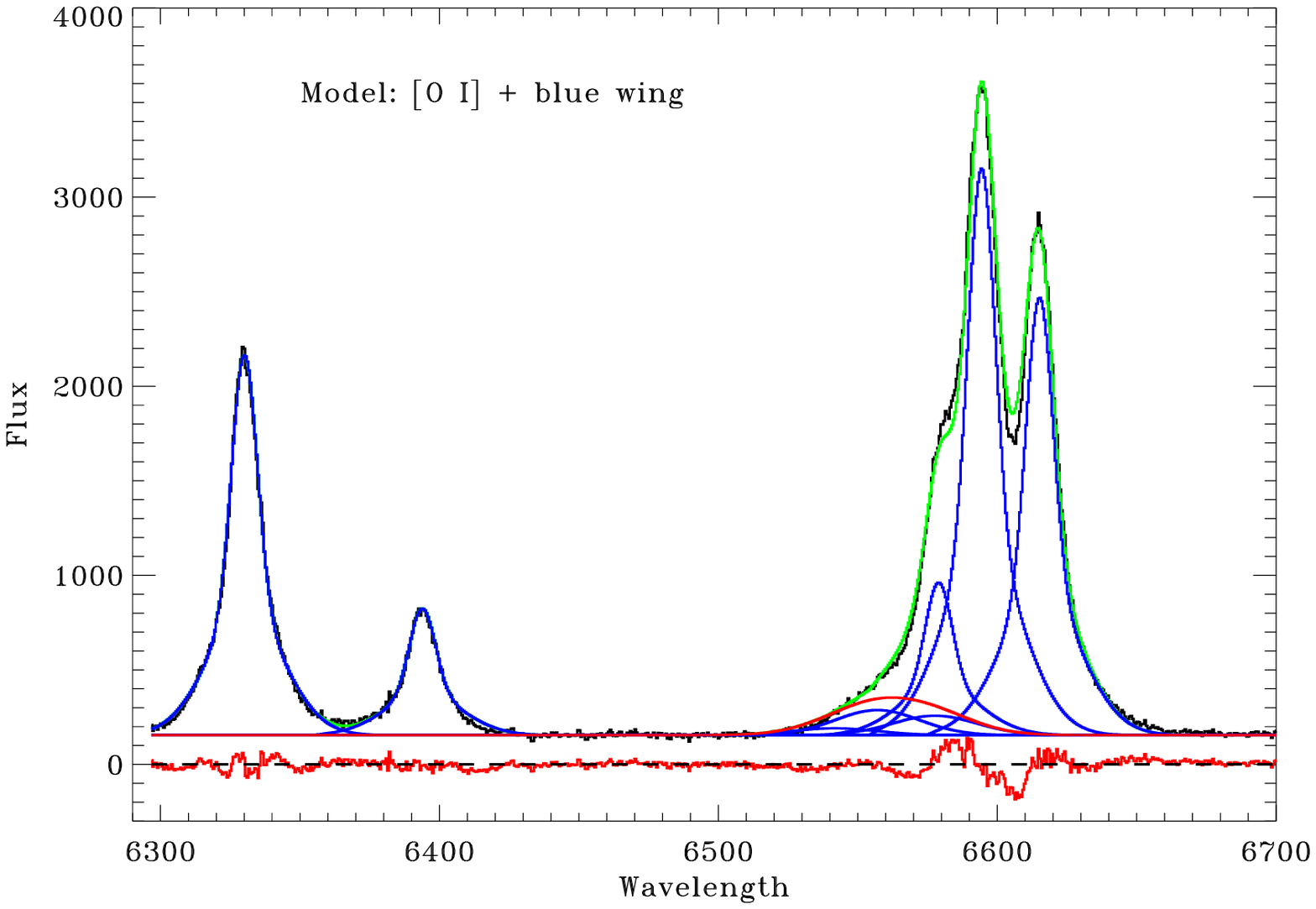}
\includegraphics[scale=0.37,angle=0]{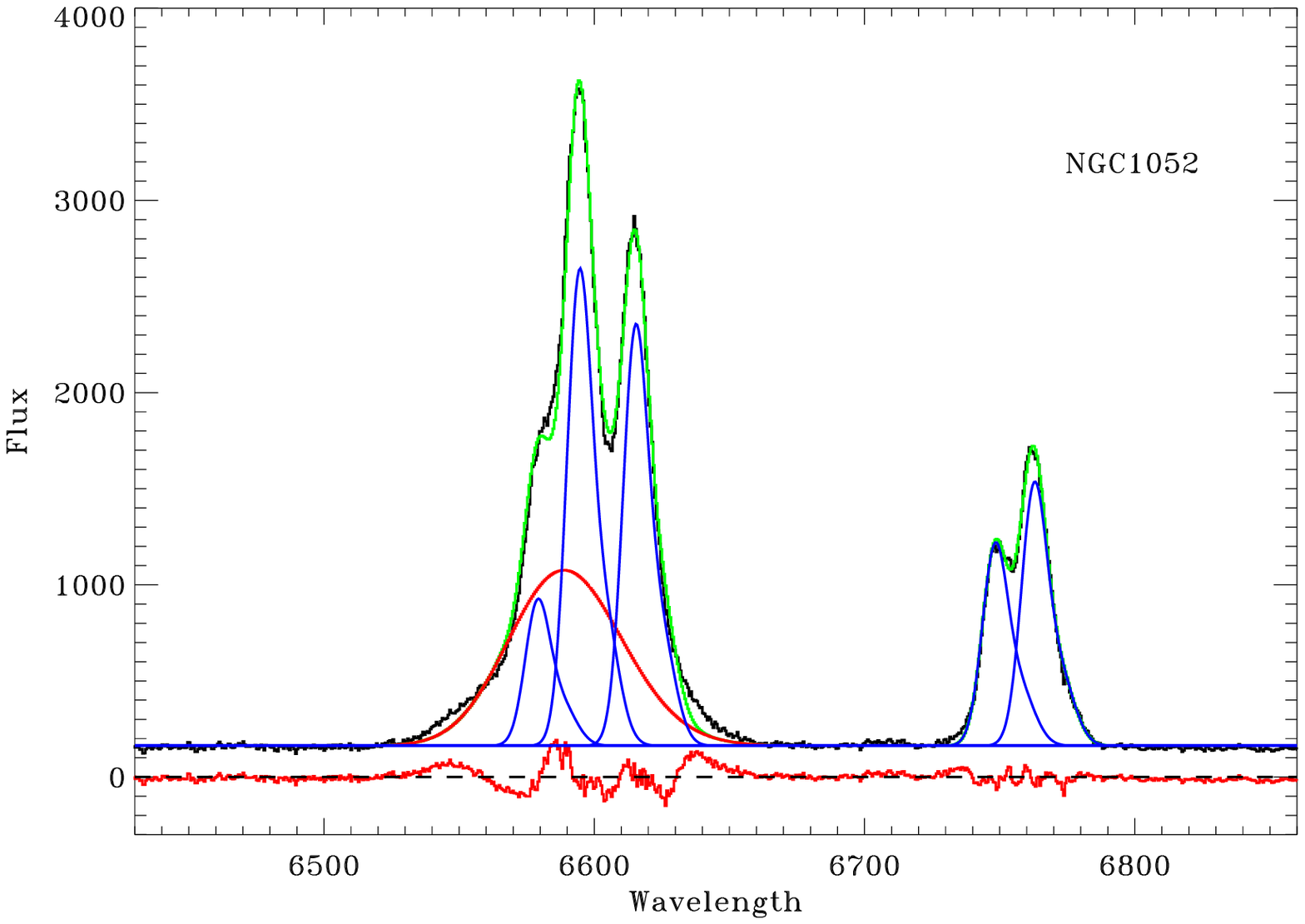}
}
\centerline{
\includegraphics[scale=0.37,angle=0]{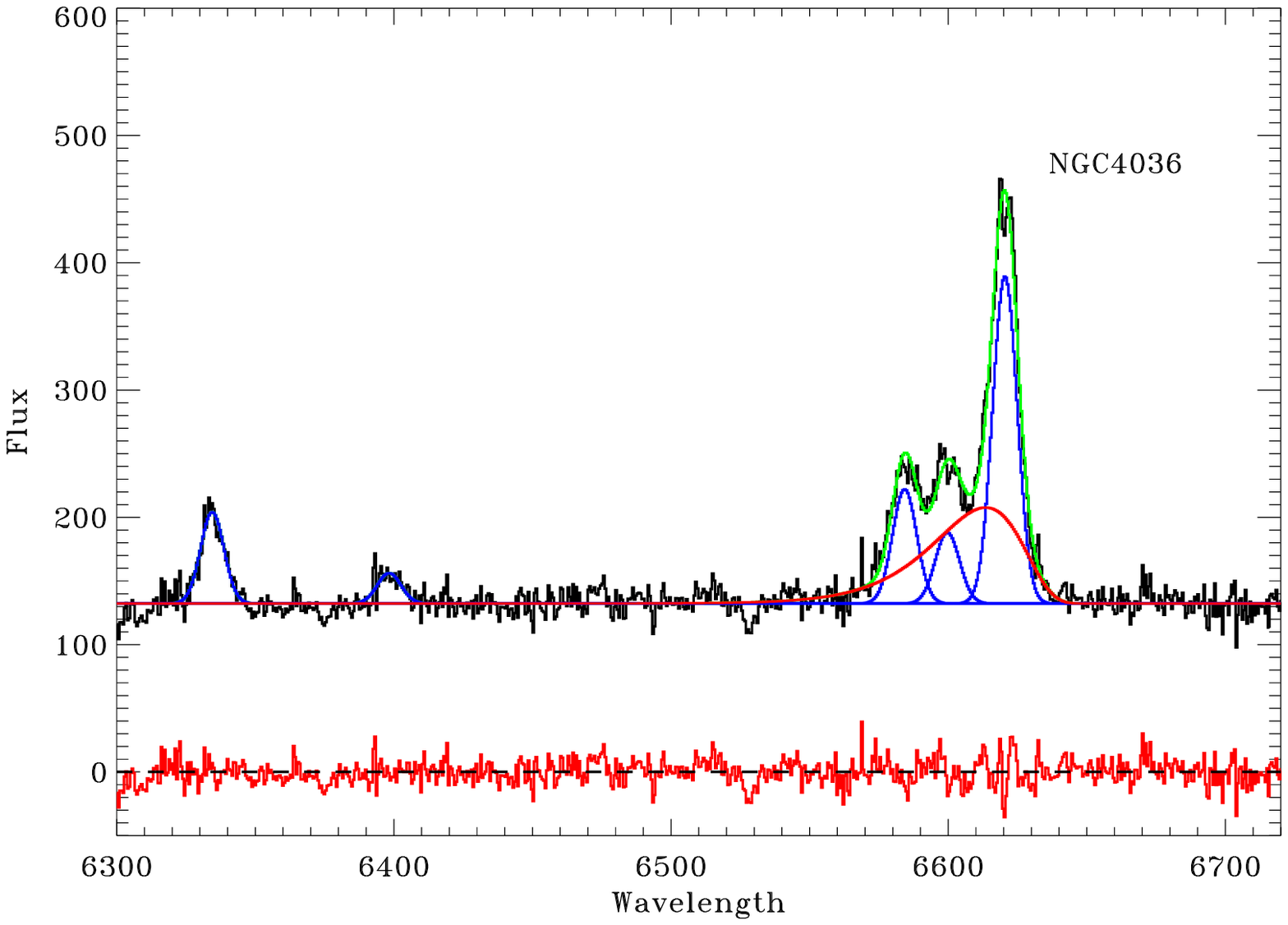}
\includegraphics[scale=0.37,angle=0]{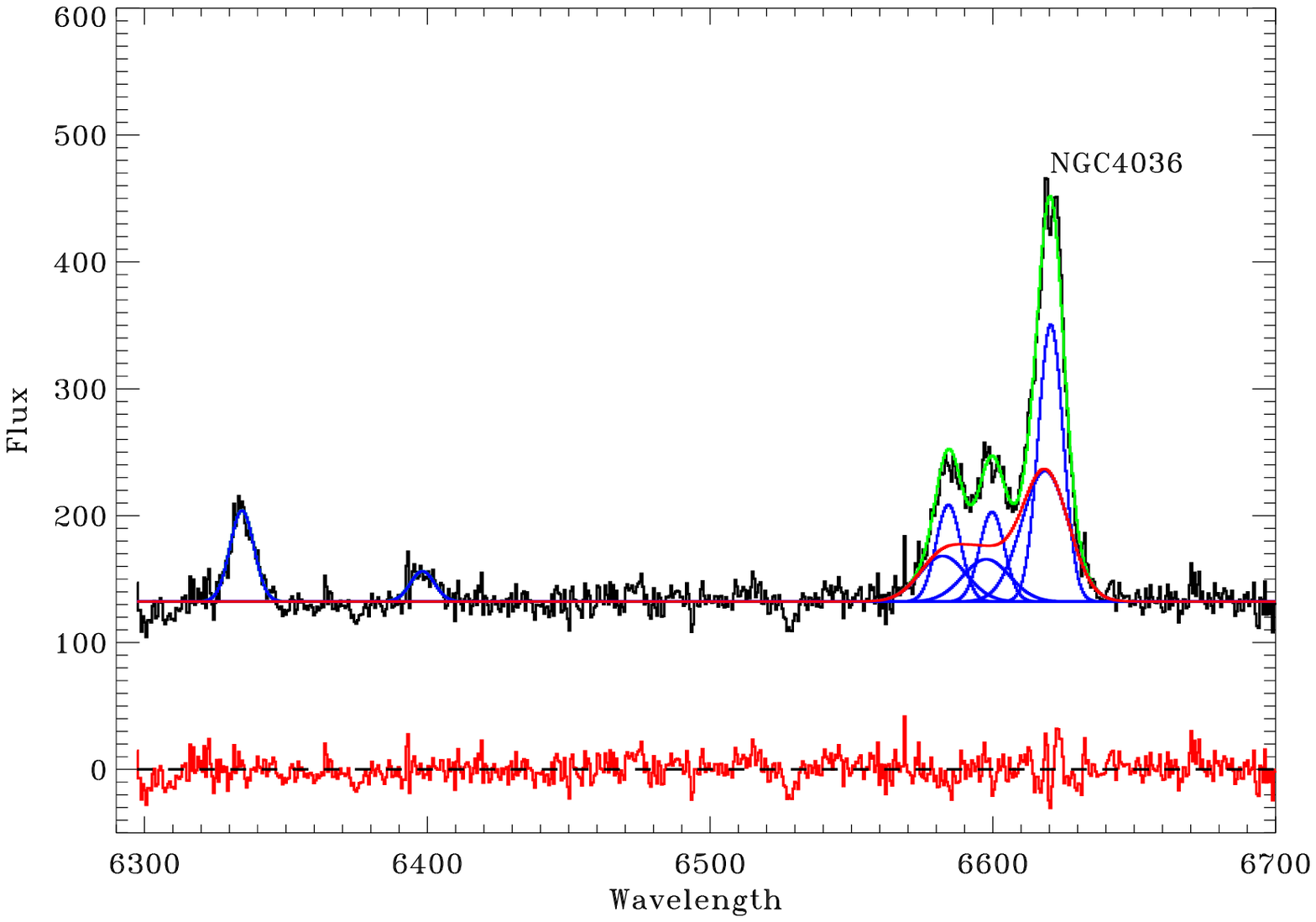}
\includegraphics[scale=0.37,angle=0]{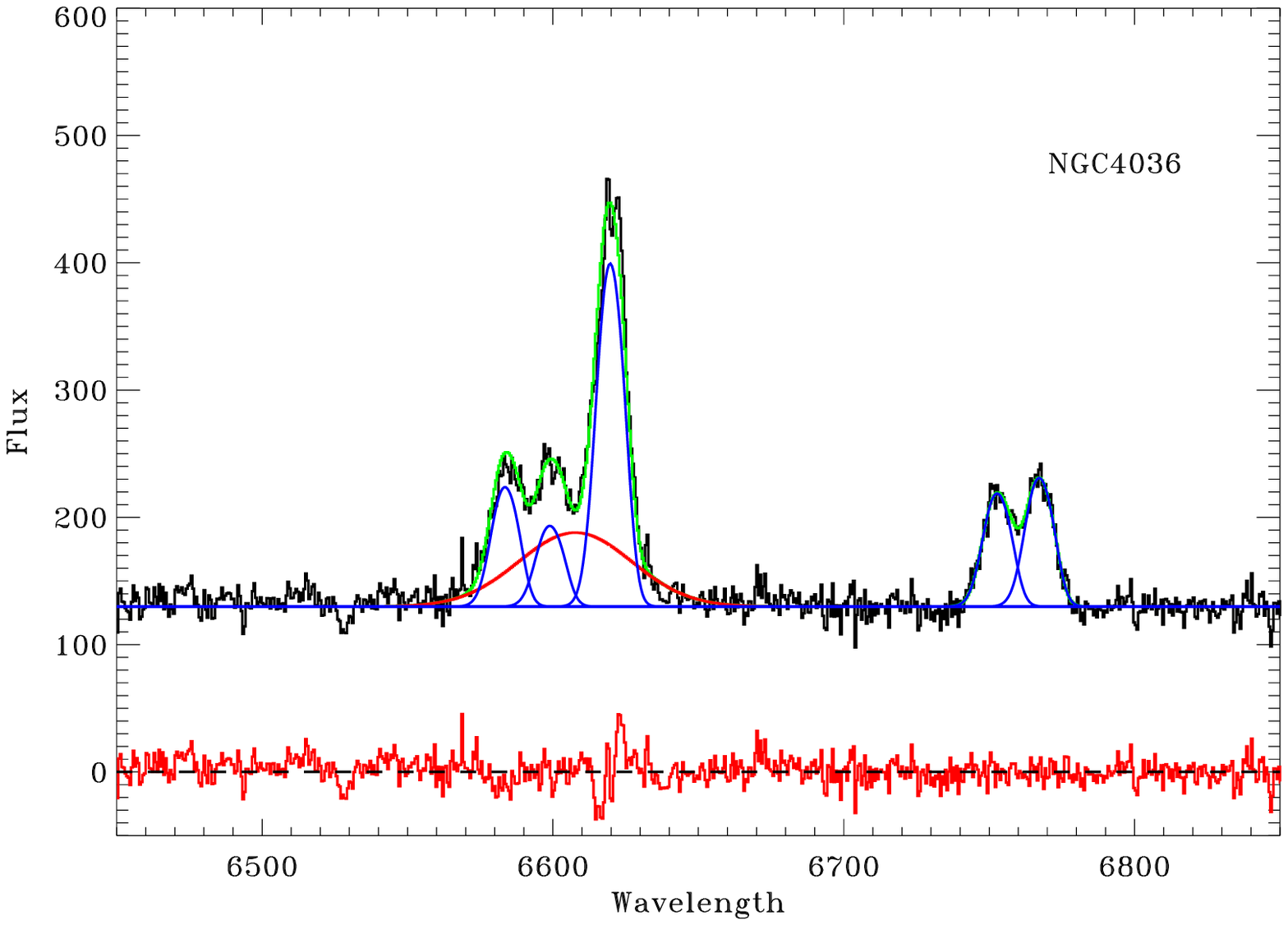}
}
\centerline{
\includegraphics[scale=0.37,angle=0]{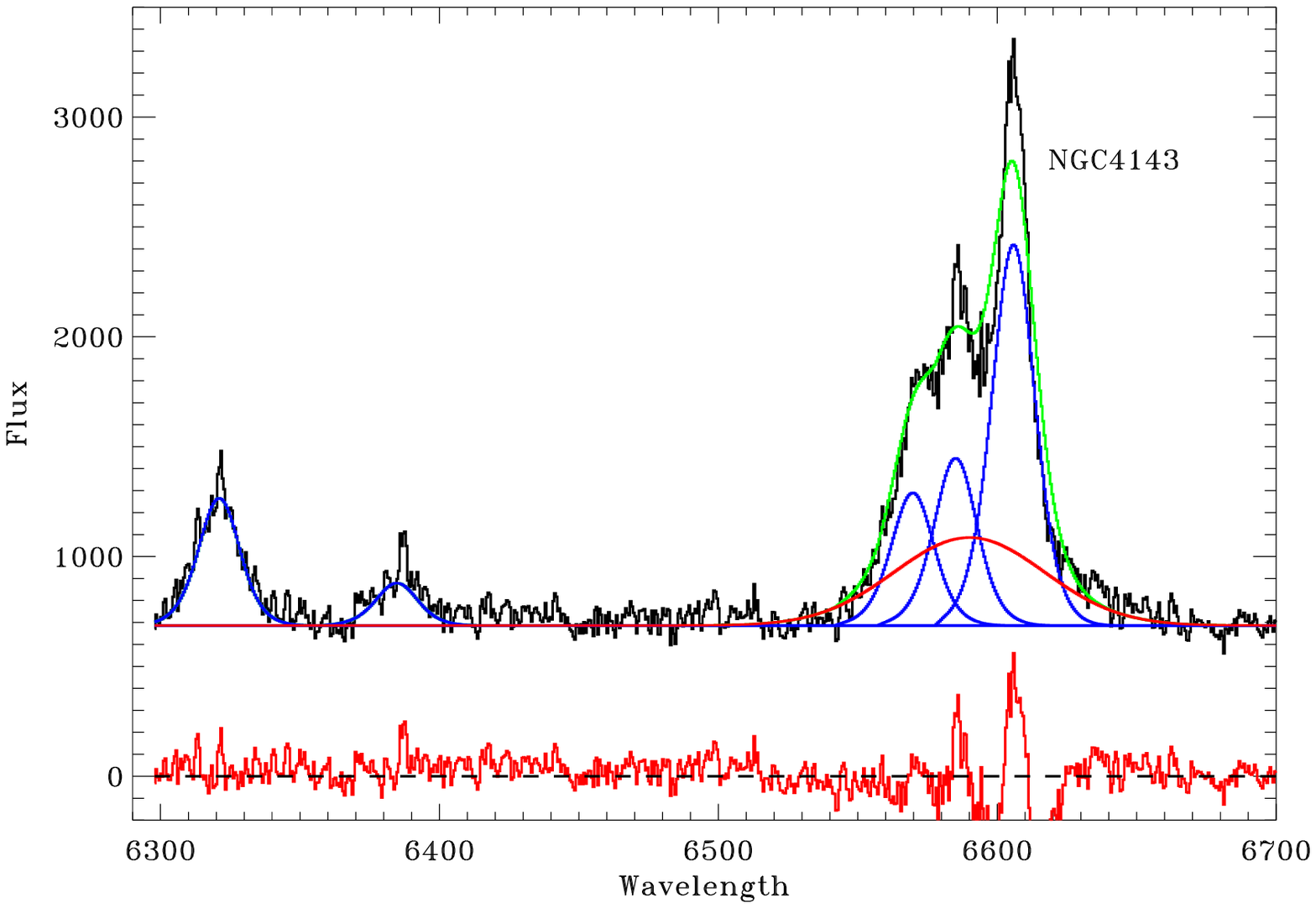}
\includegraphics[scale=0.37,angle=0]{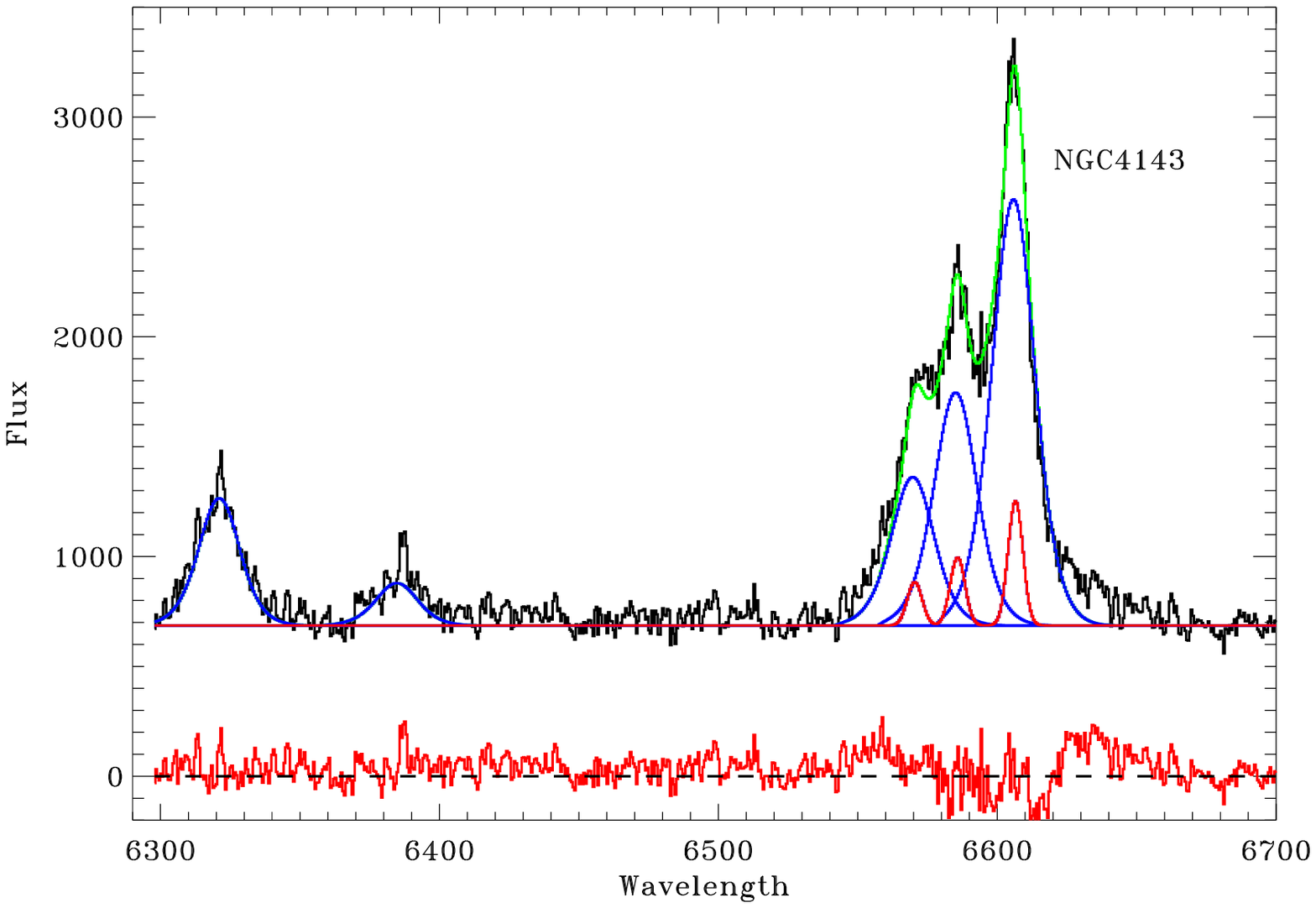}
\includegraphics[scale=0.37,angle=0]{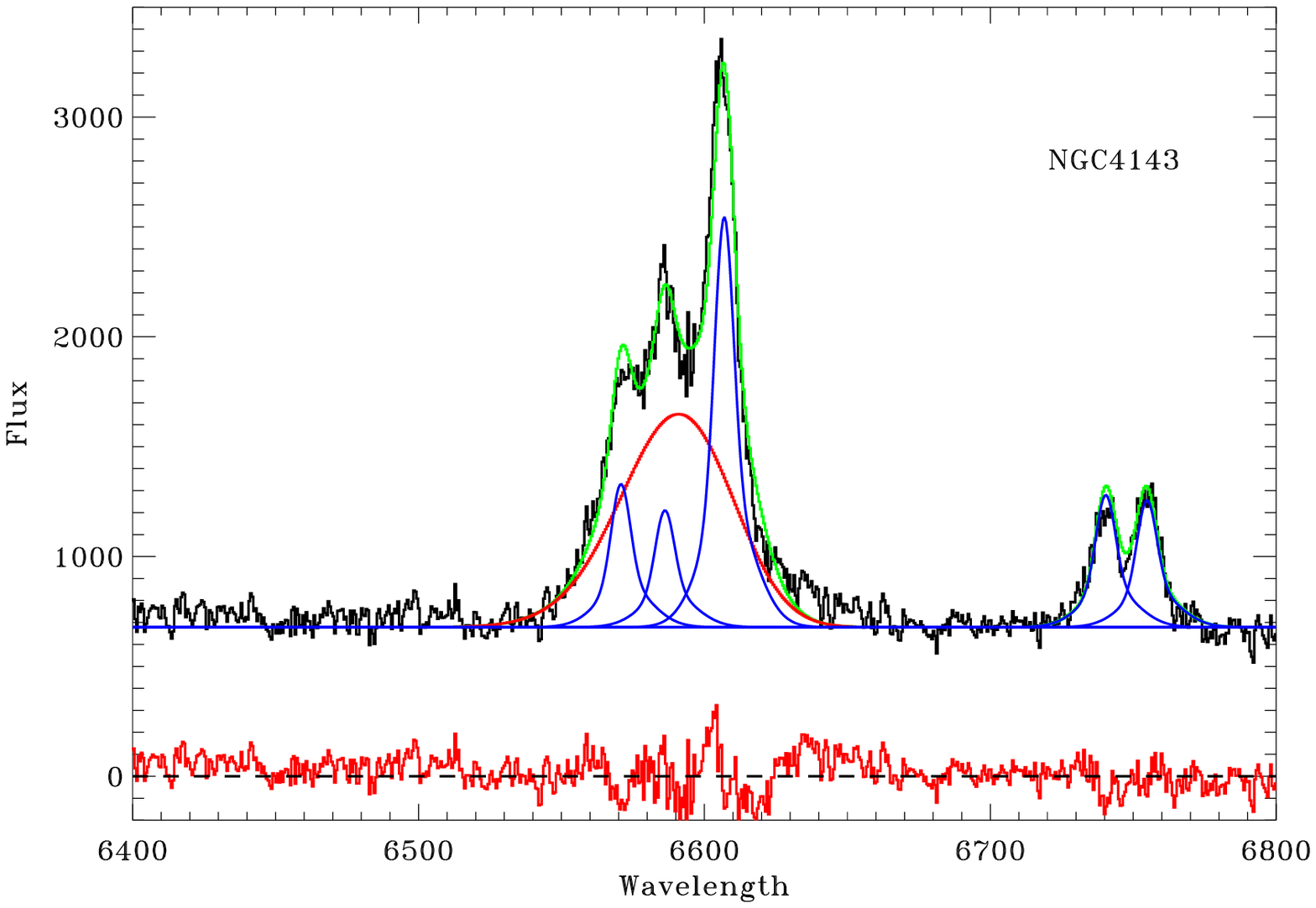}
}
\centerline{
\includegraphics[scale=0.37,angle=0]{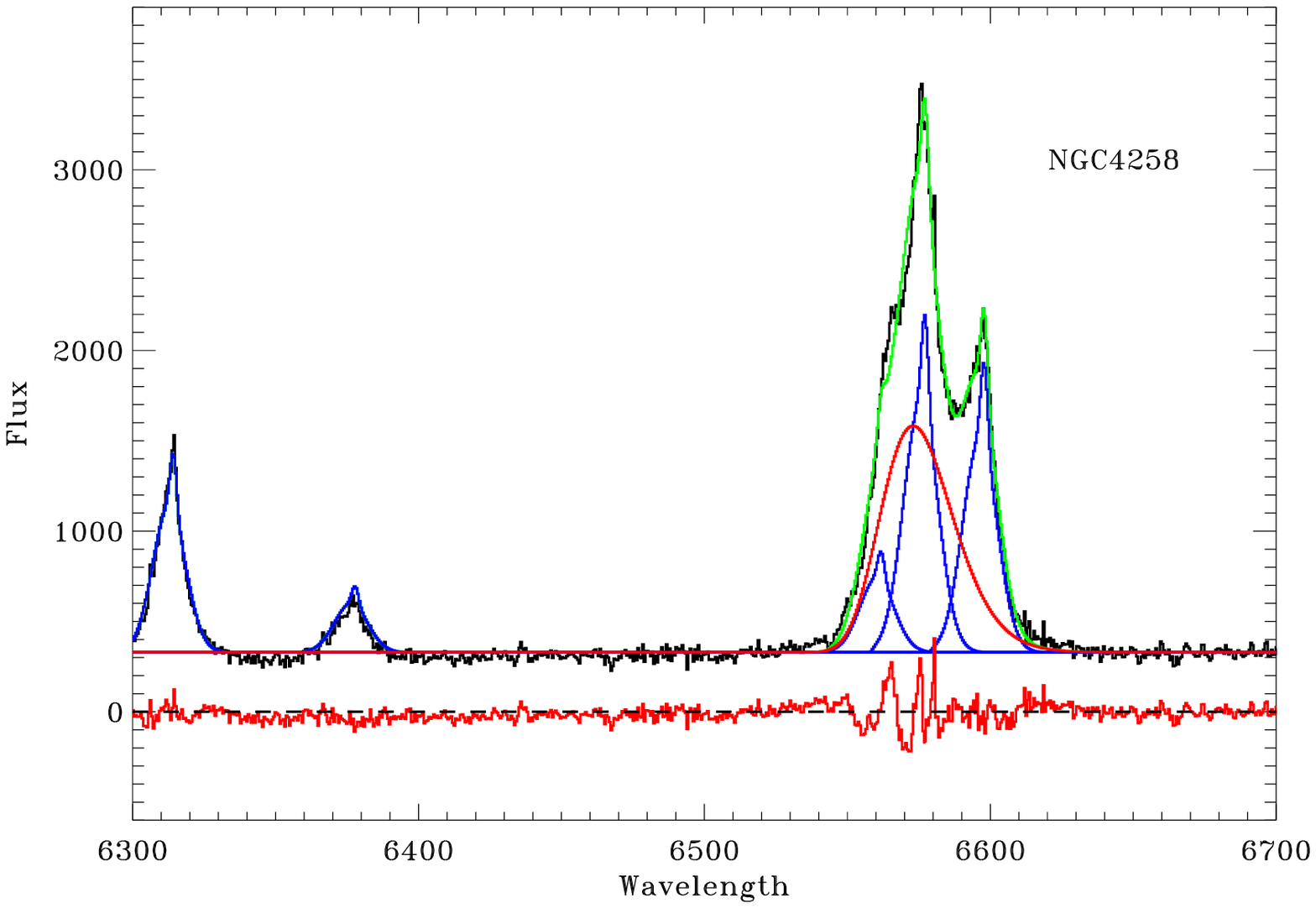}
\includegraphics[scale=0.37,angle=0]{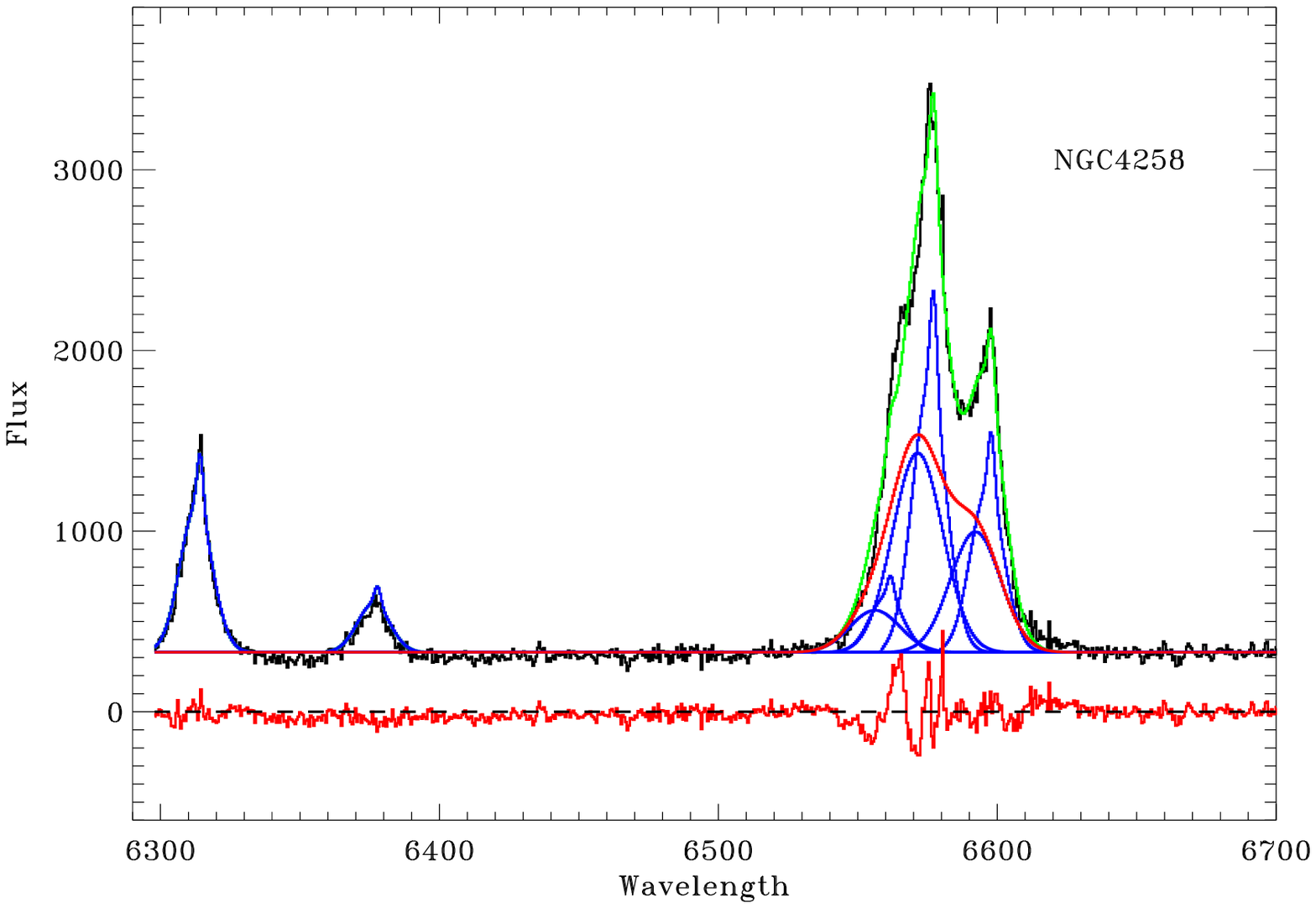}
\includegraphics[scale=0.37,angle=0]{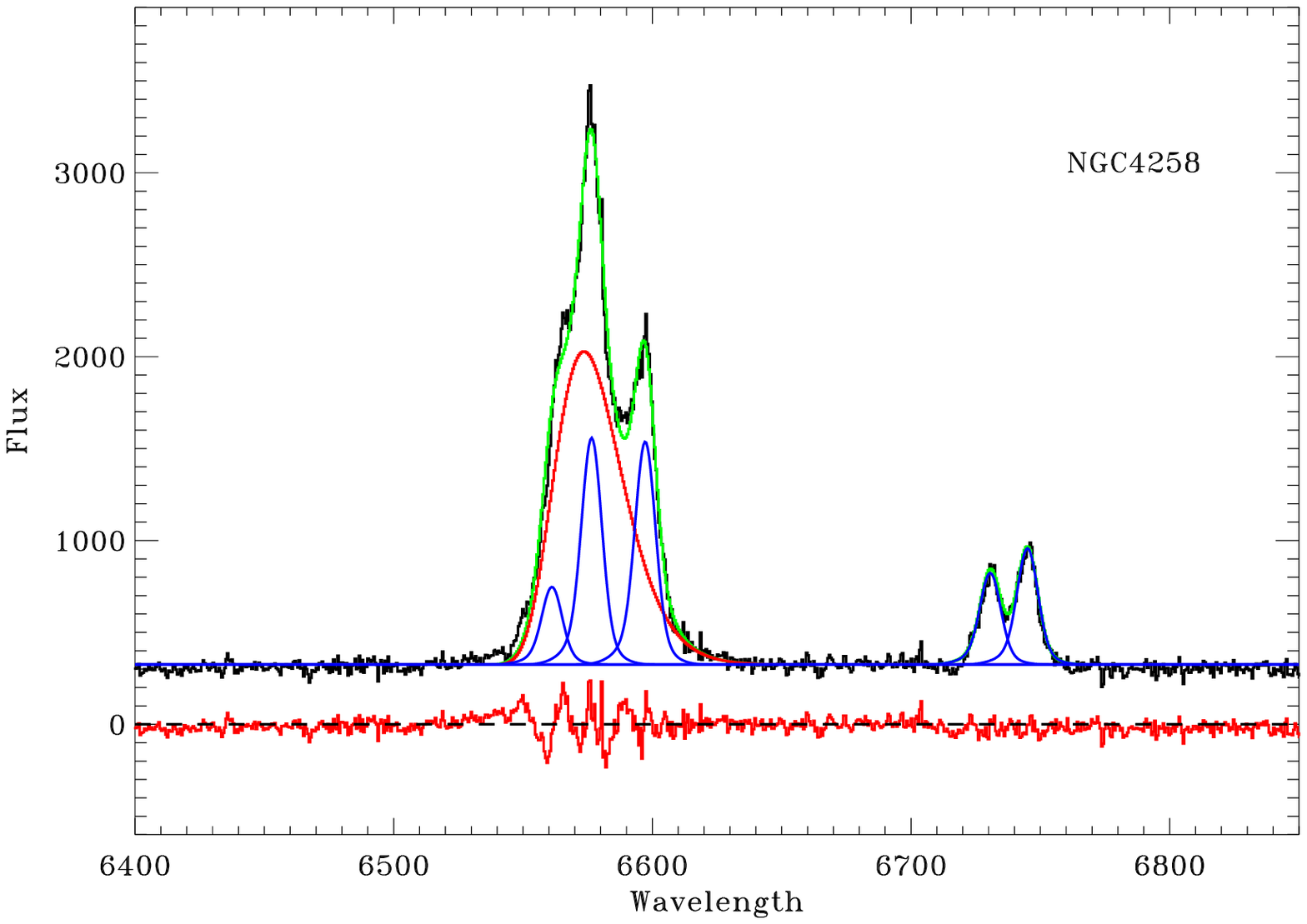}      
}
\caption{Modeling of the spectra of 4 objects without a readily visible BLR
  modeled with 3 different methods: by using the \oi\ lines as templates and
  including a broad \Ha\ component (left column), again with the \oi\ template
  but adding a wing to the \nii\ and narrow \Ha\ lines (center column), by using as
  template the \sii\ lines and including a BLR (right column). The original
  spectrum is in black, the contribution of the individual narrow lines in
  blue (their sum is in green), and in red are the residuals. The BLR
  component is in red.}
\label{clearblr1}
\end{figure*}

\begin{figure*}
\centerline{
\includegraphics[scale=0.26,angle=0]{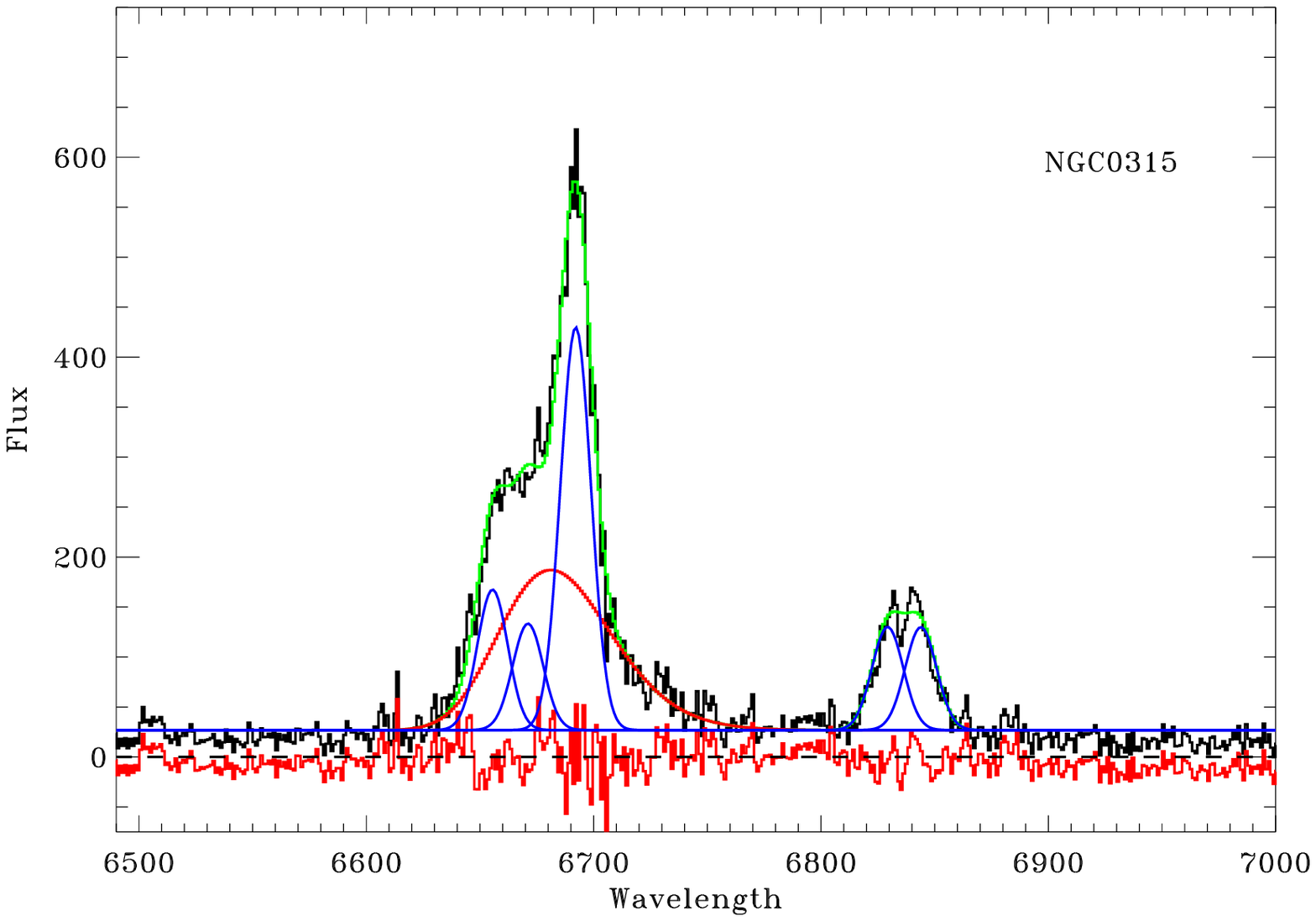}
\includegraphics[scale=0.26,angle=0]{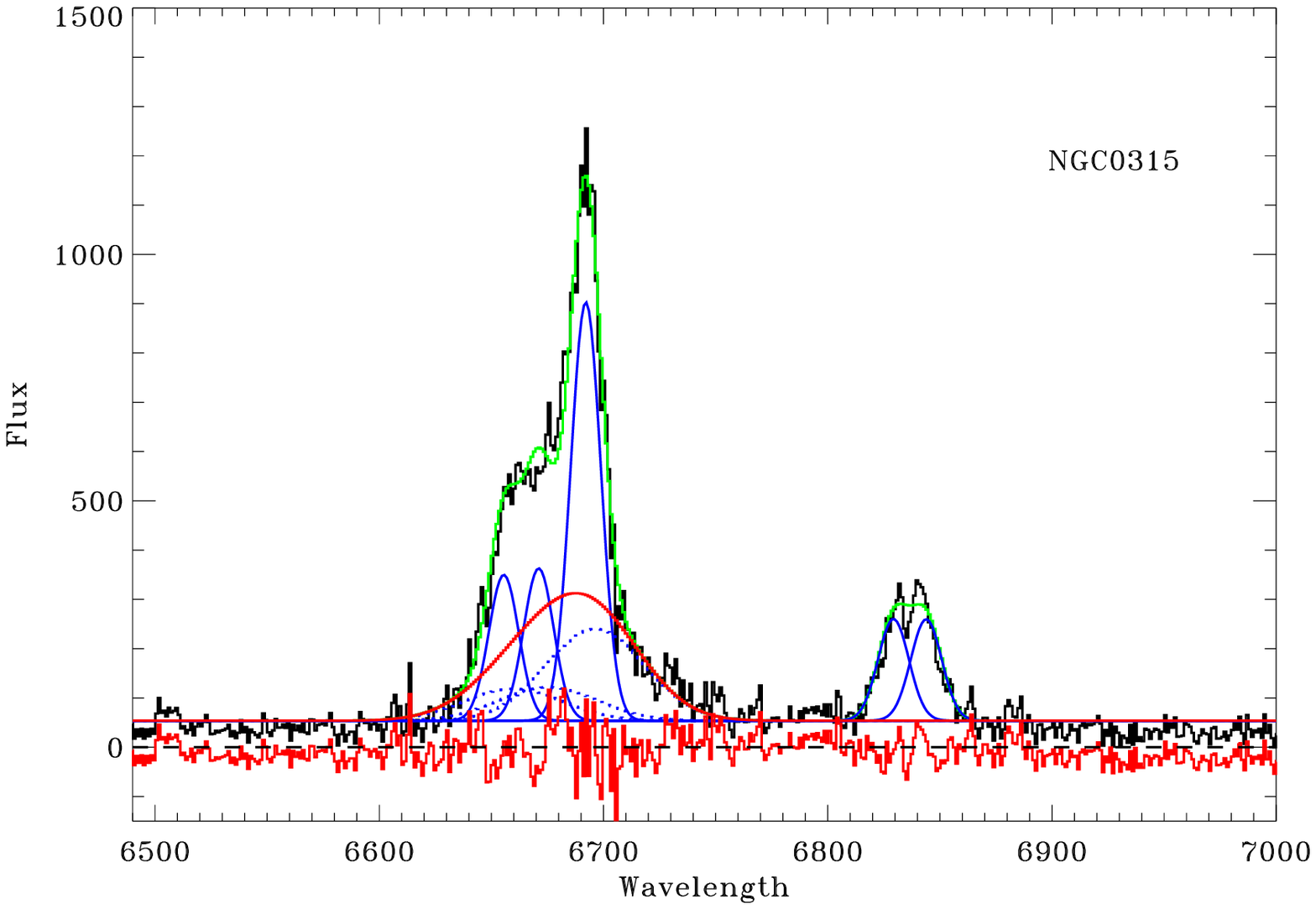}
\includegraphics[scale=0.26,angle=0]{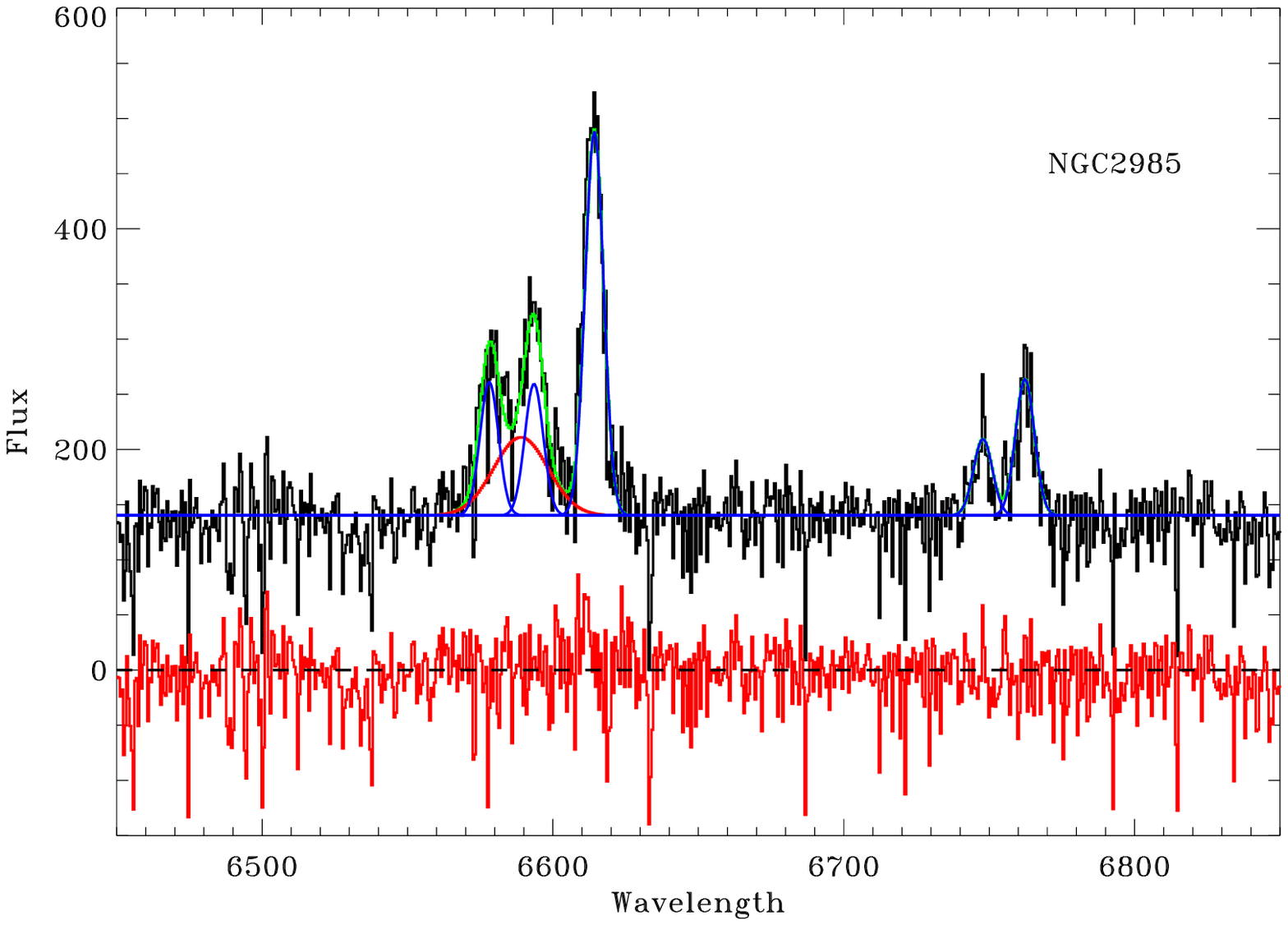}
\includegraphics[scale=0.26,angle=0]{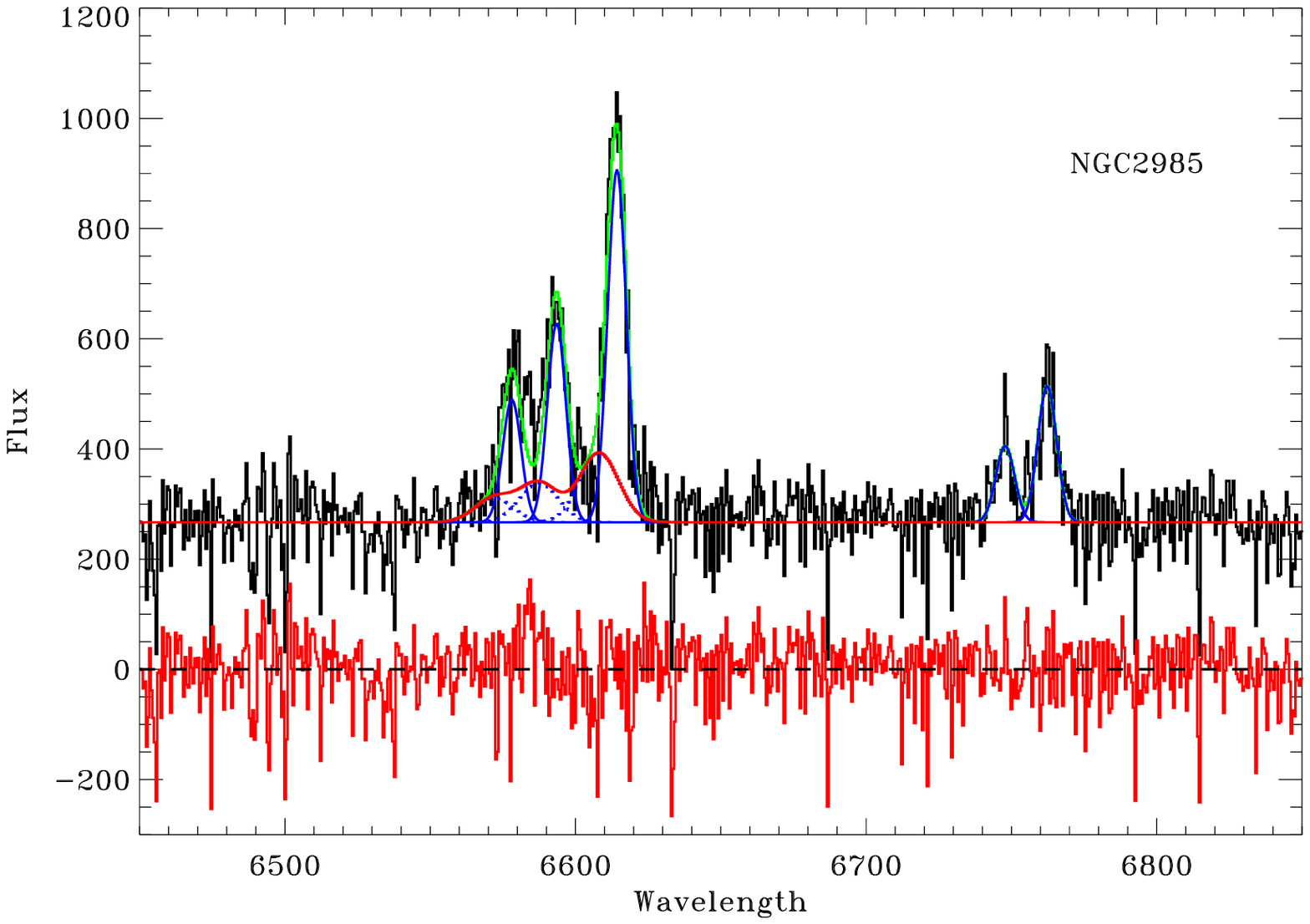}
}
\centerline{
\includegraphics[scale=0.26,angle=0]{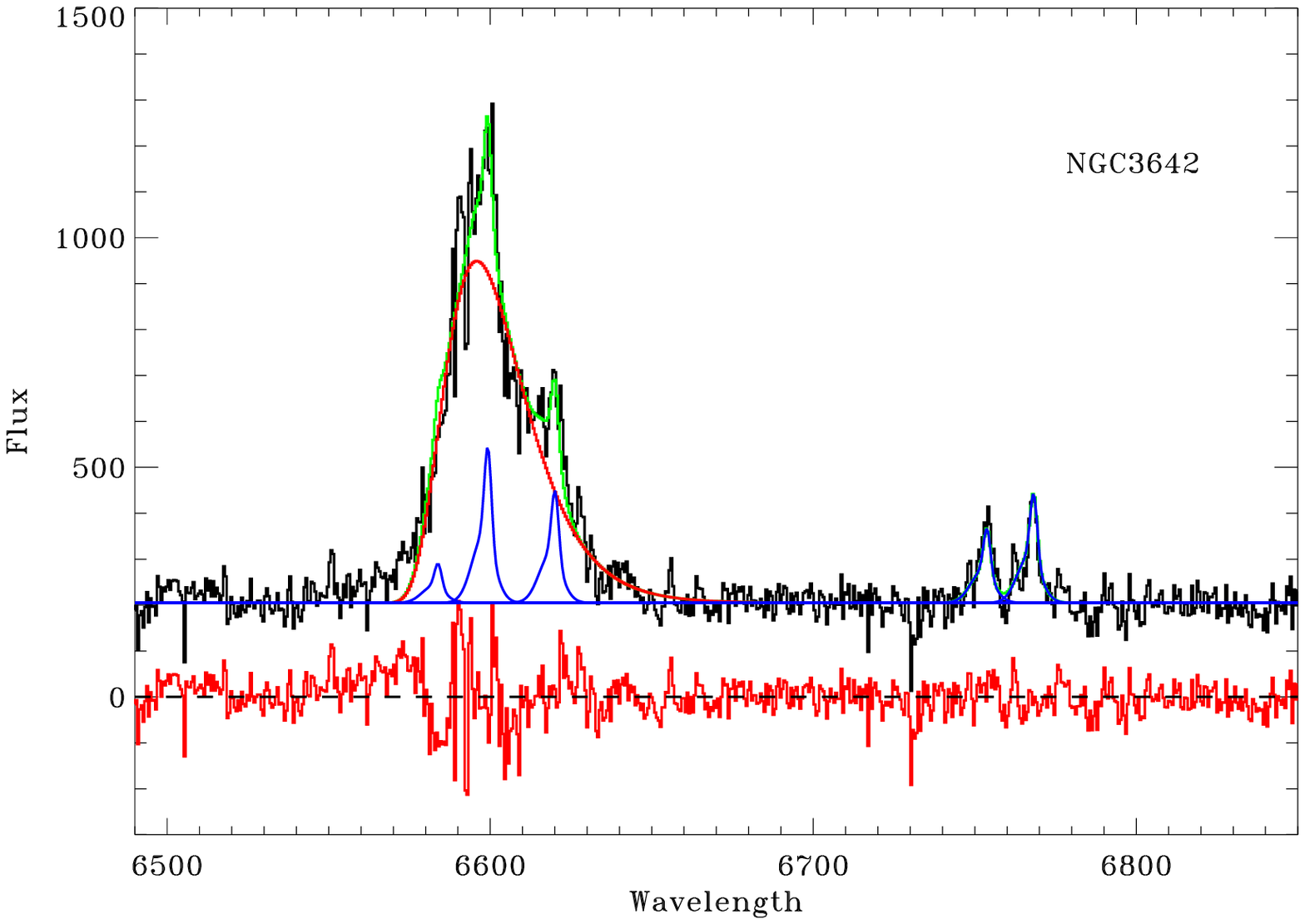}
\includegraphics[scale=0.26,angle=0]{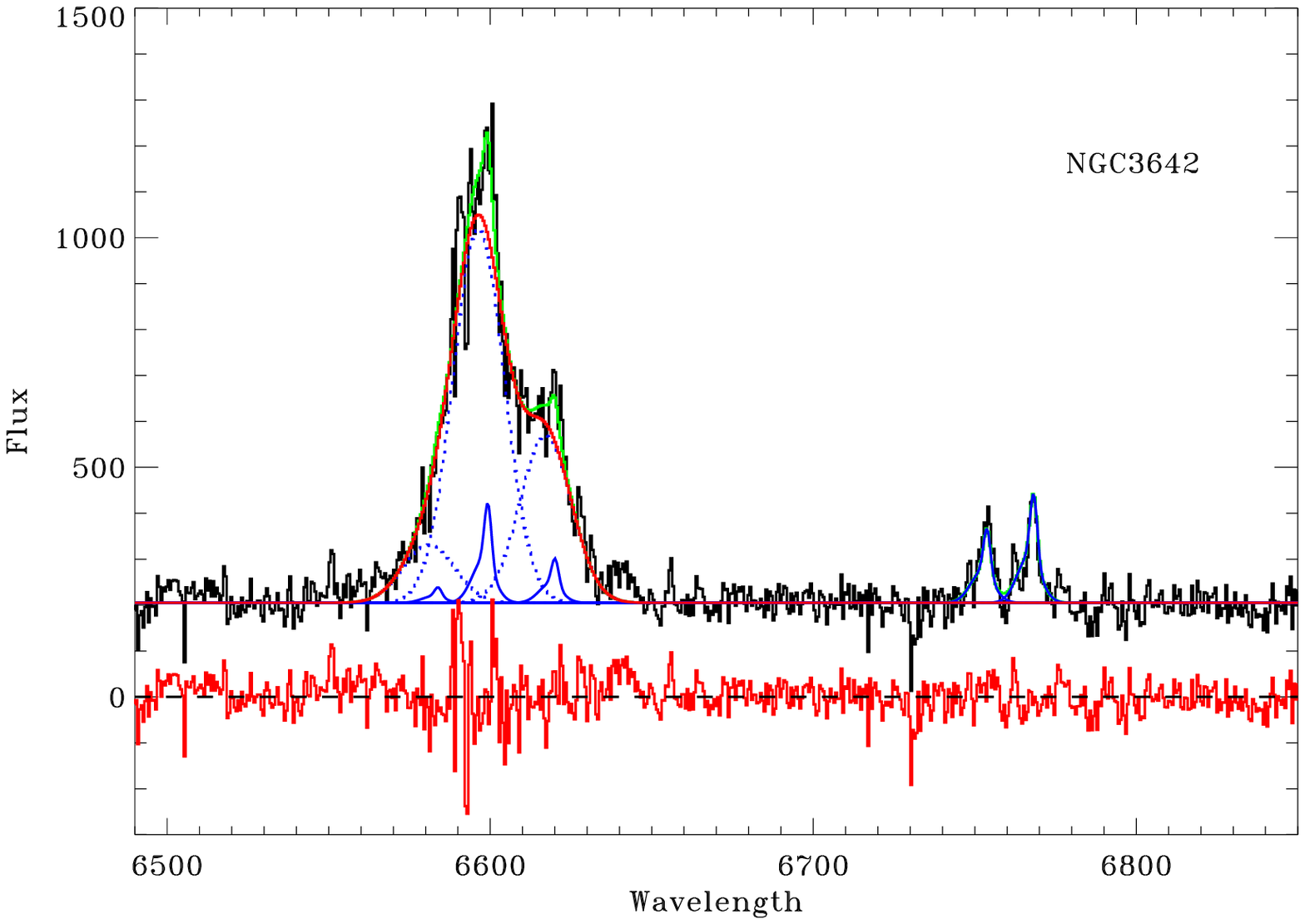}
\includegraphics[scale=0.26,angle=0]{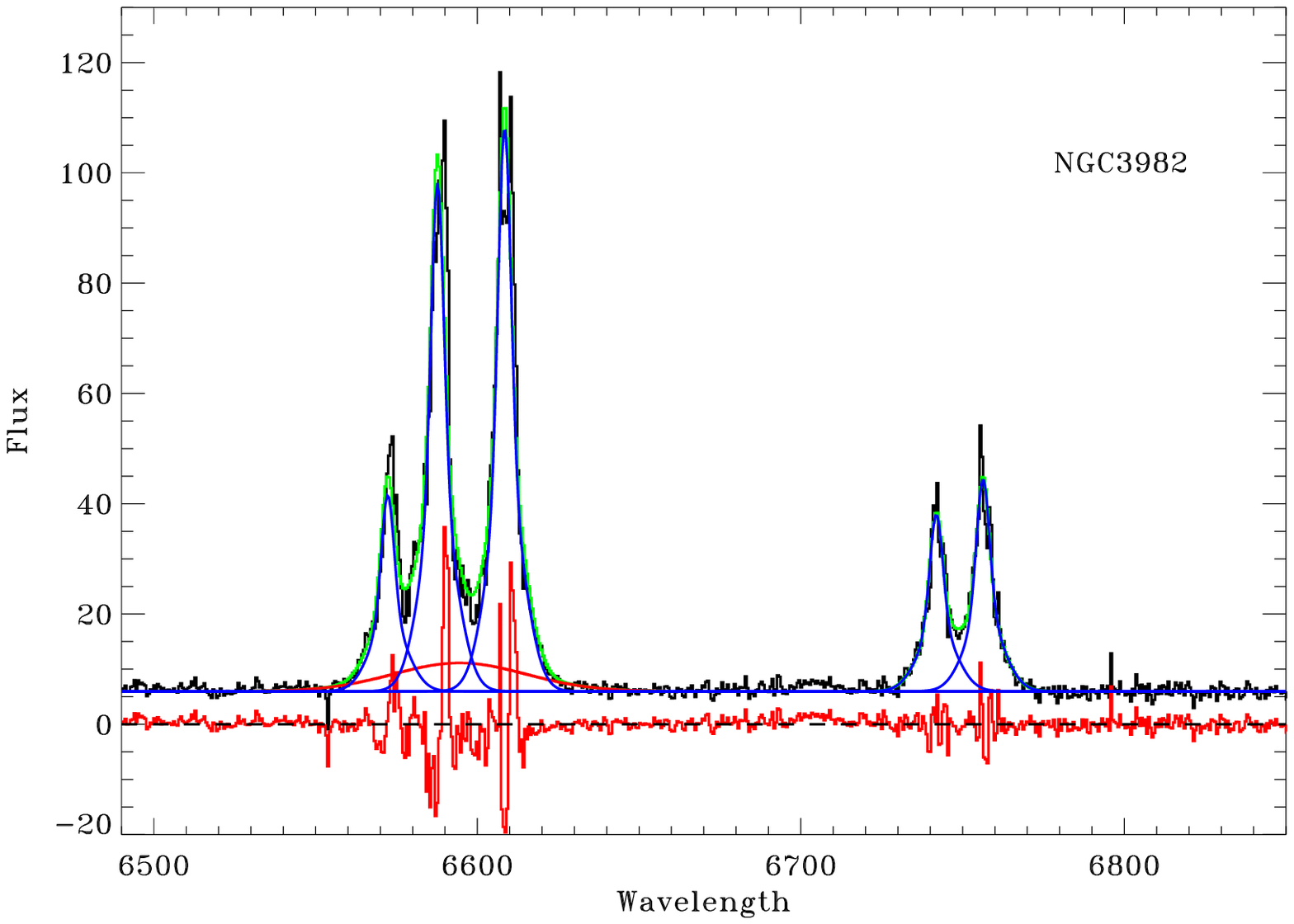}
\includegraphics[scale=0.26,angle=0]{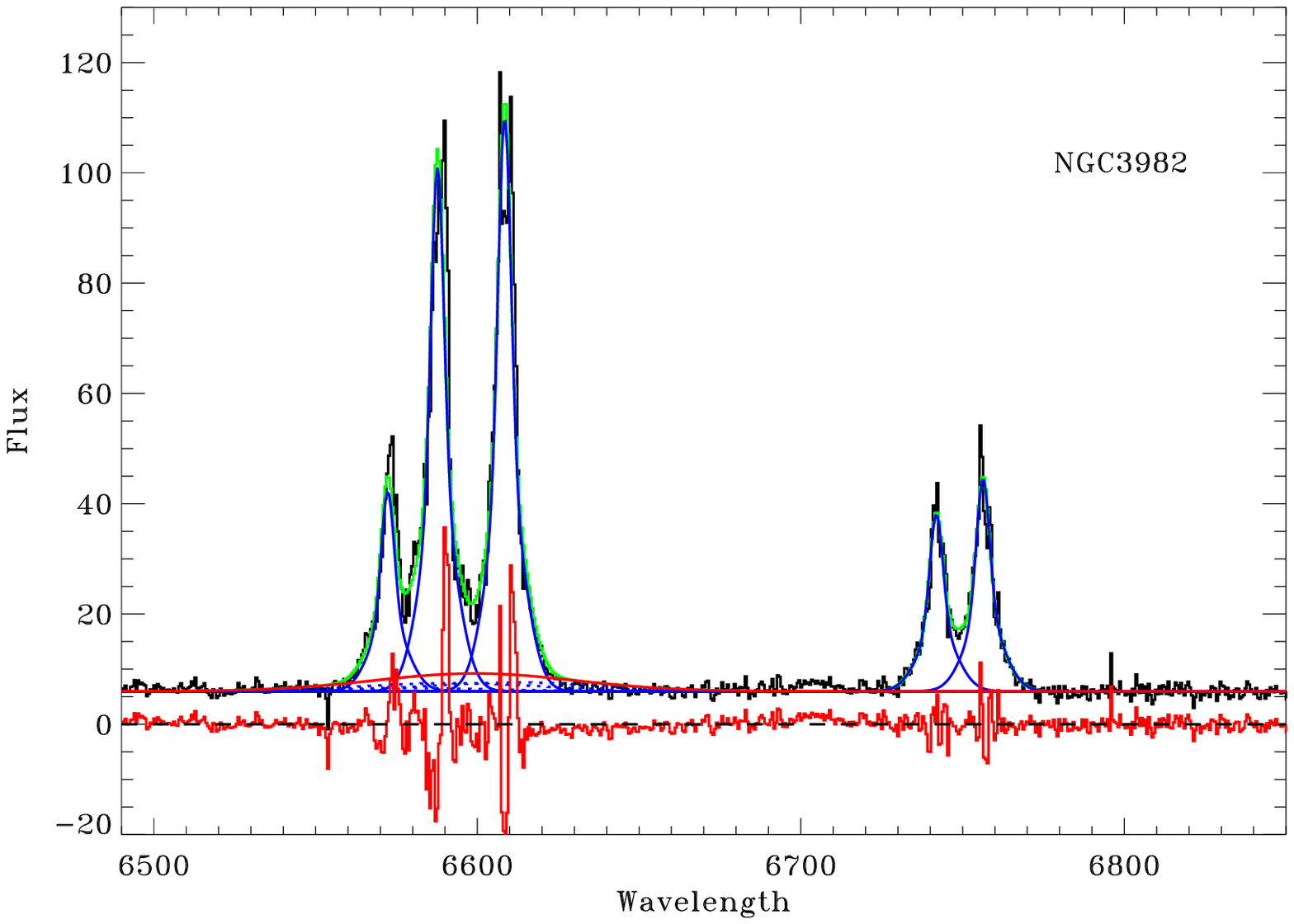}
}
\centerline{
\includegraphics[scale=0.26,angle=0]{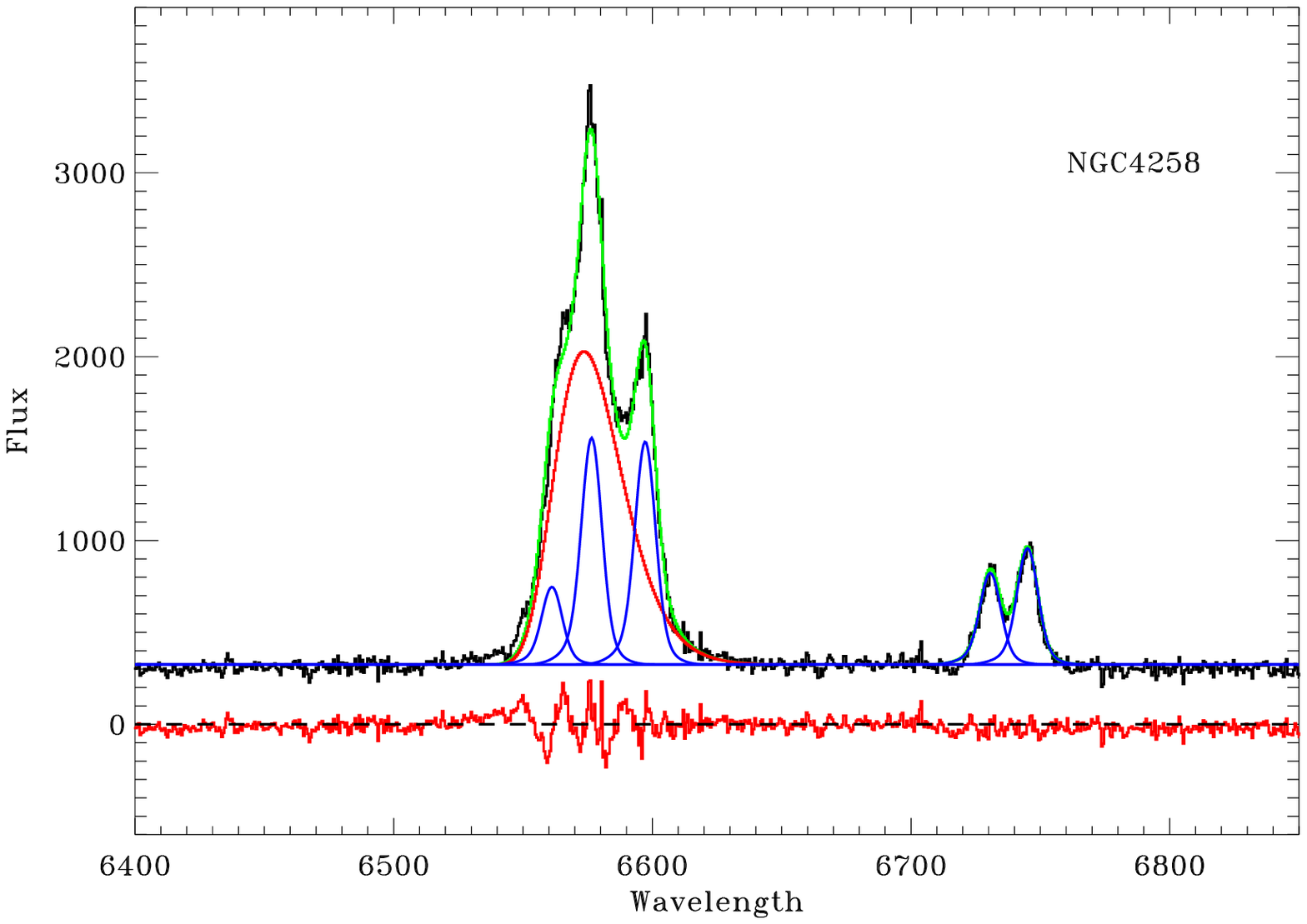}
\includegraphics[scale=0.26,angle=0]{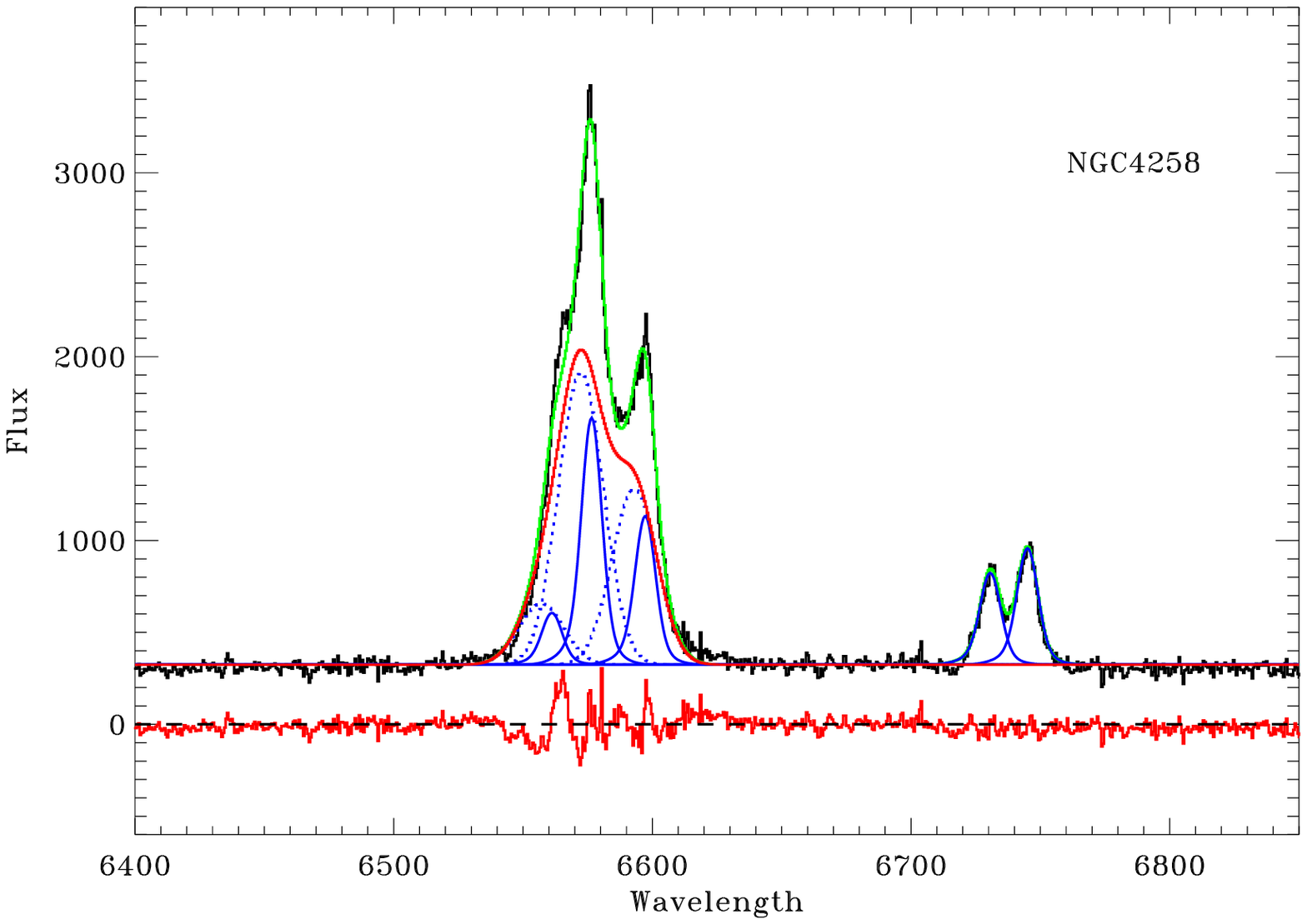}
\includegraphics[scale=0.26,angle=0]{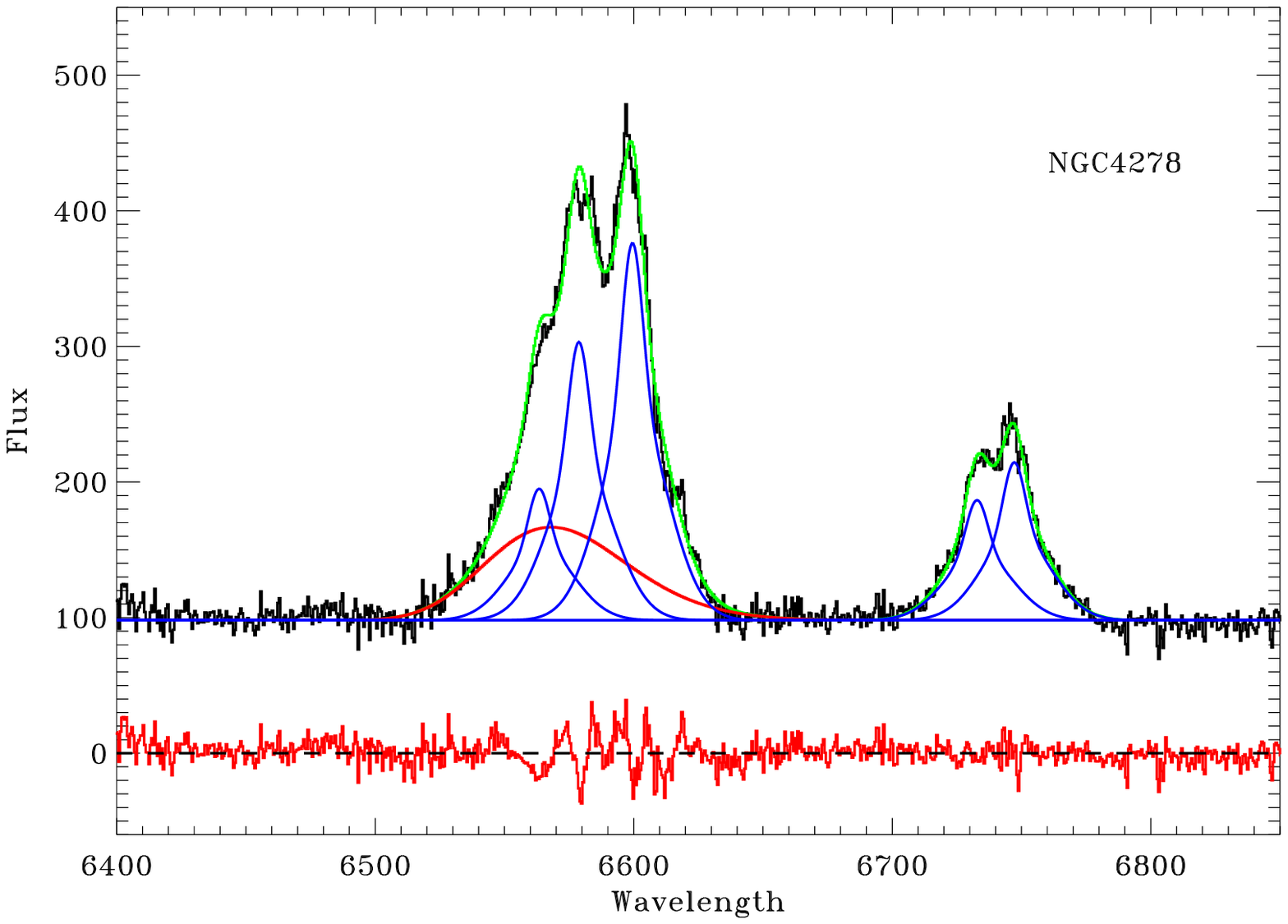}
\includegraphics[scale=0.26,angle=0]{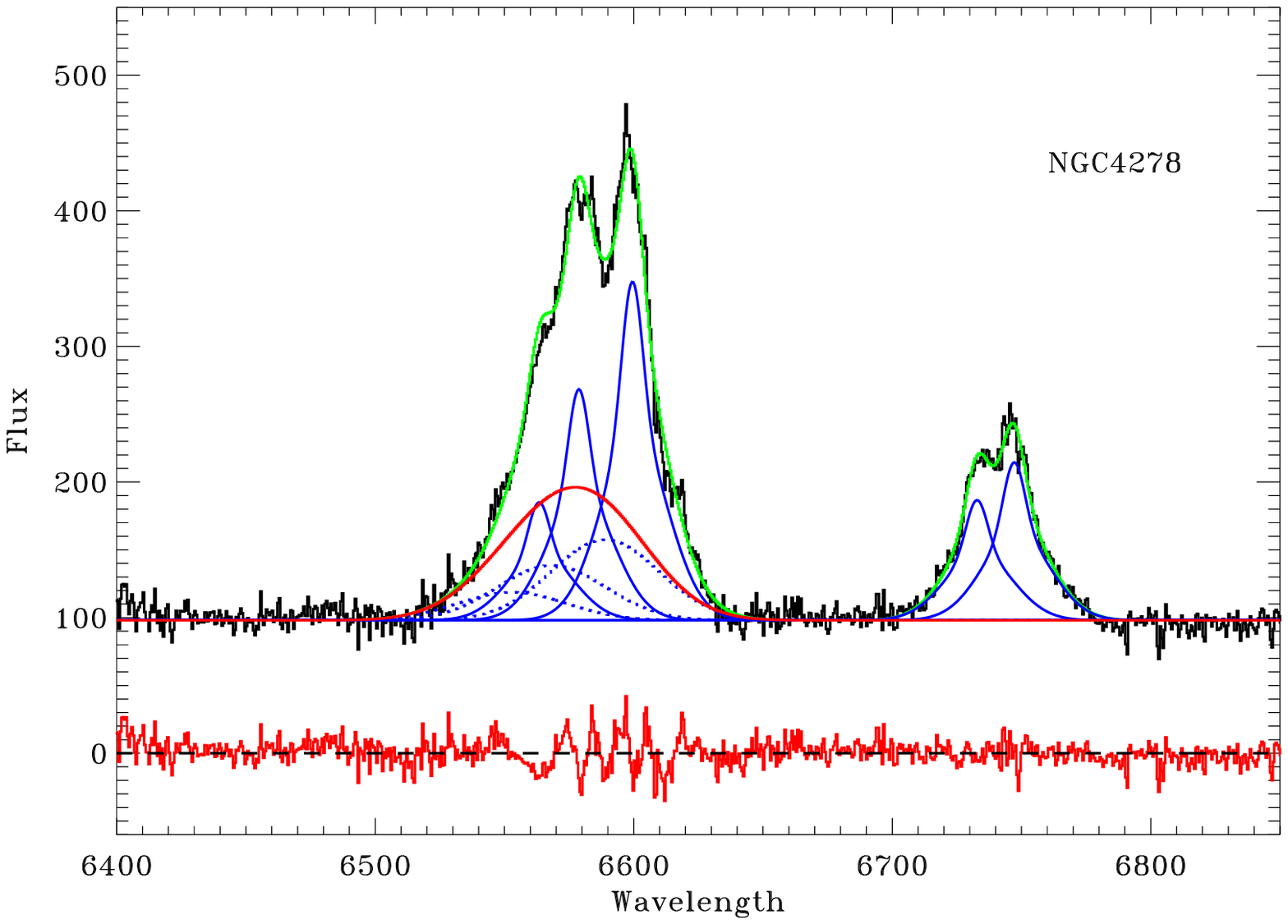}
}
\centerline{
\includegraphics[scale=0.26,angle=0]{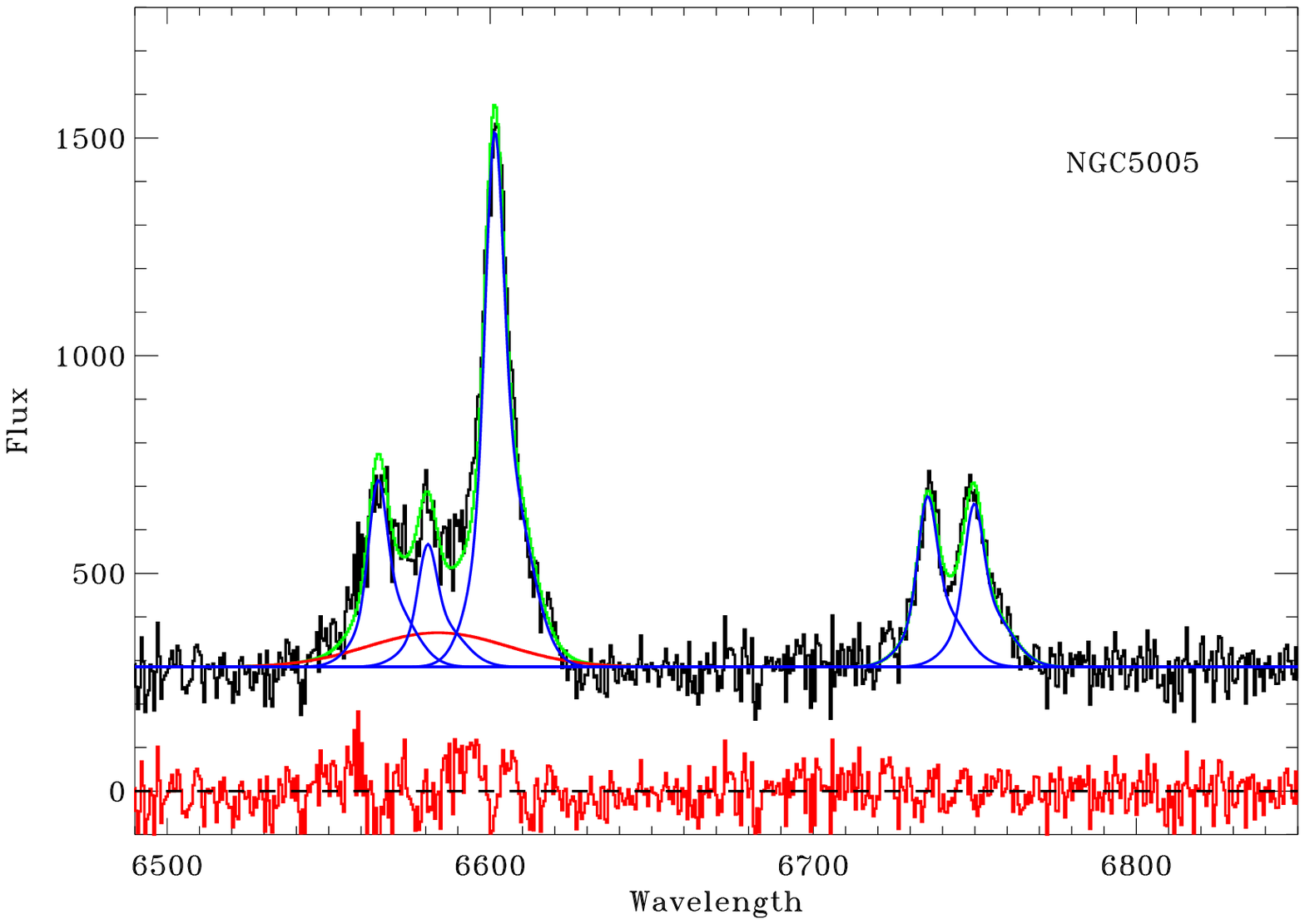}
\includegraphics[scale=0.26,angle=0]{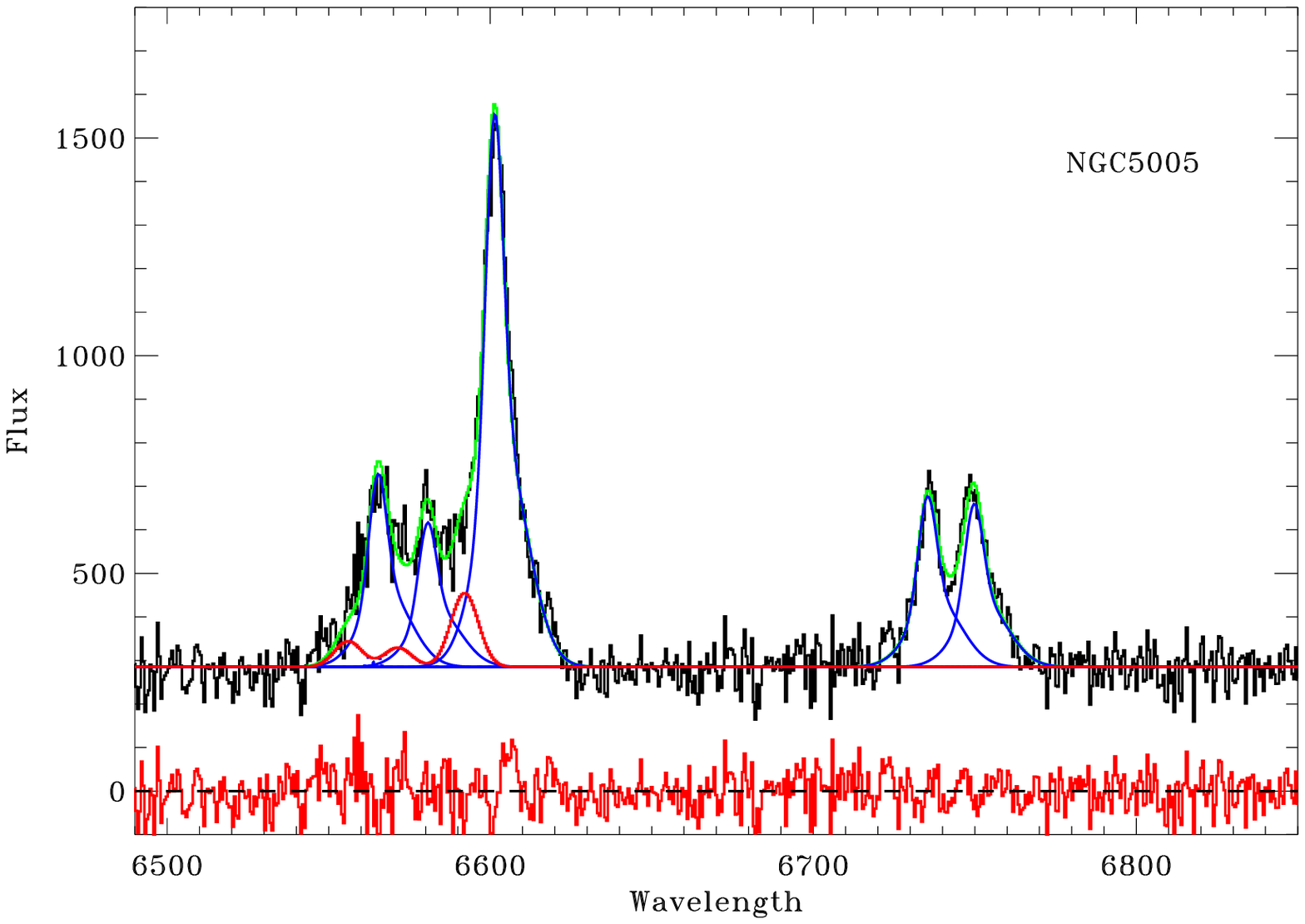}
\includegraphics[scale=0.26,angle=0]{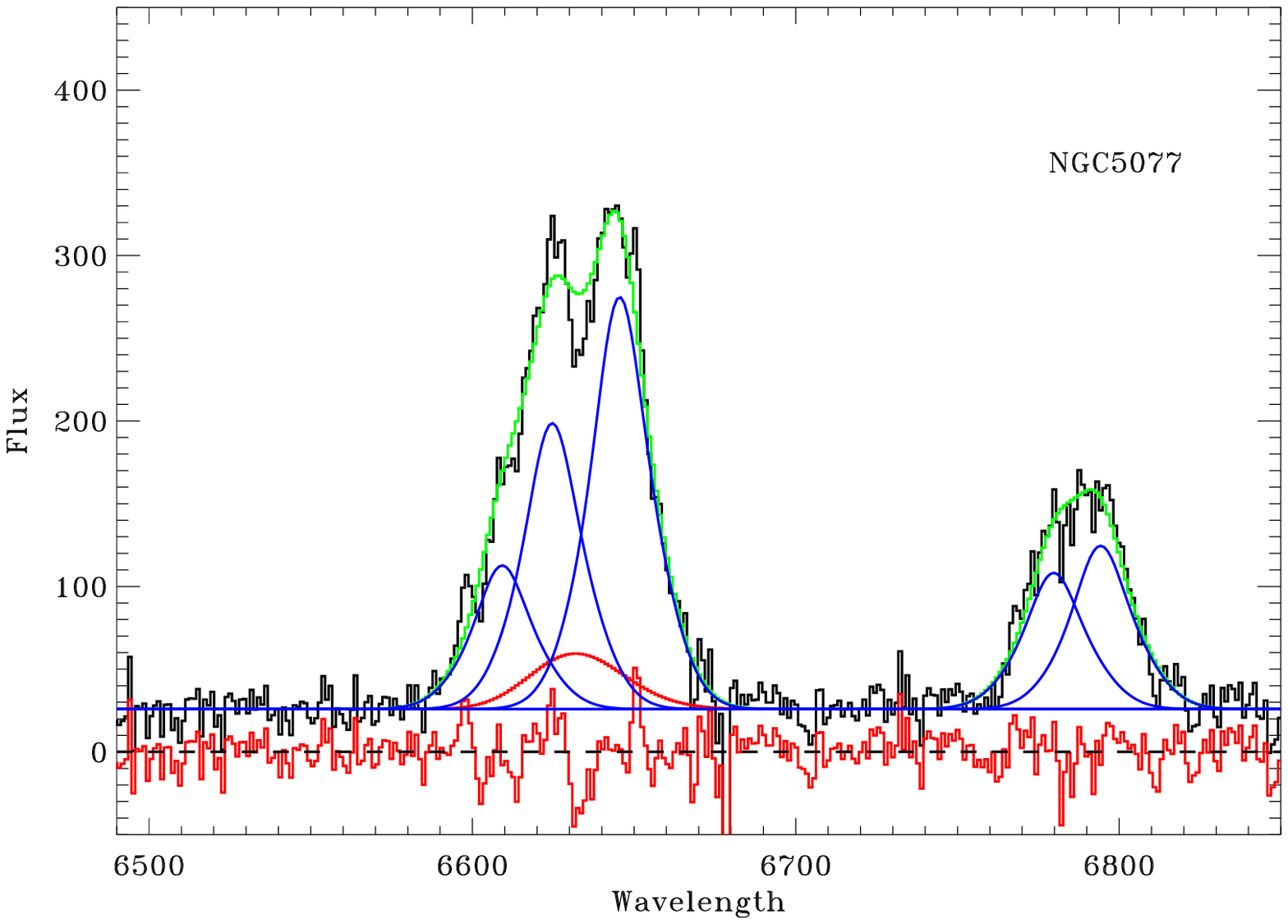}
\includegraphics[scale=0.26,angle=0]{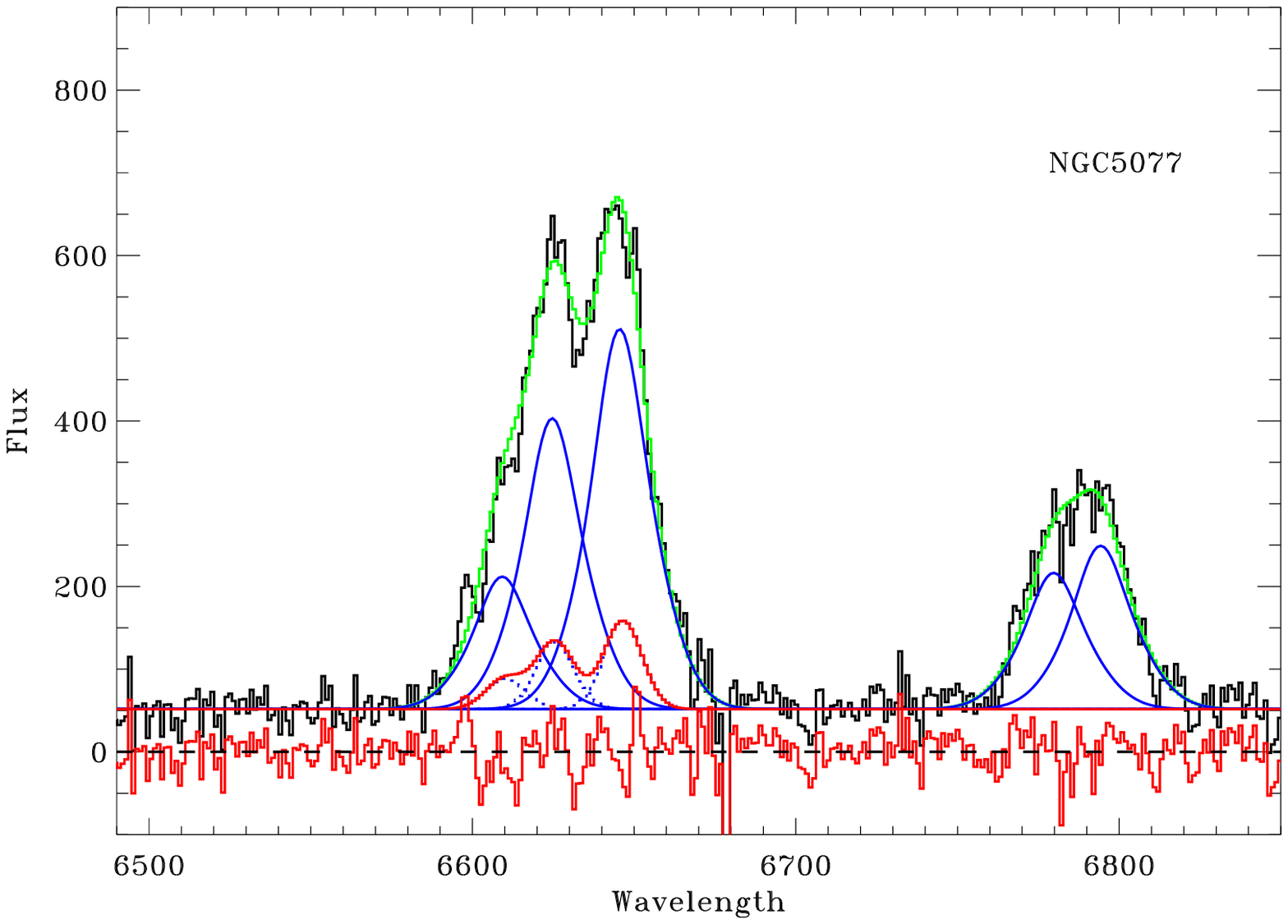}      
}
\caption{Analysis of the spectra 8 objects without a clear BLR and not
  covering the \oi\ region (or with faint \oi\ lines). Each spectrum is
  modeled with 2 methods, always using the \sii\ lines as templates and (left)
  including a broad \Ha\ component (right) or adding a wing to the \nii\ and
  narrow \Ha\ lines. The original spectrum is in black, the contribution of the
  individual narrow lines in blue (their sum is in green), and the
  residuals in red. In red we also show either the BLR component or the wings'
  contribution.}
\label{s2only2}
\end{figure*}

\end{document}